\documentclass[english,11pt, draftclsnofoot, peerreview, a4paper, oneside, onecolumn]{IEEEtran}
\usepackage[T1]{fontenc}
\usepackage[latin9]{inputenc}
\usepackage{verbatim}
\usepackage{bm}
\usepackage{amsmath}
\usepackage{amssymb}
\usepackage{graphicx}
\usepackage{esint}

\makeatletter

\providecommand{\tabularnewline}{\\}

  \newtheorem{asm}{Assumption}
  \newenvironment{asmQED}{\begin{asm}}{~\hfill\IEEEQEDclosed\end{asm}}
  \newtheorem{defQED}{Definition}
  \newenvironment{lyxDefQED}{\begin{defQED}}{~\hfill\IEEEQEDclosed\end{defQED}}
  \newtheorem{problem}{Problem}
  \newtheorem{remrk}{Remark}
  \newtheorem{thmQED}{Theorem}
  \newenvironment{lyxThmQED}{\begin{thmQED}}{~\hfill\IEEEQEDclosed\end{thmQED}}
  \newtheorem{corQED}{Corollary}
  \newenvironment{lyxCorQED}{\begin{corQED}}{~\hfill\IEEEQEDclosed\end{corQED}}
  \newtheorem{lemQED}{Lemma}
  \newenvironment{lyxLemQED}{\begin{lemQED}}{~\hfill\IEEEQEDclosed\end{lemQED}}

\usepackage{cite}
\usepackage{subfigure}

\author{Junting~Chen~\IEEEmembership{Student~Member,~IEEE}        and~Vincent~K.~N.~Lau,~\IEEEmembership{Fellow,~IEEE}\\
Dept. of Electronic and Computer Engineering \\The Hong Kong University of Science and Technology\\Clear Water Bay, Kowloon, Hong Kong\\ Email: \{eejtchen, eeknlau\}@ust.hk
}

\makeatother

\usepackage{babel}
\begin{document}

\title{Convergence Analysis of Mixed Timescale Cross-Layer Stochastic Optimization }
\maketitle
\begin{abstract}
This paper considers a cross-layer optimization problem driven by
multi-timescale stochastic exogenous processes in wireless communication
networks. Due to the hierarchical information structure in a wireless
network, a mixed timescale stochastic iterative algorithm is proposed
to track the time-varying optimal solution of the cross-layer optimization
problem, where the variables are partitioned into short-term controls
updated in a faster timescale, and long-term controls updated in a
slower timescale. We focus on establishing a convergence analysis
framework for such multi-timescale algorithms, which is difficult
due to the timescale separation of the algorithm and the time-varying
nature of the exogenous processes. To cope with this challenge, we
model the algorithm dynamics using stochastic differential equations
(SDEs) and show that the study of the algorithm convergence is equivalent
to the study of the stochastic stability of a virtual stochastic dynamic
system (VSDS). Leveraging the techniques of Lyapunov stability, we
derive a sufficient condition for the algorithm stability and a tracking
error bound in terms of the parameters of the multi-timescale exogenous
processes. Based on these results, an adaptive compensation algorithm
is proposed to enhance the tracking performance. Finally, we illustrate
the framework by an application example in wireless heterogeneous
network. \end{abstract}
\maketitle
\IEEEpeerreviewmaketitle

\section{Introduction}

Cross-layer resource optimization plays a critical role in the radio
resource management of modern wireless systems. In existing literature,
cross-layer optimization can be divided into two categories. When
the system states are slowly varying, it is desirable to have dynamic
controls adaptive to the instantaneous realizations of system state.
For example, in \cite{Chen:2006nx,Georgiadis2006}, the authors considered
dynamic power control (adaptive to the instantaneous channel fading
state) in wireless ad hoc networks. In \cite{Palomar:2003vn,Zhang:2008ys},
adaptive joint beam-forming is considered to mitigate interference
in a cellular network. In practice, it is quite difficult to obtain
real-time observations of the global system states or real-time signaling
message passing in a large scale network because of the signaling
latency%
\footnote{For example, the X2 interface in e-Node B of LTE systems has latency
of 10ms or more.%
}. As a result, it is desirable to adapt the control actions to the
system state statistics instead of real-time realizations. For example,
in \cite{Ghosh:2012kl} the author developed a limited feedback technique
that utilizes the channel distribution information (CDI) for communication
in multiuser MIMO beamforming networks. In \cite{Wajid:2009ij,Bornhorst:2012tg},
the problem of robust transmit beamforming in multi-user communication
system using the covariance-based channel information is considered.

\begin{figure}
\begin{centering}
\includegraphics[width=0.8\columnwidth]{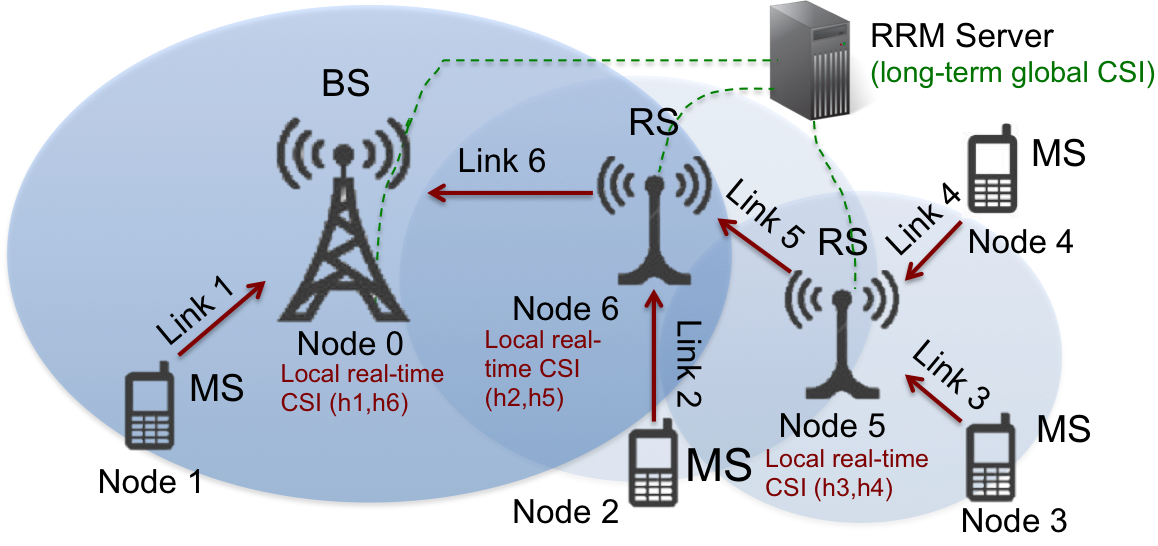}
\par\end{centering}

\caption{\label{fig:topology-relay-network} The topology of a wireless relay
network with a radio resource management (RRM) server. The BSs and
the RSs have local real-time CSI, while the RRM server has the long-term
global CSI.}
\end{figure}

In practice, system states usually evolve in mixed timescales in wireless
networks. For example, in MIMO fading channels, the channel matrix
changes in a short timescale (such as $10$ ms) but the correlation
and the path loss change in a longer timescale (such as minutes) \cite{Feng2007,mostofi2009characterization}.
Another example of mixed timescale state evolution is between the
queue length process (slower timescale) and the instantaneous link
quality (faster timescale) \cite{Munish03,Neely2008}. When the system
has a multi-timescale state evolution, it is necessary to partition
the controls in different timescales based on the \emph{information
structure} induced by the system topology. As an illustration, consider
a wireless heterogeneous network with a macro base station (BS) and
some relay BSs (RSs) as depicted in Fig. \ref{fig:topology-relay-network}.
The users transmit data flows to the macro BS with the assistance
of the RSs. The control strategies depend on the channel state which
evolves in mixed timescales. Suppose we want to adopt a cross-layer
control to the flow data rate $\mathbf{r}$ for each user and the
transmission power $\mathbf{p}$ on each link so as to maximize the
average throughput in the example wireless network with time-varying
channels. As the control policy may involve network-wise coordinations,
one good strategy is to partition the control variables into local
power control $\mathbf{p}$ and global flow control $\mathbf{r}$,
where the power control $\mathbf{p}$ adapts to the instantaneous
channel state information (CSI) locally and the flow rate $\mathbf{r}$
adapts to the CSI statistics with a global coordination. This is because,
while it is realistic for each wireless node to acquire real-time
local CSI, it is extremely difficult for the network controller to
acquire real-time global CSI. If one considers pure fast timescale
control for both the power and the flow data rate (such as in \cite{Chiang2005,Georgiadis2006}),
the policy obtained will require real-time global CSI. This is difficult
to achieve in practice and the system performance will be very sensitive
to the signaling latency in the acquisition of global CSI. On the
other hand, if one considers pure slow adaptation for both $\mathbf{p}$
and $\mathbf{r}$ (statistical adaptation), the resulting policy will
fail to exploit the instantaneous transmission opportunity observed
at each wireless node, and such an approach may be too conservative.
Therefore, it is of great importance to have a complete cross-layer
control framework with timescale separations that embraces the exogenous
mixed-timescale state evolutions and exploits the information structure
of the network topology. 

There are quite a few works that studied controls with different timescale
state evolutions \cite{Gomez:1999cr,Lin:2002oq,Chen:2006nx,Papandriopoulos:2008kl,Soldati:2009tg,Zheng:2009hc}.
However, these works handled the different timescales separately in
a heuristic manner%
\footnote{For example, the fast control and slow control are not optimizing
the same optimization objective.%
}. There are few works that considered a holistic cross layer optimization
framework exploiting mixed timescale algorithms, not to mention the
study of convergence properties of such mixed-timescale algorithms. 

In this paper, we focus on the study of general cross-layer optimization
for mixed timescales state processes. We first setup a stochastic
optimization formulation to optimize an average network utility. The
control variables are partitioned into a \emph{short-term control}
(adapts to fast timescale state processes) and a \emph{long-term control}
(adapts to slow timescale state processes) according to the information
structure induced by specific network topology. These control variables
are driven by mixed-timescale iterative algorithms to optimize the
average network utility (objective function). An important question
to the mixed timescale iterative algorithms is whether they will converge
to the optimal solution. While the convergence of some iterative algorithms
such as stochastic gradient \cite{Benaim:1999ly,Kushner2003vn} are
quite well-studied, these existing techniques considered one timescale
iteration only and the convergence behavior of mixed timescale algorithms
is highly non-trivial due to the mutual coupling between the short-term
and long-term control variables. Specifically, there are several first
order technical challenges that need to be addressed. 
\begin{itemize}
\item \textbf{Coupling in the dynamics of the long-term and short-term}
\textbf{iterations}: In most of the existing works involving multi-timescale
control variables, the iterative algorithms driving different types
of control variables are independent from each other. In other words,
the intermediate iterates of the long-term variables in the outer-loop
will not affect the convergence of the short-term variables in the
inner loop. However, this decoupling is only justified when the short-term
variables can converge to the optimal point arbitrarily fast or when
we have closed-form solutions for the short-term variables. In general,
we do not have closed-form solutions and each iteration in communication
networks may also involve explicit signaling message passing. Hence,
it may not be realistic to ignore the iteration dynamics in the short-term
variables. Due to these couplings of the iterations between the long-term
and short term variables, classical convergence proof in stochastic
gradient \cite{Benaim:1999ly,Kushner2003vn} or stochastic programming
\cite{Birge:1997ve,Santoso:2005bh,Shapiro:2009zr} cannot be applied. 
\item \textbf{Irreducible bias to the stochastic estimator}: In standard
single timescale stochastic optimization with long-term control variables
only, stochastic gradient \cite{Benaim:1999ly,Kushner2003vn} is commonly
used because the true gradient of the problem contains an expectation
operator which does not have closed form expression in most cases.
Using an \emph{unbiased stochastic gradient estimator} \cite{Benaim:1999ly,Kushner2003vn},
we do not need to compute the true gradient in every iteration. If
the original problem is strictly convex, the stochastic gradient algorithm
will converge to the optimal solution \cite{Benaim:1999ly,Kushner2003vn}.
However, in the mixed timescale iterations, the convergence errors
of the short-term control variables in the inner loop will induce
an irreducible bias to the stochastic gradient estimator of the long-term
control variables in the outer-loop. As such, standard convergence
proof in stochastic gradient \cite{Benaim:1999ly,Kushner2003vn} cannot
be applied. 
\item \textbf{Exogenous Stochastic Variations of the State Processes}: In
addition to the coupling in the mixed timescale iterations as well
as the irreducible bias due to the short-term control iterations,
the system states of the wireless system are also evolving stochastically.
For example, the long-term state process (such as the channel path
loss, the MIMO channel correlations or even the network topology)
may be time-varying due to the mobility of the users in the network
or shadowing process. As such, these exogenous variations will also
have a profound impact on the convergence behavior of the mixed timescale
iterations. 
\end{itemize}

\subsection{Related words}

In \cite{Chen2012:Saddle,Chen2013:Delay}, we have studied the convergence
behavior of iterative algorithms in wireless systems under time-varying
channel states. However, these works have focused on one-timescale
iterations and the approach cannot be applied to address the above
challenges in mixed-timescale iterations. There are also limited works
on multi-timescale stochastic optimizations. In \cite{Birge:1997ve,Santoso:2005bh,Shapiro:2009zr},
multi-stage stochastic programming has been applied for logistic and
financial planning problems. However, the problem considered has very
a special form (linear program (LP)) and the solutions cannot be applied
to our problems, which has a more general non-linear form. There are
also some application examples \cite{Gomez:1999cr,Lin:2002oq,Chen:2006nx,Papandriopoulos:2008kl,Soldati:2009tg,Zheng:2009hc}
that decompose the stochastic optimization problems into two timescale
hierarchical solutions. However, all these works did not consider
the tracking issues associated with exogenous variations of the problem
parameters. In \cite{Yin04,Yin05fk}, the authors studied the stochastic
tracking algorithms in a regime-switching environment which is driven
by a finite state two-timescale Markov chain. A dynamic step-size
selection algorithm for the tracking in a regime-switching environment
has been proposed in \cite{Costa:2007dq}. Nevertheless, a general
understanding of the convergence behavior of mixed-timescale algorithms
is still needed.

\subsection{Our contribution}

In this paper, we propose an analysis framework to study the convergence
behavior of mixed timescale iterative algorithms of cross-layer stochastic
optimization for wireless networks. Specifically, we first introduce
the cross-layer stochastic optimization framework and the variable
partitioning according to the information structure available. We
then consider a mixed timescale iterative algorithm and study the
dynamics and coupling using continuous time stochastic differential
equation (SDE). We show that the study of the convergence behavior
of the mixed timescale algorithm is equivalent to the study of \emph{stochastic
stability} of a virtual dynamic system specified by a system of SDE.
Furthermore, the optimal solution of the stochastic optimization problem
is equivalent to an \emph{equilibrium point} of the virtual dynamic
system. As a result, the convergence behavior of the algorithm is
similar to the control problem of tracking a moving target \cite{Chen2012:Saddle,Chen2013:Delay}.
Based on this insight, we derive an upper bound on the tracking error
of the mixed timescale algorithms under exogenous variations of the
short-term and long-term state processes. Based on the insights of
the impact of exogenous variations, we propose a low complexity compensation
algorithm which can substantially enhance the tracking behavior of
the mixed timescale algorithms. Finally, we apply this framework to
an example topology in wireless heterogeneous networks with relays.
Numerical simulations verified the theoretical insights obtained and
also demonstrated significant performance gain of the proposed compensation
algorithms. 

The paper is organized as follows. Section \ref{sec:system-model}
illustrates the system model, where the cross-layer stochastic optimization
framework and the mixed timescale stochastic approximation algorithm
will be introduced. Section \ref{sec:convergence-VSDS} studies the
tracking behavior of the mixed timescale algorithm using the notion
of virtual stochastic dynamic system (VSDS) and the techniques of
Lyapunov stochastic stability. In Section \ref{sec:adaptving-algorithm},
a novel compensation algorithm is proposed to enhance the tracking
performance. An application example on flow control and power allocation
in wireless relay network is illustrated in Section \ref{sec:Example}.
Section \ref{sec:numerical-results} gives the numerical results,
and we conclude the work in Section \ref{sec:conclusions}.

\emph{Notations}: For a scaler-valued function $F:\mathbb{R}^{n}\mapsto\mathbb{R}$,
$F_{x}$ denotes the vector of its partial derivatives with respect
to vector $x=(x_{1},\dots,x_{n})$, i.e., the $i$-th component of
$F_{x}$ is $F_{x}^{(i)}=\frac{\partial F}{\partial x_{i}}$. For
a vector-valued function $G:\mathbb{R}^{n}\mapsto\mathbb{R}^{m}$,
$G_{x}$ denotes the Jacobian matrix of $G=(G^{(1)},\dots,G^{(m)})$,
which is the partial derivatives of $G(\centerdot)$ with respect
to vector $x$, and whose $(i,j)$-th element is given by $\frac{\partial G^{(i)}}{x_{j}}$.
The notation $\lfloor x\rfloor$ denotes the largest integer that
is no greater than $x$. For column vectors $x=[x_{1}\,\dots\, x_{n}]^{T}$
and $y=[y_{1}\,\dots\, y_{m}]^{T}$, $(x,y)=[x_{1}\,\dots\, x_{n}\, y_{1}\,\dots\, y_{m}]^{T}$
denotes a column vector that stacks the vectors $x$ and $y$.

\section{System Model}

\label{sec:system-model}

In this section, we first introduce the cross-layer stochastic optimization
framework with mixed timescale exogenous random processes. We then
partition the control variables into short-term and long-term controls
and describe the mixed timescale iterative algorithms to track the
optimal solutions of the stochastic optimization problem. Based on
that, we elaborate the convergence issues induced by the exogenous
random processes and formally define the tracking error between the
algorithm outputs and the optimal solution.

\subsection{A Cross-Layer Stochastic Optimization Formulation}

\label{sub:two-timescale-control}

\subsubsection{Network and Mobility Model}

\label{sub:network-mobility-model}

We first discuss the system model in a wireless communication network
with node mobilities. Consider a wireless network with $N_{s}$ static
nodes and $N_{m}$ mobile nodes. The location of the static nodes
are fixed, and the mobile nodes are randomly distributed at time $t=0$.
For time $t>0$, the mobile nodes change their locations according
to a widely adopted \emph{Levy walk mobility model} \cite{Bansal:2003kl,Lee:2009uq,Rhee:2011fk}
described as follows. 
\begin{asmQED}
[Levy walk mobility model]\label{asm:Random-way-point} Starting
from time $t=0_{+}$, each mobile node chooses a random destination
in a restricted region centered at the initial location and moves
at a constant speed in $(0,v_{\max}]$. Upon reaching the destination,
the node pauses for some time and randomly chooses a new destination
and speed to go on. The travel distance and pause time at each step
follow a truncated Levy distribution \cite{Lee:2009uq,Rhee:2011fk}.
\end{asmQED}

Nodes are inter-connected through wireless links, and since the node
mobility is restricted, we assume that the network topology is fixed
despite the mobility of the mobile nodes. The network topology can
be represented by a directed graph $\mathcal{G}=(\mathcal{N},\mathcal{L})$,
where $\mathcal{N}$ is the set of nodes and $\mathcal{L}$ is the
set of wireless links that connect the transmitters and the receivers.
Fig. \ref{fig:topology-relay-network} illustrates an example wireless
network, where $\mathcal{N}$ is the set of BS, RSs and mobile users,
and $\mathcal{L}$ is the set of wireless transmission links between
them.

Define the CSI for the $j$-th link as $h_{j}\in\mathcal{H}$. We
adopt a fading model to the CSI $h_{j}$ as $h_{j}=h_{j}^{l}h_{j}^{s}$,
where $h_{j}^{l}=c_{0}D_{j}^{-\iota}\in\mathcal{H}^{l}$ is the long-term
CSI for the large-scale fading with $c_{0}$ being an antenna-gain-related
constant, and $h_{j}^{s}\in\mathcal{H}^{s}$ is the short-term CSI
for the small-scale fading \cite{Tse2005:fundamental:Wireless,mostofi2009characterization}.
$D_{j}\geq D_{\min}$ is the distance between the $j$-th link, and
$\iota$ is the path loss exponent.

\subsubsection{The CSI Dynamics}

\label{sub:CSI-model}

We specify the dynamics of the CSI $h_{j}^{s}$ and $h_{j}^{l}$ by
the exogenous random processes defined below.
\begin{lyxDefQED}
[Exogenous stochastic processes]\label{asm:two-timescale-model}
The short-term and long-term CSI processes $h_{j}^{s}(t)$ and $h_{j}^{l}(t)$
are driven by the following stochastic differential equations (SDE):
\begin{eqnarray}
dh_{j}^{s} & = & -\frac{1}{2}a_{H}h^{s}dt+\sqrt{a_{H}}dW_{t},\quad h_{j}^{s}(0)=h_{0}^{s},\forall j=1,\dots,N,\label{eq:sde-hs}\\
dh_{j}^{l} & = & -c_{0}\iota D_{j}(t)^{-\iota-1}v_{j}(t)dt,\quad h_{j}^{l}(0)=h_{0}^{l},\forall j=1,\dots,N,\label{eq:sde-hl}
\end{eqnarray}
where $a_{H}>0$, $v_{j}$ is the relative speed of the $j$-th link,
and $W_{t}$ is a standard Brownian motion.
\end{lyxDefQED}

The positive parameter $a_{H}$ specifies the time-correlation of
the short-term exogenous processes $\{h_{j}^{s}(t)\}$, which are
specified by the Ornstein-Uhlenbeck processes. $h_{j}^{s}(t)$ has
a Gaussian stationary distribution and $|h_{j}^{s}|$ follows a Rayleigh
distribution \cite{Feng2007,mostofi2009characterization}. On the
other hand, defining $\epsilon(D_{\min},v_{\max})\triangleq2c_{0}\iota D_{\min}^{-\iota-1}v_{\max}$,
we have $|dh_{j}^{l}|\leq\epsilon(D_{\min},v_{\max})dt$. The parameter
$\epsilon$, which is typically very small, represents the timescale
of the long-term processes $\{h_{j}^{l}(t)\}$.

\subsubsection{Control Variables and the Problem Formulation}

We assume the following information structure for the wireless communication
network.
\begin{asmQED}
[Signaling information structure]\label{asm:information-structure}
Each node $k\in\mathcal{N}$ has knowledge of the \emph{local CSI}
(short-term and long-term) $(h_{j}^{s},h_{j}^{l})$ for all the links
$j\in\mathcal{L}$ that connect to node $k$. On the other hand, only
the global long-term CSI $h^{l}=(h_{1}^{l},\dots,h_{N_{L}}^{l})$
is known at the central controller of the network. 
\end{asmQED}

For example, in Fig. \ref{fig:topology-relay-network}, node $5$
only has the local CSI knowledge of $h_{3}$, $h_{4}$ and $h_{5}$.
Meanwhile, the RRM server has the long-term global CSI $h_{j}^{l}$
for all the nodes $j$. Note that the above assumption is quite reasonable,
because in practical wireless communication networks, it is relatively
easy for each node to acquire local real-time CSI but it would be
difficult for the network controller to acquire the global real-time
CSI. 

According to the information structure assumption, the following defines
the mixed timescale controls in the cross-layer optimization framework.
\begin{lyxDefQED}
[Mixed Timescale Controls]\label{def:Mixed-Timescale-Controls}
The system has two sets of control variables, namely the \emph{short-term
control} and the \emph{long-term control}. The short term control
is denoted by a policy $\Omega^{s}$ which maps the realization of
CSI vector $h=(h_{1},\dots,h_{N})$ to an action $\theta_{x}\in\mathbb{R}_{+}^{N_{x}}$.
Similarly, the long-term control is denoted by a policy $\Omega^{l}$
which maps the realization of $h^{l}=(h_{1}^{l},\dots,h_{N}^{l})$
to an action $\theta_{y}\in\mathbb{R}_{+}^{N_{y}}$ .
\end{lyxDefQED}

For example, the short-term control $\theta_{x}(t)=\Omega^{s}(h(t))$
and the long-term control $\theta_{y}(t)=\Omega^{l}(h^{l}(t))$ may
correspond to the transmit power control on each wireless link and
the flow control of a wireless network, respectively. The mixed timescale
cross-layer stochastic optimization problem is given as follows.
\begin{problem}
[Mixed timescale cross-layer stochastic optimization problem]\label{pro:the-problem}

\begin{eqnarray}
\mathcal{P}_{0}(a_{H},\epsilon)=\max_{\Omega^{l},\Omega^{s}} & \mathbb{E}\left[F(\theta_{x},\theta_{y};h(t))\right]\label{eq:the-problem}\\
\mbox{subject to} & w_{i}(\theta_{x},\theta_{y};h(t))=0, & i=1,\dots,W_{e}\label{eq:the-problem-constraint-1}\\
 & w_{i}(\theta_{x},\theta_{y};h(t))\leq0, & i=W_{e}+1,\dots,W\label{eq:the-problem-constraint-2}\\
 & q_{j}(\theta_{y};h^{l}(t))\leq0, & j=1,\dots J,\forall h^{l}(t).\label{eq:the-problem-constraint-3}
\end{eqnarray}
~\hfill\IEEEQED
\end{problem}

$\mathcal{P}_{0}$ can be decomposed into a family of \emph{inner
problems} and \emph{outer problems}:\emph{ }
\begin{problem}
[The inner problem] For given $h^{s}$ and $h^{l}$, 

\begin{eqnarray}
\mathcal{P}_{1}(\theta_{y},h^{s},h^{l})=\max_{\theta_{x}\geq0} & F(\theta_{x},\theta_{y};h^{s},h^{l})\label{eq:the-problem-inner}\\
\mbox{subject to} & w_{i}(\theta_{x},\theta_{y};h(t))=0, & i=1,\dots,W_{e}\label{eq:the-problem-inner-constr-1}\\
 & w_{i}(\theta_{x},\theta_{y};h(t))\leq0, & i=W_{e}+1,\dots,W\label{eq:the-problem-inner-constr-2}
\end{eqnarray}
~\hfill\IEEEQED
\end{problem}
\begin{problem}
[The outer problem] For given $h^{l}$,\emph{ 
\begin{eqnarray}
\mathcal{P}_{2}(h^{l})=\max_{\theta_{y}\geq0} & \mathbb{E}\left[\mathcal{P}_{1}(\theta_{y},h^{s},h^{l})\big|h^{l}\right]\label{eq:the-problem-outer}\\
\mbox{subject to} & q_{j}(\theta_{y};h^{l}(t))\leq0, & j=1,\dots J.\label{eq:the-problem-outer-constr}
\end{eqnarray}
}~\hfill\IEEEQED
\end{problem}

The collection of outer-problem solution $\theta_{y}^{*}(h^{l})$
for each $h^{l}$ in (\ref{eq:the-problem-outer})-(\ref{eq:the-problem-outer-constr})
gives the optimal long-term policy $\Omega^{l*}$. Similarly, the
collection of the solution $\theta_{x}^{*}(h)$ for the inner problem
$\mathcal{P}_{1}(\theta_{y}^{*},h^{l},h^{s})$ for a given $(h^{s},h^{l})$
gives the short-term optimal policy $\Omega^{s*}$.

We have the following assumption on the optimization problem $ $$\mathcal{P}_{0}$.
\begin{asmQED}
[Properties of $ $$\mathcal{P}_{0}$]\label{asm:problem-property-P0}
We assume the following for the problem $\mathcal{P}_{0}$,

\begin{itemize}

\item \emph{Convex domain}: The constraint domain specified by (\ref{eq:the-problem-constraint-1})-(\ref{eq:the-problem-constraint-3})
is convex.

\item \emph{Concave objective}: Given any CSI variable $h\in\mathcal{H}$,
the objective function $F(\theta_{x},\theta_{y};h)$ is strictly concave
over $(\theta_{x},\theta_{y})\in\mathbb{R}_{+}^{N_{x}}\times\mathbb{R}_{+}^{N_{y}}$. 

\item\emph{ Smoothness of $\theta_{y}^{*}(h^{l})$}: Define an implicit
mapping $\psi:h^{l}\mapsto\theta_{y}^{*}$ from the long-term CSI
$h^{l}$ to the optimal solution $\theta_{y}^{*}(h^{l})$, i.e., $\psi(h^{l})$
solves the outer problem $\mathcal{P}_{2}(h^{l})$ in (\ref{eq:the-problem-outer}).
There exists a constant $\varpi<\infty$, such that $\|\frac{\partial}{\partial h^{l}}\psi(\xi)-\frac{\partial}{\partial h^{l}}\psi(\xi^{'})\|\leq\varpi\|\xi-\xi^{'}\|$
for all $\xi,\xi^{'}\in\mathcal{H}^{l}$.\textbf{ }

\end{itemize}
\end{asmQED}

Note that in the problem $\mathcal{P}_{0}$, we may exclude equality
constraints. However, one may always eliminate the equality constraints
by either substituting them into the objective function or using the
Lagrangian primal-dual method \cite{Bertsekas:1999bs,Boyd:2004kx}.
Moreover, the last assumption ensures that there is no jump in the
optimal solution $\theta_{y}^{*}(h^{l})$ along the long-term CSI
$h^{l}$.

Due to Assumption \ref{asm:problem-property-P0}, there exists a unique
optimal solution $(\Omega^{s*},\Omega^{l*})$ for the problem $\mathcal{P}_{0}$.

A strong motivation to such control variable partitioning is due to
the information structure of local real-time and global real-time
CSI observations in wireless networks. Due to the latency involved
in global signaling of wireless networks, it is not scalable to adapt
all the control variables to the fast varying short-term CSI $h^{s}$.
On the other hand, adapting the control only to the slow varying long-term
CSI $h^{l}$ would be too conservative because it fails to exploit
the real-time local CSI observations $h^{s}(t)$ for opportunistic
diversity gain \cite{Tse2005:fundamental:Wireless}. One favorable
way to strike a balance between performance and scalability is to
partition the control variables as long-term $y$ and short-term control
$\theta_{y}(t)=\Omega^{s}(h(t))$. Another motivation of the control
timescale separation is the layered control architecture widely adopted
in the wireless system \cite{Chen:2006nx,Chiang2007}. For example,
the short-term control policy may correspond to physical layer control
(e.g. power adaptation) and the long-term control may correspond to
upper layer control (such as routing and admission control).

\subsection{Example: Power and Flow Control in Wireless Relay Network}

\label{sub:example}

We illustrate the mixed timescale control with an example network
with a macro BS and $N_{R}$ RSs. A set of end users $\mathcal{E}$
each transmits one data flow to the macro BS with the assistance of
a collection of RSs $\mathcal{R}$. Fig. \ref{fig:topology-relay-network}
illustrates a specific network topology with $N_{R}=2$ RSs, $N_{m}=4$
mobile users, $N=6$ links, and $\mathcal{L}=\{1,\dots,6\}$, $\mathcal{E}=\{1,2,3,4\}$,
$\mathcal{R}=\{5,6\}$, $\mathcal{N}=\mathcal{E}\cup\mathcal{R}\cup\{0\}$,
where node $0$ denotes the macro BS. Wireless links towards a common
receiving node share the same time-frequency resource and multi-user
detection (MUD) is used to handle cross-link interference. The maximum
achievable transmission data rate at receiving node $m$ is a set
of rates $c_{k}$ that satisfy the following constraints \cite{Tse2005:fundamental:Wireless},
\begin{equation}
\sum_{k\in\mathcal{S}}c_{k}<\log\left(1+\sum_{k\in\mathcal{S}}|h_{k}|^{2}p_{k}\right),\quad\forall\mathcal{S}\subset\mathcal{L}^{+}(m)\label{eq:MUD-MAC}
\end{equation}
where $\mathcal{L}^{+}(m)$ denotes the set of links that inject data
flows to the receiving node $m$. For example, in Fig. \ref{fig:topology-relay-network},
$\mathcal{L}^{+}(5)=\{3,4\}$, $\mathcal{L}^{+}(6)=\{2,5\}$ and $\mathcal{L}^{+}(0)=\{1,6\}$. 

Denote $r_{j}$ as the flow rate from node $j\in\mathcal{E}$ and
$\Omega^{r}(\mathbf{h}^{l})$ as the flow control policy that maps
the large-scale fading variable $\mathbf{h}^{l}=(h_{1}^{l},\dots,h_{N}^{l})$
to the flow control $\mathbf{r}=(r_{1},\dots,r_{|\mathcal{E}|})$.
Denote $p_{k}$ as the power allocation on link $k\in\mathcal{L}$
and $\Omega^{p}(\mathbf{h})$ as the power allocation policy that
maps the CSI vector $\mathbf{h}=(h_{1},\dots,h_{N})$ to the transmission
power $\mathbf{p}=(p_{1},\dots,p_{N})$. Corresponding to (\ref{eq:the-problem}),
a two-timescale stochastic maximization can be formulated as follows,
\begin{problem}
[Power and flow control in wireless relay network]

\begin{eqnarray}
\max_{\Omega^{r},\Omega^{p}} & \mathbb{E}\left[\sum_{j\in\mathcal{E}}\log(r_{j})-V\sum_{k\in\mathcal{L}}p_{k}\right]\label{eq:ex-objective}\\
\mbox{subject to} & \sum_{k\in\mathcal{S}}c_{k}<\log\left(1+\sum_{k\in\mathcal{S}}|h_{k}|^{2}p_{k}\right) & \forall\mathcal{S}\subset\mathcal{L}^{+}(m),\forall m\in\mathcal{R}\cup\{0\}\label{eq:ex-constraint-capacity}\\
 & c_{k}=\sum_{j\in\mathcal{C}(k)}r_{j} & \forall k\in\mathcal{L}\label{eq:ex-constraint-capacity-2}\\
 & \sum_{k\in\mathcal{L}^{+}(m)}c_{k}-\sum_{k\in\mathcal{L}^{-}(m)}c_{k}=0 & \forall m\in\mathcal{R}\label{eq:ex-constraint-flow-balance}
\end{eqnarray}
where $\mathcal{C}(k)$ denotes the collection of data flows $r_{j}$
that routes through link $k$, and $\mathcal{L}^{-}(m)$ denotes the
set of links that carry data flows from the transmitting node $m$.
~\hfill\IEEEQED
\end{problem}

For example, in Fig. \ref{fig:topology-relay-network}, $\mathcal{C}(4)=\{r_{4}\}$,
$\mathcal{C}(5)=\{r_{3},r_{4}\}$, $\mathcal{C}(6)=\{r_{2},r_{3},r_{4}\}$,
and $\mathcal{L}^{-}(k)=k$, for $k=1,\dots,6$. We consider proportional
fair utility \cite{Tychogiorgos:2012fk} in the objective function
of (\ref{eq:ex-objective}). Constraints (\ref{eq:ex-constraint-capacity})-(\ref{eq:ex-constraint-capacity-2})
are the MUD capacity constraints, and constraints in (\ref{eq:ex-constraint-flow-balance})
are the flow balance constraints, where the incoming data flow should
be balanced with the outgoing data flow at each RS.

With timescale separation of the control $\Omega^{p}$ and $\Omega^{r}$,
the above stochastic maximization can be decomposed into families
of inner problems and outer problems. 
\begin{problem}
[Inner power control]For given flow data rate $\mathbf{r}$ and
CSI $\mathbf{h}$, 
\begin{eqnarray}
Q(\mathbf{r};\mathbf{h})=\max_{\mathbf{p}\succeq0} & -V\sum_{k\in\mathcal{L}}p_{k}\label{eq:ex-inner-prob}\\
\mbox{subject to} & \sum_{k\in\mathcal{S}}c_{k}(\mathbf{r})<\log\left(1+\sum_{k\in\mathcal{S}}|h_{k}|^{2}p_{k}\right), & \forall\mathcal{S}\subset\mathcal{L}^{+}(m),\forall m\in\mathcal{R}\cup\{0\}\label{eq:ex-inner-constraint-capcity}
\end{eqnarray}
where $c_{k}(\mathbf{r})=\sum_{j\in\mathcal{C}(k)}r_{j}$, $\forall k\in\mathcal{L}$.
~\hfill\IEEEQED
\end{problem}
\begin{problem}
[Outer flow control]For given the large-scale fading variable $\mathbf{h}^{l}$,
\begin{eqnarray}
\max_{\mathbf{r}\succeq0} & \mathbb{E}\left[\sum_{j\in\mathcal{E}}\log(r_{j})+Q(\mathbf{r};\mathbf{h})\right]\label{eq:ex-outer-prob}\\
\mbox{subject to} & \sum_{k\in\mathcal{L}^{+}(m)}c_{k}(\mathbf{r})-\sum_{k\in\mathcal{L}^{-}(m)}c_{k}(\mathbf{r})=0 & \forall m\in\mathcal{R}.\label{eq:ex-outer-constraint-flow-balance}
\end{eqnarray}
~\hfill\IEEEQED
\end{problem}

In this example, the power allocation $\mathbf{p}$ in (\ref{eq:ex-objective})
corresponds to the short-term control variable $\theta_{x}$ in (\ref{eq:the-problem}),
and the flow data rate variable $\mathbf{r}$ in (\ref{eq:ex-objective})
corresponds to the long-term control variable $\theta_{y}$. The objective
function in (\ref{eq:the-problem}) is specified by 
\[
F(\mathbf{p},\mathbf{r};\mathbf{h}^{s},\mathbf{h}^{l})=\sum_{j\in\mathcal{E}}\log(r_{j})-V\sum_{k\in\mathcal{L}}p_{k}.
\]
Meanwhile, the inner problem constraints (\ref{eq:the-problem-inner-constr-1})-(\ref{eq:the-problem-inner-constr-2})
and the outer problem constraints (\ref{eq:the-problem-outer-constr})
are specified by (\ref{eq:ex-inner-constraint-capcity}) and (\ref{eq:ex-outer-constraint-flow-balance}),
respectively. Moreover, the short-term control policy $\Omega^{s}$
in Definition \ref{def:Mixed-Timescale-Controls} is specified by
the power control $\mathbf{p}(t)=\Omega^{p}(\mathbf{h}(t))$ in this
example. Similarly, the long-term control policy $\Omega^{l}$ is
specified by $\mathbf{r}(t)=\Omega^{r}(\mathbf{h}^{l})$ here. One
can check that the objective function (\ref{eq:ex-objective}) is
concave in $(\mathbf{p},\mathbf{r})$ and the optimization domain
specified by (\ref{eq:ex-constraint-capacity})-(\ref{eq:ex-constraint-flow-balance})
is convex. In addition, as the Lagrangian function \cite{Bertsekas:1999bs,Boyd:2004kx}
of the constrained problem (\ref{eq:ex-objective})-(\ref{eq:ex-constraint-flow-balance})
is continuous in $\mathbf{h}$, the smoothness condition in Assumption
\ref{asm:problem-property-P0} is satisfied as well. Table \ref{tab:notations-example}
summarizes the associations between the example and the mixed timescale
system model in Section \ref{sub:two-timescale-control}.

\begin{table}
\begin{tabular}{|c|c|c|c|}
\hline 
\multicolumn{2}{|c|}{Components in the example} & \multicolumn{2}{c|}{Corresponding components in the model}\tabularnewline
\hline 
\hline 
$\mathbf{h}^{s}$ & Vector of small-scaling fading of the wireless links & $h^{s}$ & Short-term CSI variable\tabularnewline
\hline 
$\mathbf{h}^{l}$ & Vector of path loss variables of the wireless links & $h^{l}$ & Long-term CSI variable\tabularnewline
\hline 
$\mathbf{h}$ & The aggregated CSI & $h$ & The aggregated CSI\tabularnewline
\hline 
$\Omega^{p}$ & Power control policy & $\Omega^{s}$ & Short-term control policy\tabularnewline
\hline 
$\Omega^{r}$ & Flow control policy & $\Omega^{l}$ & Long-term control policy\tabularnewline
\hline 
$\mathbf{p}$ & Power allocation & $\theta_{x}$ & Short-term control variable\tabularnewline
\hline 
$\mathbf{r}$ & Flow rate control & $\theta_{y}$  & Long-term control variable\tabularnewline
\hline 
(\ref{eq:ex-objective}) & Objective function & $F(\centerdot)$ & Objective function\tabularnewline
\hline 
(\ref{eq:ex-inner-constraint-capcity}) & Constraints for the inner problem & (\ref{eq:the-problem-constraint-3})-(\ref{eq:the-problem-constraint-1}) & Constraints for the inner problem (\ref{eq:the-problem-inner})\tabularnewline
\hline 
(\ref{eq:ex-outer-constraint-flow-balance}) & Constraints for the outer problem & (\ref{eq:the-problem-outer-constr}) & Constraints for the outer problem (\ref{eq:the-problem-outer})\tabularnewline
\hline 
\end{tabular}

\caption{\label{tab:notations-example}Problem associations between the example
and the mixed timescale system model.}
\end{table}

The control variable partitioning is motivated by the information
structure of the wireless relay network. On one hand, the CSI $h_{k}$
is available at some corresponding receiving node $m=\{k:k\in\mathcal{L}^{+}(m)\}$.
Hence the inner problem is solved locally at each receiving node $m\in\mathcal{R}\cup\{0\}$
based on real-time CSI $\{\mathbf{h}_{k}:\forall k\in\mathcal{L}^{+}(m)\}$.
On the other hand, solving the outer problem with the flow balance
constraints (\ref{eq:ex-constraint-flow-balance}) (or (\ref{eq:ex-outer-constraint-flow-balance}))
requires a global coordination. Updating the variable $\mathbf{r}$
needs to have the global knowledge of $Q(\mathbf{r};\mathbf{h})$
and the statistics of the long-term CSI $\mathbf{h}^{l}$. It also
needs to handle the global coupling induced by the flow balance constraints
(\ref{eq:ex-constraint-flow-balance}) (or (\ref{eq:ex-outer-constraint-flow-balance})).
As a result, explicit message passing is involved and the update can
only be done in a longer timescale.

\subsection{Iterative Solution of the Mixed-Timescale Stochastic Optimization
Problem }

\label{sub:two-timescale-algorithm}

In this section, we discuss a stochastic gradient based two-timescale
algorithm for solving $\mathcal{P}_{0}(a_{H},\epsilon)$, which consists
of an inner iteration and an outer iteration. We are interested in
the case where the inner and outer problems $\mathcal{P}_{1}$ and
$\mathcal{P}_{2}$ do not have closed form solutions and iterations
are needed to find the optimal solution. 

\begin{figure}
\begin{centering}
\includegraphics[width=0.8\columnwidth]{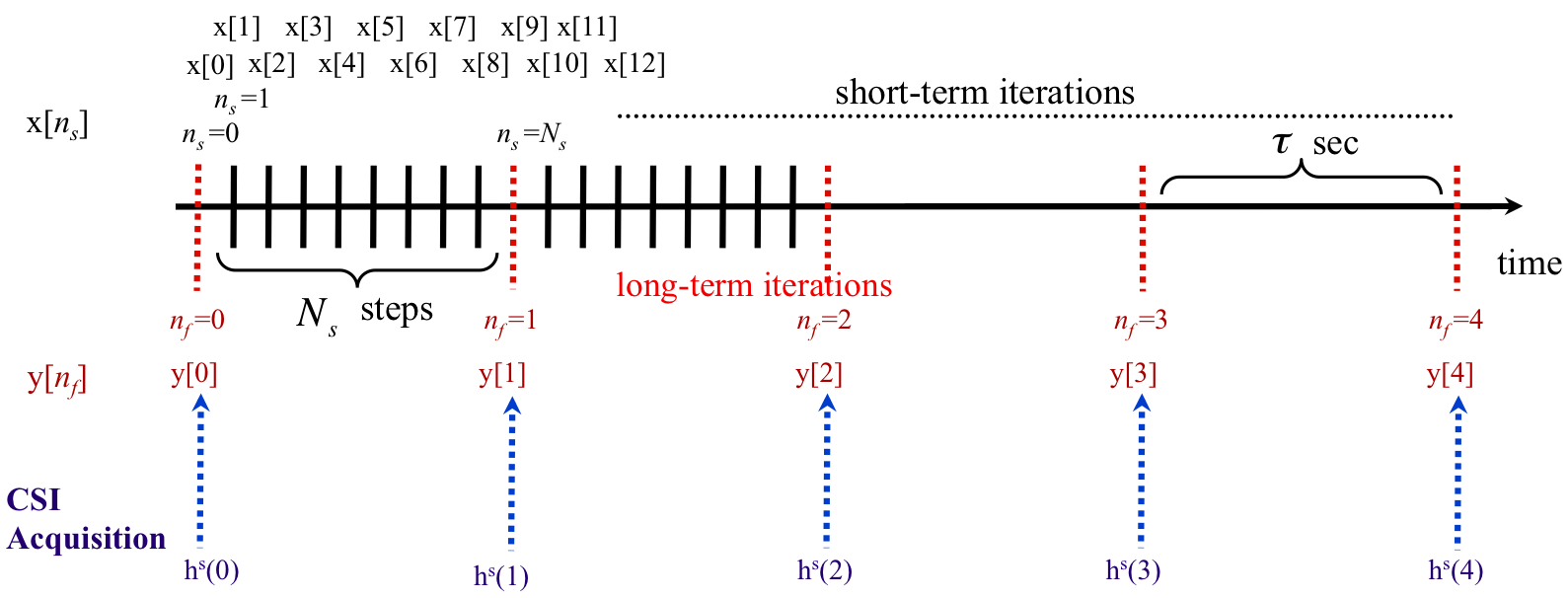}
\par\end{centering}

\caption{\label{fig:timescale} An illustration of the two-timescale algorithm. }
\end{figure}

Let $x=(\theta_{x},\lambda_{x})$ be the variable for the inner iteration,
where $\theta_{x}$ is the short-term control variable in (\ref{eq:the-problem-inner})
and $\lambda_{x}$ is an algorithm specific auxiliary variable%
\footnote{In Lagrange primal-dual methods, $\lambda_{x}$ is the Lagrange multiplier.%
}. Similarly, let $y=(\theta_{y},\lambda_{y})$ be the variable for
the outer iteration, where $\theta_{y}$ is the long-term control
variable in (\ref{eq:the-problem-outer}) and $\lambda_{y}$ is the
auxiliary variable. The time is partitioned into \emph{frames} with
duration $\tau$ and slots as illustrated in Fig. \ref{fig:timescale}.
One frame consists of $N_{s}$ slots and local real-time CSI is acquired
at each node at the beginning of each frame. During the $n_{s}$-th
slot and the $n_{f}$-th frame ($n_{f}=\lfloor n_{s}/N_{s}\rfloor$),
the short-term variable $x_{n_{s}}$ and long-term variable $y_{n_{f}}$
are updated according to the following mixed-timescale iterations, 

\begin{eqnarray}
x_{n_{s}} & = & \mathcal{P}_{\mathcal{X}(y)}\left[x_{n_{s}-1}+\gamma_{n_{s}}G(x_{n_{s}-1},y_{n_{f}};h^{s}(n_{f}\tau),h^{l}(n_{f}\tau))\right]\label{eq:alg-xn}\\
y_{n_{f}} & = & \mathcal{P}_{\mathcal{Y}}\left[y_{n_{f}-1}+\mu_{n_{f}}K(x_{n_{s}},y_{n_{f}-1};h^{s}(n_{f}\tau),h^{l}(n_{f}\tau))\right]\label{eq:alg-yn}
\end{eqnarray}
where $\gamma_{n_{s}}$ and $\mu_{n_{f}}$ are the step size sequences,
$\mathcal{P}_{\mathcal{D}}\left[\centerdot\right]$ is an Euclidian
projection onto the domain $\mathcal{D}$, and $\mathcal{X}(y)$ and
$\mathcal{Y}$ are the domains related to the problem constraints
(\ref{eq:the-problem-inner-constr-1})-(\ref{eq:the-problem-inner-constr-2})
and (\ref{eq:the-problem-outer-constr}). Fig. \ref{fig:timescale}
illustrates the timescales of the iterations in (\ref{eq:alg-xn})-(\ref{eq:alg-yn}).
Each node acquires local CSI at each frame boundary and updates the
short-term control variables $x_{n_{s}}$ once at each slot according
to (\ref{eq:alg-xn}). The centralized controller updates the long-term
variable $y_{n_{f}}$ once every frame according to (\ref{eq:alg-yn}). 

In addition, we consider the following assumptions on the iterations
(\ref{eq:alg-xn})-(\ref{eq:alg-yn}).
\begin{asmQED}
[Properties of the iterations]\label{asm:Property-iteration} Denote
$\mathcal{M}(\centerdot)\triangleq(G(\centerdot),K(\centerdot))$
as the mapping of the joint iteration vector $(x_{n_{s}},y_{n_{f}})$.
We assume the following properties:

\begin{itemize}

\item \emph{Definiteness of the iteration mappings}: The Jacobian
matrices of the iteration mappings $G_{x}(\centerdot)\triangleq\nabla_{x}G(\centerdot)$,
$K_{y}(\centerdot)\triangleq\nabla_{y}K(\centerdot)$, and $\nabla\mathcal{M}$
have the following properties: There exists $ $ $\alpha_{x},\alpha_{y},\alpha>0$,
such that $x^{T}G_{x}x\leq-\alpha_{x}\|x\|^{2}$, $y^{T}K_{y}y\leq-\alpha_{y}\|y\|^{2}$,
and $(x,y)^{T}\nabla\mathcal{M}(x,y)\leq-\alpha\left(\|x\|^{2}+\|y\|^{2}\right)$
for all $(x,y)\in\left(\cup_{y\in\mathcal{Y}}\mathcal{X}(y)\right)\times\mathcal{Y}$,
given any $(h^{s},h^{l})\in\mathcal{H}^{s}\times\mathcal{H}^{l}$.

\item\emph{ Lipschtiz continuous and bounded growth}: There exist
positive constants $l_{x}$ and $l_{y}$, such that $\|K(x,y_{1};\centerdot)-K(x,y_{2};\centerdot)\|\leq l_{y}\|y_{1}-y_{2}\|$,
for all $y_{1}$ and $y_{2}$, and $\|K_{x}(x,y;\centerdot)x\|\leq l_{x}\|x\|$,
for all $x$.

\end{itemize}
\end{asmQED}

Note that the first assumption is easily satisfied in a compact domain.
It is needed to guarantee the iterations (\ref{eq:alg-xn})-(\ref{eq:alg-yn})
to be stable under exogenous variation of $h(t)$, where the parameters
$\alpha_{x}$ and $\alpha_{y}$ indicate the convergence speed of
the inner iteration and the outer iteration, respectively, and $\alpha$
indicates the convergence speed of the whole algorithm (under one-timescale).
The second assumption is a standard assumption for studying the convergence.
In this paper, we are interested in the convergence of (\ref{eq:alg-xn})-(\ref{eq:alg-yn})
to the stationary points defined as follows.
\begin{lyxDefQED}
[Stationary points of (\ref{eq:alg-xn})-(\ref{eq:alg-yn})] Given
$y$ and $(h^{s},h^{l})$, the partial stationary point $\hat{x}(y;h^{s},h^{l})$
of (\ref{eq:alg-xn}) is given by the solution of 
\[
x-\mathcal{P}_{\mathcal{X}(y)}\left[x+G(x,y;h^{s},h^{l})\right]=0.
\]
Similarly, given $h^{s}$ and $h^{l}$, the stationary point $(x^{*}(h),y^{*}(h^{l}))$
of (\ref{eq:alg-xn})-(\ref{eq:alg-yn}) is given by the solution
of 
\begin{eqnarray*}
x-\mathcal{P}_{\mathcal{X}(y)}\left[x+G(x,y;h^{s},h^{l})\right] & = & 0\\
y-\mathcal{P}_{\mathcal{Y}}\left[y+\mathbb{E}K(x,y;h^{s},h^{l})\right] & = & 0
\end{eqnarray*}

\end{lyxDefQED}

Under Assumption \ref{asm:Property-iteration}, the partial stationary
point and the stationary point defined above is unique.
\begin{remrk}
[Interpretation of the projection] The projection $\mathcal{P}_{\mathcal{D}}(\theta)$
is to find the nearest point $x\in\mathcal{D}$ from $\theta$, i.e.,
$x=\mathcal{P}_{\mathcal{D}}(\theta)$ is the solution to the minimization
problem $\min_{x}\|x-\theta\|_{2}^{2}$, subject to $x\in\mathcal{D}$.
Note that if the constraint set is a hyper-rectangle, i.e., $\mathcal{D}=\Pi_{i=1}^{N}[a_{i},b_{i}]$,
the projection can be computed by restricting each component as $a_{i}\leq x^{(i)}\leq b_{i}$.
When a general constraint set $\mathcal{D}$ is considered, the projection
can be computed by Lagrange multipliers\cite{Bertsekas:1999bs,Boyd:2004kx}.
~\hfill\IEEEQED
\end{remrk}
\begin{remrk}
[Examples of iterative algorithms] Different choices of the iteration
mappings $G(\centerdot)$ and $K(\centerdot)$ yield different variants
of the stochastic algorithm. We review a few commonly used algorithms
in the following.

\begin{itemize}

\item\emph{ Stochastic projected gradient \cite{Benaim:1999ly,Kushner2003vn}}:
The mappings $G(\centerdot)$ and $K(\centerdot)$ are the gradients
of the objective function $ $$F(\centerdot)$ in (\ref{eq:the-problem}).
Specifically, 
\begin{eqnarray}
G(x_{n_{s}-1},y_{n_{f}};h^{s}(t_{n_{f}}),h^{l}(t_{n_{f}})) & = & \Gamma_{x}\nabla_{x}F(x_{n_{s}-1},y_{n_{f}},h^{s}(n_{f}\tau),h^{l}(n_{f}\tau))\label{eq:alg-stochastic-projected-x}\\
K(x_{n_{s}},y_{n_{f}-1};h^{s}(t_{n_{f}}),h^{l}(t_{n_{f}})) & = & \Gamma_{y}\nabla_{y}F(x_{n_{s}},y_{n_{f}-1},h^{s}(n_{f}\tau),h^{l}(n_{f}\tau))\label{eq:alg-stochastic-projected-y}
\end{eqnarray}
where $\Gamma_{x}$ and $\Gamma_{y}$ are positive definite scaling
matrices to accelerate the convergence. The iteration variables $x_{n_{s}}$
and $y_{n_{f}}$ in (\ref{eq:alg-xn})-(\ref{eq:alg-yn}) correspond
to $\theta_{x}$ and $\theta_{y}$ in (\ref{eq:the-problem}), i.e.,
$\theta_{x}(n_{s})=x_{n_{s}}$ and $\theta_{y}(n_{f})=y_{n_{f}}$.
In addition, the projection domains in (\ref{eq:alg-xn})-(\ref{eq:alg-yn})
are specified by 
\begin{equation}
\mathcal{X}(y)=\{\theta_{x}\in\mathbb{R}_{+}^{N_{x}}:\eqref{eq:the-problem-inner-constr-1}-\eqref{eq:the-problem-inner-constr-2}\mbox{ are satisfied}\},\quad\mathcal{Y}=\{\theta_{y}\in\mathbb{R}_{+}^{N_{y}}:\eqref{eq:the-problem-outer-constr}\mbox{ is satisfied}\}.\label{eq:alg-stochastic-projected-domain}
\end{equation}
Note that $K(\centerdot)$ is a stochastic estimator of the desired
gradient descent direction \\$\nabla_{y}\mathbb{E}\big[\max_{x}F(x,y_{n_{f}-1};h(n_{f}\tau))\big|h^{l}(n_{f}\tau)\big]$
(c.f. \cite{benveniste1990adaptive,Kushner2003vn}). 

\item\emph{ Stochastic primal-dual algorithm \cite{Boyd:2004kx,Zhang:2007fk}}:
We first form a Lagrangian function of the problem in (\ref{eq:the-problem}),
\[
L(\theta_{x},\theta_{y},\lambda_{x},\lambda_{y};h)=F(\theta_{x},\theta_{y};\centerdot)-\sum_{i}\lambda_{x,i}w_{i}(\theta_{x},\theta_{y};\centerdot)-\sum_{j}\lambda_{y,j}q_{j}(\theta_{y};\centerdot)
\]
where $\lambda_{x}\geq0$ and $\lambda_{y}\geq0$ are the Lagrange
multipliers for the primal variables $\theta_{x}$ and $\theta_{y}$,
respectively. Let $x=(\theta_{x},\lambda_{x})$ and $y=(\theta_{y},\lambda_{y})$.
The mappings $G(\centerdot)$ and $K(\centerdot)$ for stochastic
primal-dual algorithm are given by 
\begin{equation}
G(\centerdot)=\left(\begin{array}{c}
\nabla_{\theta_{x}}L(\theta_{x},\theta_{y},\lambda_{x},\lambda_{y};h)\\
-\nabla_{\lambda_{x}}L(\theta_{x},\theta_{y},\lambda_{x},\lambda_{y};h)
\end{array}\right),\quad\mbox{and}\quad K(\centerdot)=\left(\begin{array}{c}
\nabla_{\theta_{y}}L(\theta_{x},\theta_{y},\lambda_{x},\lambda_{y};h)\\
-\nabla_{\lambda_{y}}L(\theta_{x},\theta_{y},\lambda_{x},\lambda_{y};h)
\end{array}\right).\label{eq:alg-stochastic-primal-dual}
\end{equation}
In addition, the projection domains are given by 
\begin{equation}
\mathcal{X}(y)=\mathbb{R}_{+}^{N_{x}}\times\mathbb{R}_{+}^{W},\quad\mbox{and}\quad\mathcal{Y}=\mathbb{R}^{N_{y}}\times\mathbb{R}_{+}^{J}.\label{eq:alg-stochastic-primal-dual-domain}
\end{equation}

\end{itemize}

~\hfill\IEEEQED
\end{remrk}

Under Assumption \ref{asm:Property-iteration}, the convergence of
(\ref{eq:alg-xn})-(\ref{eq:alg-yn}) can be established from standard
techniques \cite{benveniste1990adaptive,Kushner2003vn} for static
CSI $h^{s}$ and $h^{l}$, as summarized below. 
\begin{lyxThmQED}
[Convergence under static CSI $h^{s}$ and $h^{l}$]\label{thm:static-CSI}
Consider $G(\centerdot)$ and $K(\centerdot)$ are given by (\ref{eq:alg-stochastic-projected-x})-(\ref{eq:alg-stochastic-projected-y})
(or (\ref{eq:alg-stochastic-primal-dual})), with the projection domains
$\mathcal{X}$ and $\mathcal{Y}$ given by (\ref{eq:alg-stochastic-projected-domain})
(or (\ref{eq:alg-stochastic-primal-dual-domain})). If the step size
sequences $\gamma_{n_{s}}$ and $\mu_{n_{f}}$ satisfy $\sum\gamma_{n_{s}}=\infty,\sum\mu_{n_{f}}=\infty$
and $\sum\gamma_{n_{s}}^{2}<\infty,\sum\mu_{n_{f}}^{2}<\infty$, then
the iteration $(x_{n_{s}},y_{n_{f}})$ in (\ref{eq:alg-xn})-(\ref{eq:alg-yn})
converges to the stationary point $(x^{*}(h^{s},h^{l}),y^{*}(h^{l}))$.
Furthermore, $(\theta_{x}^{*}(h^{s},h^{l}),\theta_{y}^{*}(h^{l}))$
solves the problem in (\ref{eq:the-problem}).
\end{lyxThmQED}

However, when the CSI $h^{s}(t)$ and $h^{l}(t)$ are time-varying,
the above convergence is not guaranteed. This is because, on one hand,
the inner iteration $x_{n_{s}}$ may not converge and hence induce
bias to the estimator $K(\centerdot)$ for updating $y_{n_{f}}$.
On the other hand, the optimal target $y^{*}(h^{l}(t))$ is time-varying
as well, and existing convergence results \cite{Benaim:1999ly,Kushner2003vn}
fail to apply. In this paper, we shall focus on investigating the
convergence behavior of the mixed-timescale iterations (\ref{eq:alg-xn})-(\ref{eq:alg-yn})
when the CSI $h^{l}(t)$ and $h^{s}(t)$ have mixed-timescale stochastic
time-variations.

\subsection{Impact of Exogenous Variations and Tracking Errors}

\label{sub:model-of-convergence-tracking-error}

With the presence of the exogenous variations of $h(t)$, the mixed
timescale iterations in (\ref{eq:alg-xn})-(\ref{eq:alg-yn}) are
continuously perturbed and the convergence to the stationary point
target $(x^{*}(h(t)),y^{*}(h^{l}(t)))$ is not guaranteed. The impact
of the exogenous variations can be summarized as follows,
\begin{itemize}
\item \emph{Impact on the inner iteration $x_{n_{s}}$}: Since $h(t)$ is
time-varying, $x_{n_{s}}$ needs to track the time-varying optimal
point $x^{*}(h(t))$. The convergence error may depend on the relative
variation speed between the iteration dynamics (\ref{eq:alg-xn})
and the exogenous variations of $h(t)$.
\item \emph{Impact on the outer iteration $y_{n_{f}}$}: Likewise, the optimal
point $y^{*}(h^{l}(t))$ is time-varying, and $y_{n_{f}}$ may hardly
reach $y^{*}(h^{l}(t))$. Moreover, the convergence error of the inner
problem yields $\hat{\mathcal{P}}_{1}-\mathcal{P}_{1}\neq0$, which
induces a bias to the estimator $K(\centerdot)$ in (\ref{eq:alg-yn}).
\end{itemize}

We formally define the tracking error for the iterations (\ref{eq:alg-xn})-(\ref{eq:alg-yn}).
\begin{lyxDefQED}
[Tracking error]\label{def:Tracking-error_ex-ey} The mean square
tracking error for the short-term control variable $x$ is defined
as \textbf{\emph{}} 
\begin{equation}
e_{x}=\lim\sup_{n_{f}\to\infty}\frac{1}{n_{f}}\sum_{i=1}^{n_{f}}\mathbb{E}\left[\|x_{iN_{s}}-\hat{x}(y_{i},h^{s}(i\tau),h^{l}(i\tau))\|^{2}\right]\label{eq:tracking-error-ex}
\end{equation}
whereas, the tracking error for the long-term control variable $y$
is defined as
\begin{equation}
e_{y}=\lim\sup_{n_{f}\to\infty}\frac{1}{n_{f}}\sum_{i=1}^{n_{f}}\mathbb{E}\left[\|y_{i}-y^{*}(h^{l}(i\tau))\|^{2}\right].\label{eq:tracking-error-ey}
\end{equation}

\end{lyxDefQED}

In the rest of the paper, we shall study the tracking errors $e_{x}$
and $e_{y}$ under the mixed-timescale CSI $h^{s}(t)$ and $h^{l}(t)$.

\section{Virtual Dynamic Systems for Convergence Analysis}

\label{sec:convergence-VSDS}

In this section, we derive the \emph{virtual dynamic systems} for
studying the tracking error of the iterations (\ref{eq:alg-xn})-(\ref{eq:alg-yn})
under time-varying CSI. We first consider a mean continuous time dynamic
system (MCTS) which captures the mean behavior of the mixed-timescale
iterations in (\ref{eq:alg-xn})-(\ref{eq:alg-yn}) under static $h^{l}$.
Using SDE approximations, we then extend the results to consider the
impact of time-varying $h^{l}(t)$ on the overall tracking errors
and derive the VSDS.

\subsection{Case 1: The Mean Continuous-Time Dynamic System (MCTS) for Static
$h^{l}$ and Time-varying $h^{s}(t)$}

\label{sub:case-1-static-h^l}

In this subsection, we consider the case for static $h^{l}$ and time-varying
$h^{s}(t)$. Under time-varying $h^{s}(t)$, constant step size $\gamma$
should be used in the inner iteration (\ref{eq:alg-xn}) to track
the time-varying partial stationary point $\hat{x}(t)$. On the other
hand, since the stationary point $y^{*}(h^{l})$ is static, a diminishing
step size $\mu_{n_{f}}$ can be used in the outer iteration (\ref{eq:alg-yn})
to assist the convergence. We derive a mean continuous-time dynamic
system (MCTS) to characterize the ``mean'' behavior of the algorithm
trajectories for (\ref{eq:alg-xn})-(\ref{eq:alg-yn}). The MCTS is
defined as follows.
\begin{lyxDefQED}
[Mean continuous-time dynamic system (MCTS)]The state trajectory
of mean continuous-time dynamic systems (MCTS) $x_{c}(t)$ and $y_{c}(t)$
are defined as the solutions to the following Skorohod reflective
ordinary differential equations (ODEs) \cite{Skorohod61},
\begin{eqnarray}
\dot{x}_{c} & = & G(x_{c},y_{c},h^{s},h^{l})+z_{x}\label{eq:ode-xc}\\
\dot{y}_{c} & = & k(y_{c},h^{l})+z_{y}.\label{eq:ode-yc}
\end{eqnarray}
where $\dot{x}_{c}\triangleq\frac{d}{dt}x_{c}(t)$, $\dot{y}_{c}\triangleq\frac{d}{dt}y_{c}(t)$.
The terms $z_{x}(t)$ and $z_{y}(t)$ are the reflection terms to
keep the trajectories $x$ and $y$ inside their domains $\mathcal{X}(y)$
and $\mathcal{Y}$, respectively. The function $k(\centerdot)$ is
defined as 
\begin{equation}
k(y,h^{l})\triangleq\lim_{n_{f}\to\infty}\mathbb{E}\left[K(\hat{x}(y,h(n_{f}\tau)),y,h(n_{f}\tau))\right]\label{eq:def-k(y,h)}
\end{equation}
where $K(\centerdot)$ is the iteration mapping specified in (\ref{eq:alg-yn}).
\end{lyxDefQED}

Note that since the short-term CSI process $h^{s}(t)$ in (\ref{eq:sde-hs})
is ergodic and stationary, the limit in (\ref{eq:def-k(y,h)}) always
exists.

\begin{figure}
\begin{centering}
\includegraphics[scale=0.7]{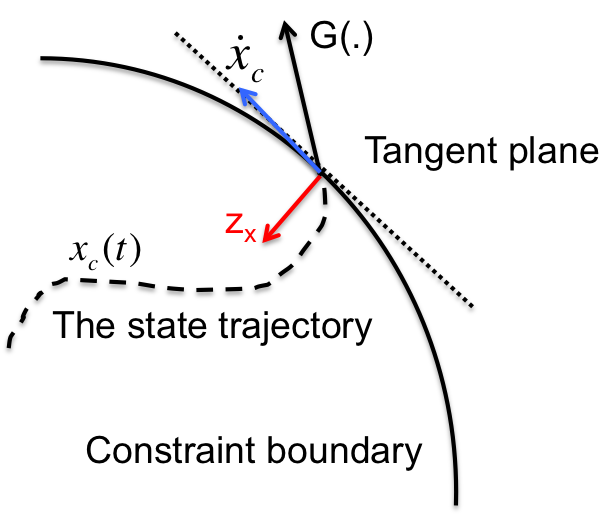}
\par\end{centering}

\caption{\label{fig:Reflection-term} Illustration of the reflection term $z_{x}(t)$
when the virtual state trajectory $x_{c}(t)$ reaches the boundary
of the constrain domain.}
\end{figure}

\begin{remrk}
[Interpretation of the reflection] The reflection term is the minimum
force to restrict the trajectory inside the constraint domain. Taking
$z_{x}$ for example, for $x_{c}\in\mathring{\mathcal{X}}$, in the
interior of $\mathcal{X}$, $z_{x}=0$. For $x\in\partial\mathcal{X}$,
the boundary of $\mathcal{X}$, $z_{x}$ lies in the convex cone generated
by the inward normals on the surface of the boundary as illustrated
in Fig. \ref{fig:Reflection-term}. The magnitude of $z_{x}$ is such
that $G(\centerdot)+z_{x}$ lies in the tangent of the surface. Therefore,
when the trajectory reaches the boundary, it can only go along on
the boundary. Note that $z_{x}(t)$ and $z_{y}(t)$ can be computed
using Lagrange multipliers when the constraint domain are not hyper-rectangles.
Please refer to Appendix \ref{app:derivation-reflection-term} for
a derivation of the reflection terms.~\hfill\IEEEQED
\end{remrk}

In the following, we illustrate the connection between the algorithm
trajectory (\ref{eq:alg-xn})-(\ref{eq:alg-yn}) and the MCTS (\ref{eq:ode-xc})-(\ref{eq:ode-yc}).
We first derive the property of the \emph{equilibrium} of MCTS.
\begin{lyxDefQED}
[Equilibrium and partial equilibrium] The point $(x_{c}^{*},y_{c}^{*})$
is an equilibrium of the MCTS (\ref{eq:ode-xc})-(\ref{eq:ode-yc})
under $(h^{s},h^{l})\in\mathcal{H}^{s}\times\mathcal{H}^{l}$, if
$\dot{x}_{c}=\dot{y}_{c}=0$, i.e., $G(x_{c}^{*},y_{c}^{*},h^{s},h^{l})+z_{x}=0$
and $k(y_{c}^{*},h^{l})+z_{y}=0$. In addition, for any $y_{c}\in\mathcal{Y}$,
$\hat{x}_{c}(y_{c},h)$ is a partial equilibrium of the MCTS in (\ref{eq:ode-xc})
if $\dot{\hat{x}}_{c}=0$, i.e., $G(\hat{x}_{c},y_{c},h^{s},h^{l})+z_{x}=0$.
\end{lyxDefQED}

There is a strong connection between the iteration trajectory and
the MCTS as summarized in the following theorem.
\begin{lyxThmQED}
[Connection between the algorithm trajectory and the MCTS]\label{thm:Connection-discrete-alg-MCTS}
Assume $\epsilon=0$ (static long-term CSI $h^{l}$). Consider $G(\centerdot)$
and $K(\centerdot)$ are given by (\ref{eq:alg-stochastic-projected-x})-(\ref{eq:alg-stochastic-projected-y})
(or (\ref{eq:alg-stochastic-primal-dual})), with the projection domains
$\mathcal{X}$ and $\mathcal{Y}$ given by (\ref{eq:alg-stochastic-projected-domain})
(or (\ref{eq:alg-stochastic-primal-dual-domain})), and the step size
sequences satisfy (i) $\gamma_{n_{f}}=\gamma$ for some small enough
$\gamma>0$, and (ii) $\sum\mu_{n_{f}}=\infty$ and $\sum\mu_{n_{f}}^{2}<\infty$.
In addition, the short-term CSI timescale is much slower than the
algorithm timescale, i.e., $a_{H}\tau\ll\gamma$. Then the algorithm
iteration trajectory $(x_{n_{s}(t)},y_{n_{f}(t)})$ in (\ref{eq:alg-xn})-(\ref{eq:alg-yn})
converges to the virtual state trajectory of the MCTS $(x_{c}(t),y_{c}(t))$
in (\ref{eq:ode-xc})-(\ref{eq:ode-yc}) in probability, i.e., for
any $\eta>0$, 
\[
\lim\sup_{t\to\infty}\mbox{Pr}\left\{ \|x_{n_{s}(t)}-x_{c}(t)\|>\eta\right\} =0,\;\mbox{and }\;\lim\sup_{t\to\infty}\mbox{Pr}\left\{ \|y_{n_{f}(t)}-y_{c}(t)\|>\eta\right\} =0
\]
where $n_{s}(t)=\lfloor tN_{s}/\tau\rfloor$ and $n_{f}(t)=\lfloor t/\tau\rfloor$.
\end{lyxThmQED}

The theorem is established using stochastic approximation \cite{Benaim:1999ly,Kushner2003vn}.
We sketch the proof in Appendix \ref{app:thm-connection-alg-MCTS}.

From Theorem \ref{thm:Connection-discrete-alg-MCTS}, the study of
the algorithm convergence is equivalent to the study of the \emph{stability}%
\footnote{A deterministic dynamic system $\dot{x}=f(t,x)$ is asymptotically
stable at the equilibrium $x^{*}$, if there is a $\delta>0$, such
that for any $\|x(0)-x^{*}\|\leq\delta$, $x(t)\to x^{*}$, as $t\to\infty$
\cite{Khalil1996}. %
} of the MCTS. The MCTS in (\ref{eq:ode-xc})-(\ref{eq:ode-yc}) can
be viewed as the desired ``mean'' trajectory in the continuous-time
counter-part of the discrete-time algorithm iterations (\ref{eq:alg-xn})-(\ref{eq:alg-yn}),
which has filtered the noisy perturbation induced by the exogenous
variation of $h^{s}(t)$ in the estimator $K(\centerdot)$. Hence,
the stability of the MCTS provides a necessary and sufficient condition
for the convergence of the original iterations (\ref{eq:alg-xn})-(\ref{eq:alg-yn})
under static $h^{l}$.

Note that, under Assumption \ref{asm:Property-iteration}, the MCTS
(\ref{eq:alg-yn}) is asymptotically stable \cite{Khalil1996}, i.e.,
$y_{c}(t)\to y^{*}$ as $t\to\infty$. Therefore, from Theorem \ref{thm:Connection-discrete-alg-MCTS},
we have the following result on the convergence under static $h^{l}$
and time-varying $h^{s}(t)$ when the CSI timescale is sufficiently
slower than the algorithm timescale.
\begin{lyxCorQED}
[Convergence under static $h^{l}$ and slow time-varying $h^{s}(t)$]\label{cor:convergence-slow-varying}
For sufficiently fast algorithm iteration, i.e., $a_{H}\tau\ll\gamma$,
the iteration $(x_{n_{s}(t)},y_{n_{f}(t)})$ in (\ref{eq:alg-xn})-(\ref{eq:alg-yn})
converges to $(x^{*}(t),y^{*})$ almost surely, i.e., $\|x_{n_{s}(t)}-x^{*}(t)\|\to0$
and $\|y_{n_{f}(t)}-y^{*}\|\to0$ almost surely, as $t\to\infty$.
\end{lyxCorQED}
%

\begin{figure}
\begin{centering}
\includegraphics[scale=0.5]{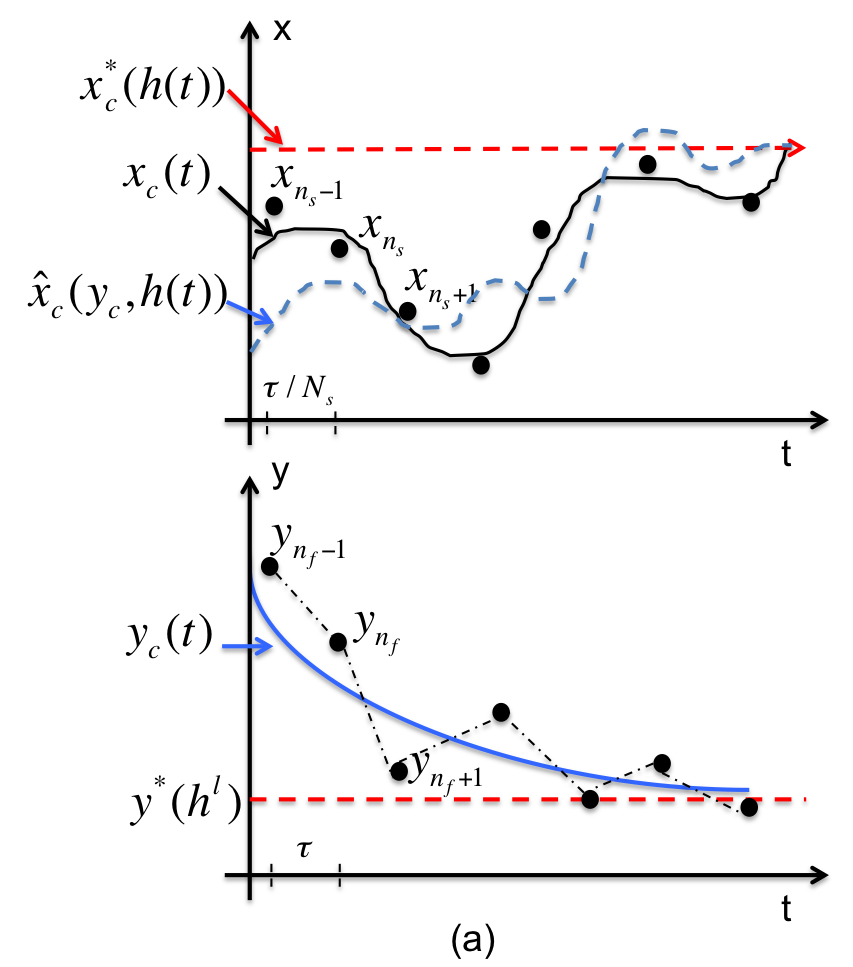}
\par\end{centering}

\begin{centering}
\includegraphics[scale=0.5]{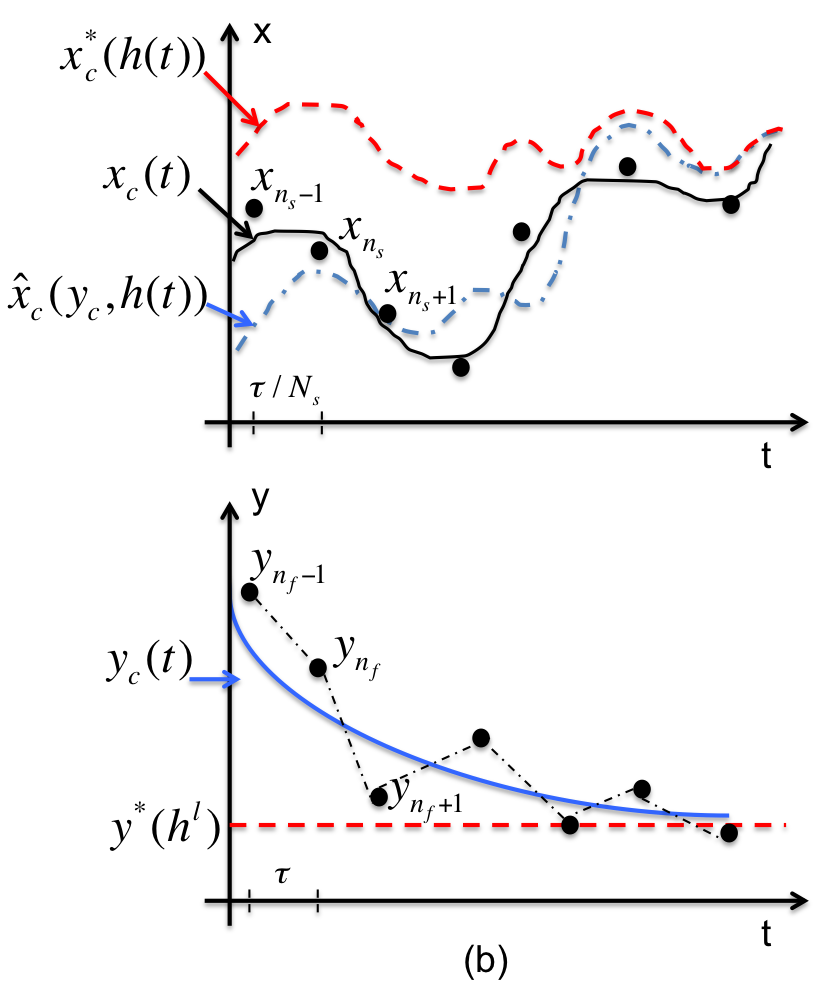}\includegraphics[scale=0.5]{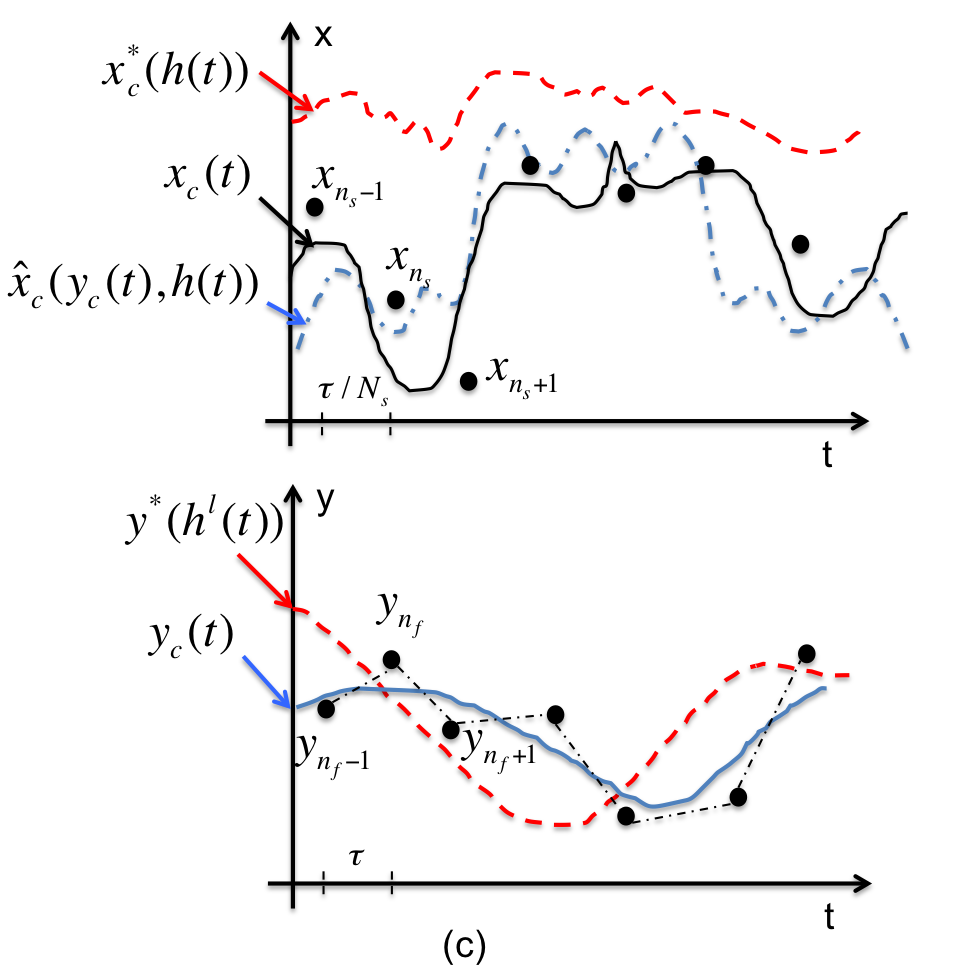}
\par\end{centering}

\caption{\label{fig:Trajectory} Illustrations of the algorithm iterations,
the virtual dynamic system MCTS and the moving equilibrium of the
MCTS. (a) illustrates the case for static $h^{l}$ and $h^{s}$, (b)
illustrates the case for static $h^{l}$ and time-varying $h^{s}(t)$,
and (c) illustrates the case of time-varying $h^{l}(t)$ and $h^{s}(t)$. }
\end{figure}

Fig. \ref{fig:Trajectory}.a) illustrates the relationships between
the algorithm iterations, the virtual dynamic system MCTS and the
moving equilibrium of the MCTS for the case when both $h^{s}$ and
$h^{l}$ are static. The virtual system MCTS as well as the algorithm
iterations $x_{n_{s}}$ and $y_{n_{f}}$ would eventually converge
to the static equilibrium $(x^{*}(h),y^{*}(h^{l}))$. Fig. \ref{fig:Trajectory}.b)
illustrates the case for static $h^{l}$ and slowly time-varying $h^{s}(t)$.
For variable $y$, there is a static equilibrium $y^{*}(h^{l})$ of
the virtual dynamic system MCTS, and the virtual system state $y_{c}(t)$
converges to the static target $y^{*}(h^{l})$. For variable $x$,
the trajectory of the partial equilibrium $\hat{x}(y_{c}(t),h(t))$
is driven by the time-varying $y_{c}(t)$ and $h(t)$, and it eventually
converges to the trajectory of the moving equilibrium $x^{*}(h(t))$,
as $y_{c}(t)$ converges to $y^{*}(h^{l})$. Meanwhile, the dynamics
of the MCTS tracks the moving partial equilibrium $\hat{x}(y_{c}(t),h(t))$.
For both $x$ and $y$, the algorithm iterations $x_{n_{s}}$ and
$y_{n_{f}}$ in (\ref{eq:alg-xn})-(\ref{eq:alg-yn}) roughly follow
the dynamics of the virtual system MCTS. However, the case is different
under time-varying $h^{l}(t)$ and $h^{s}(t)$, as illustrated in
Fig. \ref{fig:Trajectory}.c). The equilibrium $y^{*}(h^{l}(t))$
of the virtual dynamic system MCTS moves around and $y_{c}(t)$ never
converges. Affected by the time-varying $y^{*}(t)$ and the error
gap $y_{c}(t)-y^{*}(t)$, the dynamics of the partial equilibrium
$\hat{x}(y_{c}(t),h(t))$ cannot converge to the optimal trajectory
$x^{*}(h(t))$. Nevertheless, the algorithm iterations $x_{n_{s}}$
and $y_{n_{f}}$ in (\ref{eq:alg-xn})-(\ref{eq:alg-yn}) still follow
the behavior of the virtual dynamic system MCTS.

In the next section, we extend the MCTS equivalence framework to the
case with time-varying $h^{l}(t)$ and $a_{H}\tau\approx\gamma$.

\subsection{Case 2: Virtual Stochastic Dynamic System for Time-Varying $h^{s}(t)$
and $h^{l}(t)$}

\label{sub:case-2-time-varying-h^l}

When $h^{l}(t)$ is time-varying, the algorithm iterations $(x_{n_{s}},y_{n_{f}})$
in (\ref{eq:alg-xn})-(\ref{eq:alg-yn}) should continuously track
the time-varying partial optimal point $(\hat{x}(t),y^{*}(t))$, which
is a moving target as illustrated in Fig. \ref{fig:Trajectory}.b).
As such, we consider constant step size $\gamma_{n_{s}}=\mu_{n_{f}}=\gamma$
instead of diminishing step size. 

We need to first quantify the dynamics of the moving partial equilibrium
$(\hat{x}(t),y^{*}(t))$ from the MCTS under the variation of $h^{l}(t)$.
Using the CSI timescale separation property $\epsilon\ll a_{H}$ for
$h^{s}(t)$ and $h^{l}(t)$, the dynamics of the moving target $(\hat{x}(t),y^{*}(t))$
is given by:
\begin{lyxLemQED}
[Dynamics of the moving partial equilibrium]\label{lem:Dynamics-moving-equilibrium}
Define $\widetilde{G}(x,y,h^{s},h^{l})=G(x,y,h^{s},h^{l})+z_{x}$.
The dynamics of $\hat{x}_{c}(t)$ and $y_{c}^{*}(t)$ are given by
\begin{eqnarray}
d\hat{x}_{c} & = & -\widetilde{G}_{x}^{-1}(\hat{x}_{c}(y_{c},\centerdot),y_{c},h^{s},h^{l})\left[\widetilde{G}_{h^{s}}(\hat{x}_{c}(y_{c},\centerdot),y_{c},h^{s},h^{l})dh^{s}+\widetilde{G}_{y}(\hat{x}_{c}(y_{c},\centerdot),y_{c},h^{s},h^{l})dy\right]\label{eq:sde-moving-optimum-x}\\
dy_{c}^{*} & = & \psi_{h^{l}}(h^{l})dh^{l}\label{eq:sde-moving-optimum-y}
\end{eqnarray}
where $\widetilde{G}_{x}(\centerdot)\triangleq\frac{\partial}{\partial x}\widetilde{G}(x_{c},y_{c},h^{s},h^{l})$,
$\widetilde{G}_{y}(\centerdot)\triangleq\frac{\partial}{\partial y}\widetilde{G}(x_{c},y_{c},h^{s},h^{l})$,
$\widetilde{G}_{h^{s}}(\centerdot)\triangleq\frac{\partial}{\partial h^{s}}\widetilde{G}(x_{c},y_{c},h^{s},h^{l})$
and $\psi_{h^{l}}=\frac{\partial}{\partial h^{l}}\psi(h^{l})$ as
defined in Assumption \ref{asm:Property-iteration}.
\end{lyxLemQED}
\begin{proof}
Please refer to Appendix \ref{app:pf-lem-dynamic-moving} for the
proof.
\end{proof}

With the notion of MCTS and the dynamic of the moving equilibrium
in (\ref{eq:sde-moving-optimum-x})-(\ref{eq:sde-moving-optimum-y}),
the instantaneous tracking errors can be expressed as 
\begin{eqnarray}
x_{n_{s}(t)}-\hat{x}_{c}(t) & = & (x_{n_{s}(t)}-x_{c}(t))+(x_{c}(t)-\hat{x}_{c}(t))\nonumber \\
 & = & \sqrt{\gamma}\widetilde{x}_{c}^{\gamma}(t)+\widetilde{x}_{c}^{e}(t)\label{eq:tracking-error-decomposition-x}
\end{eqnarray}
and 
\begin{eqnarray}
y_{n_{f}(t)}-y_{c}^{*}(t) & = & (y_{n_{f}(t)}-y_{c}(t))+(y_{c}(t)-y_{c}^{*}(t))\nonumber \\
 & = & \sqrt{\gamma}\widetilde{y}_{c}^{\gamma}(t)+\widetilde{y}_{c}^{e}(t)\label{eq:tracking-error-decomposition-y}
\end{eqnarray}
where $\widetilde{x}_{c}^{\gamma}(t)\triangleq\frac{1}{\sqrt{\gamma}}\left(x_{n_{s}(t)}-x_{c}(t)\right)$
and $\widetilde{y}_{c}^{\gamma}(t)\triangleq\frac{1}{\sqrt{\gamma}}\left(y_{n_{f}(t)}-y_{c}(t)\right)$
are the scaled error gap between the iteration trajectory (\ref{eq:alg-xn})-(\ref{eq:alg-yn})
and the MCTS, and $\widetilde{x}_{c}^{e}(t)\triangleq x_{c}(t)-\hat{x}_{c}(t)$
and $\widetilde{y}_{c}^{e}(t)\triangleq y_{c}(t)-y_{c}^{*}(t)$ are
the scaled tracking error from the MCTS to the moving partial equilibrium
(\ref{eq:sde-moving-optimum-x})-(\ref{eq:sde-moving-optimum-y}). 

Even though it is very hard to quantify the tracking errors $x_{n_{s}(t)}-\hat{x}_{c}(t)$
and $y_{n_{f}(t)}-y_{c}^{*}(t)$, we can try to study the distributions
of the decomposed error states $\widetilde{x}_{c}^{\gamma}(t)$, $\widetilde{x}_{c}^{e}(t)$,
$\widetilde{y}_{c}^{\gamma}(t)$, and $\widetilde{y}_{c}^{e}(t)$
with the help of a virtual dynamic system defined in the following.

Define a joint virtual state $u(t)=(\widetilde{x}_{c}(t),\widetilde{y}_{c}(t),\widetilde{x}_{c}^{e}(t),\widetilde{y}_{c}^{e}(t),\widetilde{h}^{s}(t))\in\mathbb{R}^{2N_{x}+2N_{y}+N}$,
where $\widetilde{x}_{c},\widetilde{x}_{c}^{e}\in\mathbb{R}^{N_{x}}$,
$\widetilde{y}_{c},\widetilde{y}_{c}^{e}\in\mathbb{R}^{N_{y}}$ and
$\widetilde{h}^{s}(t)$ is a short-term virtual CSI state with initial
value $\widetilde{h}^{s}(0)=h^{s}(0)$. Correspondingly, define a
virtual long-term CSI state $\widetilde{h}^{l}(t)$ as the solution
of $d\widetilde{h}^{l}=-\frac{\tau}{N_{s}\gamma}H_{L}(t)dt$, with
initial value $\widetilde{h}^{l}(0)=h^{l}(0)$, where $H_{L}(t)$
is an $N\times N$ diagonal matrix, with the $j$-th diagonal element
being $c_{0}\iota D_{j}(t)^{-\iota-1}v_{j}(t)$. We use a short hand
notation $\widetilde{G}_{x}^{-1}\widetilde{G}_{h^{s}}(\centerdot)$
to stand for the matrix $\widetilde{G}_{x}^{-1}(x_{c},y_{c},h^{s},h^{l};t)\widetilde{G}_{h^{s}}(x_{c},y_{c},h^{s},h^{l};t)$
from (\ref{eq:sde-moving-optimum-y}). The VSDS is defined as follows. 
\begin{lyxDefQED}
[Virtual stochastic dynamic system (VSDS)]\label{def:VSDS} The
virtual state trajectory of the VSDS $u(t)$ is characterized by the
following SDE: 
\begin{equation}
du=U_{1}(t,u)dt+U_{2}(t,u)dW_{t}+dZ_{u}\label{eq:VSDS}
\end{equation}
where 
\[
U_{1}(t,u)\triangleq\left(\begin{array}{c}
G_{x}(x_{c},y_{c},\widetilde{h}^{s},\widetilde{h}^{l})\widetilde{x}_{c}+G_{y}(x_{c},y_{c},\widetilde{h}^{s},\widetilde{h}^{l})\widetilde{y}_{c}\\
N_{s}^{-1}K_{x}(\hat{x}_{c},y_{c},\centerdot)(\widetilde{x}_{c}+\widetilde{x}_{c}^{e})+N_{s}^{-1}K_{y}(\hat{x}_{c},y_{c},\centerdot)\widetilde{y}_{c}\\
G(x_{c},y_{c},\widetilde{h}^{s},\widetilde{h}^{l})-\frac{1}{2}\frac{a_{H}\tau}{N_{s}\gamma}\widetilde{G}_{x}^{-1}\widetilde{G}_{h^{s}}(\centerdot)\widetilde{h}^{s}+\widetilde{G}_{x}^{-1}\widetilde{G}_{y}(\centerdot)N_{s}^{-1}k(y_{c},\widetilde{h}^{l})\\
N_{s}^{-1}k(y_{c},\widetilde{h}^{l})+\frac{\tau}{N_{s}\gamma}\psi_{h^{l}}(\widetilde{h}^{l})H_{L}(t)\\
-\frac{1}{2}\frac{a_{H}\tau}{N_{s}\gamma}\widetilde{h}^{s}
\end{array}\right),
\]
and
\[
U_{2}(t,u)\triangleq\left(\begin{array}{ccccc}
\mathbf{0}_{N_{x}\times N_{x}} &  &  & \ldots & \mathbf{0}_{N_{x}\times N}\\
 & \sqrt{\tau N_{s}^{-1}}\Sigma^{\frac{1}{2}}(y_{c};\widetilde{h}^{l}) &  &  & \vdots\\
 &  & \mathbf{0}_{N_{x}\times N_{x}} &  & \sqrt{\frac{a_{H}\tau}{N_{s}\gamma}}\widetilde{G}_{x}^{-1}\widetilde{G}_{h^{s}}\\
\vdots &  &  & \mathbf{0}_{N_{y}\times N_{y}} & \mathbf{0}_{N_{x}\times N_{x}}\\
\mathbf{0} & \dots &  &  & \sqrt{\frac{a_{H}\tau}{N_{s}\gamma}}\mathbf{I}_{N}
\end{array}\right).
\]
Moreover, $W_{t}$ is a $(2N_{x}+2N_{y}+N)$-dimensional Brownian
motion, $dZ_{u}=\big(\mathbf{0}_{N_{x}},dZ_{x}+dZ_{y},$\\$\widetilde{G}_{x}^{-1}\widetilde{G}_{y}(\centerdot)dZ_{y},dZ_{y},\mathbf{0}_{N}\big)$
is the reflection term, and $\Sigma(y_{c};\widetilde{h}^{l})$ is
the covariance matrix for the stochastic estimator $K(\centerdot)$
in (\ref{eq:alg-yn}).
\end{lyxDefQED}

As is summarized in the following theorem, the VSDS provides a weak
convergence limit (convergence in distribution, c.f. \cite{benveniste1990adaptive,Kushner2003vn})
to the dynamics of the error gaps $\widetilde{x}_{c}^{\gamma}(t)$,
$\widetilde{x}_{c}^{e}(t)$, $\widetilde{y}_{c}^{\gamma}(t)$, and
$\widetilde{y}_{c}^{e}(t)$, when $\gamma\to0$. 
\begin{lyxThmQED}
[Algorithm Tracking Errors and VSDS]\label{thm:Weak-convergence}
Assuming CSI timescale separation for $h^{l}(t)$ and $h^{s}(t)$,
i.e., $\epsilon\ll a_{H}$, the joint state $(\widetilde{x}_{c}^{\gamma},\widetilde{y}_{c}^{\gamma},\widetilde{x}_{c}^{e},\widetilde{y}_{c}^{e},\widetilde{h}^{s})$
weakly converges to $u(t)$, as $\gamma\to0$, which is the solution
to the VSDS in (\ref{eq:VSDS}).
\end{lyxThmQED}
\begin{proof}
Please refer to Appendix \ref{app:proof-weak-conv} for the proof.
\end{proof}

The above results allow us to work with the VSDS in (\ref{eq:VSDS})
to study the convergence of the mixed timescale tracking algorithm
(\ref{eq:alg-xn})-(\ref{eq:alg-yn}). The VSDS provides statistical
dynamics for the decomposed tracking error states. We formally summarize
the connection between the VSDS and the tracking errors for the mixed-timescale
iterations (\ref{eq:alg-xn})-(\ref{eq:alg-yn}) in the following
theorem.
\begin{lyxCorQED}
[Connections between the VSDS and the iterations]\label{cor:connection-VSDS}
Assuming CSI timescale separation for $h^{l}(t)$ and $h^{s}(t)$,
i.e., $\epsilon\ll a_{H}$, the tracking errors for the algorithm
iterations (\ref{eq:alg-xn})-(\ref{eq:alg-yn}) defined in (\ref{eq:tracking-error-ex})
and (\ref{eq:tracking-error-ey}) can be upper bounded from $u(t)$,
i.e., 
\[
e_{x}\leq\lim\sup_{t\to\infty}\frac{1}{t}\max\left(\gamma,1\right)\int_{0}^{t}\mathbb{E}\left[\|\widetilde{x}_{c}(s)\|^{2}+\|\widetilde{x}_{c}^{e}(s)\|^{2}\right]ds
\]
and 
\[
e_{y}\leq\lim\sup_{t\to\infty}\frac{1}{t}\max\left(\gamma,1\right)\int_{0}^{t}\mathbb{E}\left[\|\widetilde{y}_{c}(s)\|^{2}+\|\widetilde{y}_{c}^{e}(s)\|^{2}\right]ds
\]
where $\widetilde{x}_{c}(t)$, $\widetilde{x}_{c}^{e}(t)$, $\widetilde{y}_{c}(t)$,
and $\widetilde{y}_{c}^{e}(t)$ are the components of the joint state
$u(t)$ in the VSDS (\ref{eq:VSDS}).
\end{lyxCorQED}

Corollary \ref{cor:connection-VSDS} can be seen from the tracking
errors (\ref{eq:tracking-error-decomposition-x})-(\ref{eq:tracking-error-decomposition-y})
and the results in Theorem \ref{thm:Weak-convergence}. With Theorem
\ref{cor:connection-VSDS}, we can focus on studying the \emph{stochastic
stability} (formally defined in Section \ref{sub:Lyapunov-stochastic-stability})
of the VSDS in (\ref{eq:VSDS}) in order to understand the convergence
behavior of the mixed timescale iterations in (\ref{eq:alg-xn}) and
(\ref{eq:alg-yn}). 

Moreover, the VSDS suggests that the tracking errors consist of two
parts: (i) the \emph{steady state error} $\gamma\|\widetilde{x}_{c}\|^{2}$
and $\gamma\|\widetilde{y}_{c}\|^{2}$, which are the mean square
error gaps between the iteration trajectory $(x_{n_{s}},y_{n_{f}})$
in (\ref{eq:alg-xn})-(\ref{eq:alg-yn}) and the MCTS $(x_{c}(t),y_{c}(t))$
in (\ref{eq:ode-xc})-(\ref{eq:ode-yc}), and (ii) the \emph{mean
tracking error} $\|\widetilde{x}_{c}^{e}(t)\|^{2}$ and $\|\widetilde{y}_{c}^{e}(t)\|^{2}$,
which are the mean square distances between the MCTS and the target
moving partial equilibrium $(\hat{x}(t),y^{*}(t))$ in (\ref{eq:sde-moving-optimum-x})-(\ref{eq:sde-moving-optimum-y}).
Note that when $h^{l}$ is static, i.e., $H_{L}(t)\equiv0$, the tracking
error $\widetilde{y}_{c}^{e}(t)$ in the VSDS converges to 0, and
hence, there is only steady state error $\gamma\|\widetilde{y}_{c}\|^{2}$
(due to constant step size) for the long-term variable $y$.

\section{Convergence Analysis of Mixed Timescale Iterations}

\label{sec:convergence-stability}

In this section, we derive the tracking error bound of the mixed timescale
iteration (\ref{eq:alg-xn})-(\ref{eq:alg-yn}) by studying the VSDS
obtained from Section \ref{sec:convergence-VSDS}. We first briefly
review the Lyapunov stochastic stability techniques. By studying the
stability of the VSDS, we then derive a sufficient condition for the
convergence of the mixed-timescale algorithm. Moreover, we obtain
a tracking error bound in terms of the parameters of the exogenous
process $h(t)$.

\subsection{Preliminary Results on the Lyapunov Stochastic Stability}

\label{sub:Lyapunov-stochastic-stability}

It is very hard to derive the exact solutions for the VSDS. Instead,
we are interested in the expected upper bound value of the joint state
$\|u(t)\|$, which represents the aggregated tracking error $e_{x}+e_{y}$
of the iterations. This is captured by the \emph{stochastic stability}
in mean square defined as follows.
\begin{lyxDefQED}
[Stochastic stability in mean square]\label{def:Stochastic-Stability-1}
Given any initial state $u(0)\in\mathcal{U}$, the stochastic process
$u(t)$ is globally stochastically stable in mean square, if there
exists $0\leq\delta<\infty$, such that $\lim\sup_{t\to\infty}\frac{1}{t}\int_{0}^{t}\mathbb{E}\left\Vert u(\tau)\right\Vert ^{2}d\tau\leq\delta.$
\end{lyxDefQED}

We use a Lyapunov method to study the stochastic stability of $u(t)$.
Define a non-negative function $V(u)=\frac{1}{2}u^{T}u$ along the
trajectory of $u(t)$. The function has the property that $V(u)\sim\|u\|^{2}$,
which plays the role of an energy function, where a larger $\|u\|$
gives a larger function value. We summarize the main techniques of
stochastic stability analysis as follows.
\begin{lyxDefQED}
[Lyapunov drift operator]\label{def:lyapunov-drift} Consider a
stochastic process $u(t)$ and a real-valued Lyapunov function $V(u)$.
The Lyapunov drift operator is an infinitesimal estimator on $V(\centerdot)$
defined as $\widetilde{\mathcal{L}}V(u)=\lim_{\delta\downarrow0}\frac{1}{\delta}\left[\mathbb{E}\left[V(u(t+\delta)|V(u(t)\right]-V(u(t))\right].$
\end{lyxDefQED}
\begin{lyxLemQED}
[Stochastic Stability from Lyapunov Drift]\label{lem:stoch-stability-region-z}
Consider a function $f(u)$ that satisfies $f(u)\geq a\|u\|^{r}$
for all $u\in\mathcal{U}$, and some $a,r>0$. Suppose the stochastic
Lyapunov drift of the process $u(t)$ has the following property 
\begin{equation}
\mathcal{\widetilde{L}}V(u)\leq-f(u)+g(s)\label{eq:Lya-criteria-property}
\end{equation}
for all $u\in\mathcal{U}$, where $s(t)$ is a stochastic process
that satisfies $\lim\sup_{t\to\infty}\frac{1}{t}\int_{0}^{t}\mathbb{E}\left[g(s(\tau))\right]d\tau\leq d$
for some function $g(s)$ and $d<\infty$. Then the process $u(t)$
is stochastically stable, and 
\[
\lim\sup_{t\to\infty}\frac{1}{t}\int_{0}^{t}\mathbb{E}\|u(\tau)\|^{r}d\tau\leq\frac{d}{a}.
\]

\end{lyxLemQED}

The above result is based on the Foster-Lyapunov criteria for continuous
time processes in \cite{Meyn93:StabilityIII-Foster} and is a simple
extension of the results in \cite[Theorem 2]{Chen2013:Delay}. The
advantage of the Lyapunov method enables a qualitative analysis of
the SDE without explicitly solving it. Using such a technique, we
derive the stability results for the mixed timescale algorithm in
the following.

\subsection{Stability of the Mixed-Timescale Algorithm}

\label{sub:stability-algorithm}

Corresponding to stability of a random process, we define the iteration
stability as follows.
\begin{lyxDefQED}
[Stability of the iteration]\label{def:stability-iteration} The
iterations (\ref{eq:alg-xn}) and (\ref{eq:alg-yn}) are stable if
the corresponding tracking error defined in (\ref{eq:tracking-error-ex})
and (\ref{eq:tracking-error-ey}) are bounded, i.e., there exists
a $B<\infty$, such that $e_{x}+e_{y}\leq B$.
\end{lyxDefQED}

Note that due to the stochastic iterations, the tracking errors defined
in (\ref{eq:tracking-error-ex}) and (\ref{eq:tracking-error-ey})
may be unbounded statistically. To study the algorithm stability,
we can equivalently investigate the stability of the VSDS $u(t)$.
Towards this end, we first construct a Lyapunov drift $\mathcal{\widetilde{L}}V(u)$
on the trajectory of the VSDS in (\ref{eq:VSDS}). We then proceed
to find a function $g(s)$ that satisfies condition (\ref{eq:Lya-criteria-property}).
Finally, we use Theorem \ref{cor:connection-VSDS} to obtain the stability
result. 

We summarize the sufficient conditions of the stability of the mixed
timescale algorithm as follows.
\begin{lyxThmQED}
[Sufficient conditions for the algorithm stability]\label{thm:sufficient-cond-stability-alg}
Assume CSI timescale separation for $h^{l}(t)$ and $h^{s}(t)$, i.e.,
$\epsilon\ll a_{H}$. Suppose that there exist $0<v_{H},v_{y}<\infty$,
such that $\|G_{x}^{-1}G_{h^{s}}\|\leq v_{H}$ and $\|G_{x}^{-1}G_{y}\|\leq v_{y}$.
Then the sufficient condition for the algorithm to be stable is given
by 
\begin{equation}
\alpha N_{s}\left(8\alpha_{x}-\frac{a_{H}\tau}{N_{s}\gamma}v_{H}^{2}\right)-2l_{x}^{2}-2l_{y}^{2}v_{y}^{2}>0.\label{eq:rho-sufficient-condition}
\end{equation}

\end{lyxThmQED}
\begin{proof}
Please refer to Appendix \ref{app:pf-thm-sufficient-cond} for the
proof.
\end{proof}

The above results have several implications on the convergence. 
\begin{itemize}
\item \emph{Convergence of the inner problem}: The term $N_{s}\left(8\alpha_{x}-\frac{a_{H}\tau}{N_{s}\gamma}v_{H}^{2}\right)$
specifies the convergence behavior of the inner problem. Recall that
the parameter $a_{H}$ controls the variation speed of the fast changing
CSI $h^{s}(t)$, $\alpha_{x}$ represents the convergence rate of
the inner problem, $\tau$ is the frame duration, $N_{s}$ is the
number of slots per frame, and $\gamma$ is the step size. As such,
for a given $a_{H}$, we need to have sufficiently fast inner iterations
($\alpha_{x}$) or sufficient number of slots per frame ($N_{s}$)
in order to have bounded tracking error. 
\item \emph{Convergence of the outer problem and the coupling} effect: The
convergence of the inner problem and outer problem is coupled together.
The stability of the inner problem (a positive $8\alpha_{x}-\frac{a_{H}\tau}{N_{s}\gamma}v_{H}^{2}$)
is a premise of the stability of the whole algorithm. To achieve the
stability, we desire small $v_{H}$ and $l_{x}$, which represent
the sensitivity of the partial stationary point $\hat{x}$ w.r.t.
the change of $h^{s}(t)$. On the other hand, we also want the parameters
$l_{y}$ and $v_{y}$ to be small, which means that $y^{*}$ shall
not be quite sensitive to the bias induced by the tracking error $x_{n_{s}}-\hat{x}(y_{n_{f}},\centerdot)$. 
\item \emph{Impact from the iteration timescale}: One can reduce the frame
duration $\tau$, increase the number of slots $N_{s}$ per frame,
or increase the step size $\gamma$ to enhance the stability of the
overall algorithm. However, shortening the frame duration $\tau$
may result in a larger amount of signaling overhead to update the
long-term variable $y_{n_{f}}$ and the acquisition of local CSI $h^{s}(n_{f}\tau)$;
increasing the number of slots $N_{s}$ yields a higher computational
complexity; and moreover, a large step size $\gamma$ may give larger
steady state error $\mathcal{O}(\gamma)$ for the discrete-time trajectory. 
\end{itemize}

\subsection{Upper Bound of the Tracking Error}

Stability is only a weak result of convergence. We are interested
in the tracking error bound of the algorithm. Under the sufficient
condition specified in (\ref{eq:rho-sufficient-condition}), using
the Lyapunov technique in Lemma \ref{lem:stoch-stability-region-z},
we study the result on the upper bound of the tracking errors $e_{x}$
and $e_{y}$.
\begin{lyxThmQED}
[Upper bound of the tracking error]\label{thm:upper-bound-error}
Assume the conditions in Theorem \ref{thm:sufficient-cond-stability-alg}.
If $\overline{\Sigma}\triangleq\lim\sup_{t\to\infty}\frac{1}{t}\int_{0}^{t}\mbox{tr}\left(\Sigma(y^{*}(h^{l}(\tau)))\right)d\tau<\infty$,
the tracking errors $e_{x}$ and $e_{y}$ are given by: 
\begin{equation}
e_{x}+e_{y}\leq\frac{\eta}{\rho}\left(\tau\overline{\Sigma}+C\right)\label{eq:tracking-error-upper-bound}
\end{equation}
where $\rho=\mathcal{O}\left(N_{s}\alpha^{2}\alpha_{x}\alpha_{y}\right)$,
$\eta=\mathcal{O}\left(\sqrt{N_{s}^{2}\alpha_{x}^{2}+\alpha^{2}}\right)$,
$ $$C=\frac{a_{H}\tau}{\gamma}N(1+v_{H}^{2})+\mathcal{O}(\epsilon^{2}\varpi^{2}\tau^{2}\gamma^{-2})$,
and $N$ is the dimension of the CSI vector $h^{s}$.
\end{lyxThmQED}
\begin{proof}
Please refer to Appendix \ref{app:thm-upper-bound} for the proof.
\end{proof}

The above result shows that the upper bound of the tracking error
depends on several important parameters, such as the timescale parameter
$a_{H}$ for $h^{s}(t)$, the sensitivities $v_{H}$ and $v_{y}$
of the stationary points $\hat{x}$ and $y^{*}$, respectively, as
well as the sensitivities $\varpi$ of $y^{*}(h^{l})$ over $h^{l}$.
A faster time-varying scenario corresponds to larger $a_{H}$, which
result in a larger tracking error bound. In addition, we can observe
the followings.
\begin{itemize}
\item \emph{Special case for static }$h^{l}$ and $h^{s}$: Under static
CSI, where the CSI timescale parameters $a_{H}=\epsilon=0$, we have
the term $C=0$ in the error bound. The tracking error is governed
by the steady state error $\frac{\eta}{\rho}\tau\overline{\Sigma}$
due to the constant step size $\gamma$. Note that if diminishing
step size is used for the outer iteration (\ref{eq:alg-yn}), i.e.,
$\mu_{n_{f}}\to0$, the tracking error bound in (\ref{eq:tracking-error-upper-bound})
becomes $0$. This corresponds to the case in Fig. \ref{fig:Trajectory}.a).
\item \emph{Special case for static $h^{l}$ and time-varying $h^{s}(t)$}:
Under static $h^{l}$, where long-term CSI timescale parameter $\epsilon=0$,
the term $C$ decreases, because the term $\mathcal{O}(\epsilon^{2}\varpi^{2}\tau^{2}\gamma^{-2})$
becomes $0$. In particular, if diminishing step size is used for
the outer iteration (\ref{eq:alg-yn}), i.e., $\mu_{n_{f}}\to0$,
the error bound becomes $\frac{\eta}{\rho}C$ and is mainly contributed
by the tracking error in the inner iteration. This corresponds to
the case in Fig. \ref{fig:Trajectory}.b).
\item \emph{Impact of the algorithm parameters}: When the CSI $(h^{s},h^{l})$
is fast changing, i.e., the CSI timescale parameters $a_{H}$ and
$\epsilon$ are large, one can reduce the frame duration $\tau$,
increase the number of slots $N_{s}$ per frame, or increase the step
size $\gamma$ to reduce the tracking error, with the price of larger
signaling overhead, higher computational complexity and larger steady
state error $\mathcal{O}(\gamma)$ as discussed in Section \ref{sub:stability-algorithm}.
\end{itemize}

\section{Compensation for Mixed Timescale Iterations}

\label{sec:adaptving-algorithm}

In the previous sections, we have analyzed the convergence behavior
of the iteration (\ref{eq:alg-xn})-(\ref{eq:alg-yn}) under mixed
timescale time-varying CSI $h(t)$. In this section, we shall enhance
the algorithm for better tracking performance. Since the convergence
of the outer long-term variable $y$ depends on the convergence of
the inner problem, it is essential to accelerate the convergence of
the short-term variable $x_{n_{s}}$. Towards this end, we introduce
a compensation term in the algorithm (\ref{eq:alg-xn}) to offset
the exogenous disturbance to the VSDS in (\ref{eq:VSDS}).

\subsection{Adaptive Compensations for the Time-varying CSI}

Recall that in the mixed timescale iterations (\ref{eq:alg-xn})-(\ref{eq:alg-yn}),
the inner iteration tracks the moving target $\hat{x}(t)$ driven
by the time-varying $h(t)$ and $y(t)$. On the other hand, the outer
iteration tracks the moving target $y^{*}(t)$ driven by $h^{l}(t)$.
As a result, when $h(t)$ and $y(t)$ are time-varying, they generate
disturbance to the tracking iterations (\ref{eq:alg-xn}) and (\ref{eq:alg-yn}).
This can be seen from the dynamics of the error states $\widetilde{x}_{c}^{e}(t)$
and $\widetilde{y}_{c}^{e}(t)$ in the VSDS in (\ref{eq:VSDS}) (c.f.
equations (\ref{eq:sde-t_xe}) and (\ref{eq:sde-t_ye}))

\begin{eqnarray}
d\widetilde{x}_{c}^{e} & = & G(x_{c},y_{c},\widetilde{h}^{s},\widetilde{h}^{l})dt\underbrace{+\widetilde{G}_{x}^{-1}\widetilde{G}_{h^{s}}(\centerdot)d\widetilde{h}^{s}}_{\mbox{\scriptsize\ exogenous disturbance from }\widetilde{h}^{s}(t)}\underbrace{+\widetilde{G}_{x}^{-1}\widetilde{G}_{y}(\centerdot)dy_{c}}_{\mbox{\scriptsize\ distrubance from }y(t)}\label{eq:sde-xe-1}\\
d\widetilde{y}_{c}^{e} & = & N_{s}^{-1}k(y_{c},\widetilde{h}^{l})dt\underbrace{-\psi_{h^{l}}(h^{l})d\widetilde{h}^{l}}_{\mbox{\scriptsize\ exogenous disturbance from }\widetilde{h}^{l}(t)}.\label{eq:sde-ye-1}
\end{eqnarray}

\begin{figure}
\begin{centering}
\includegraphics[scale=0.5]{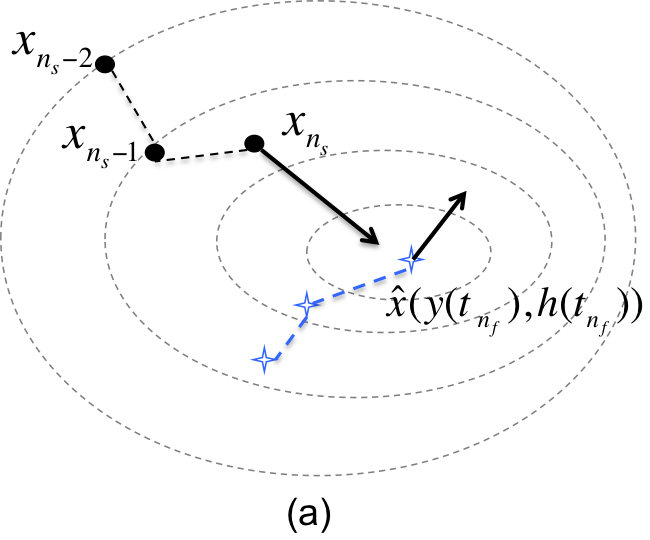} $\qquad\qquad$\includegraphics[scale=0.5]{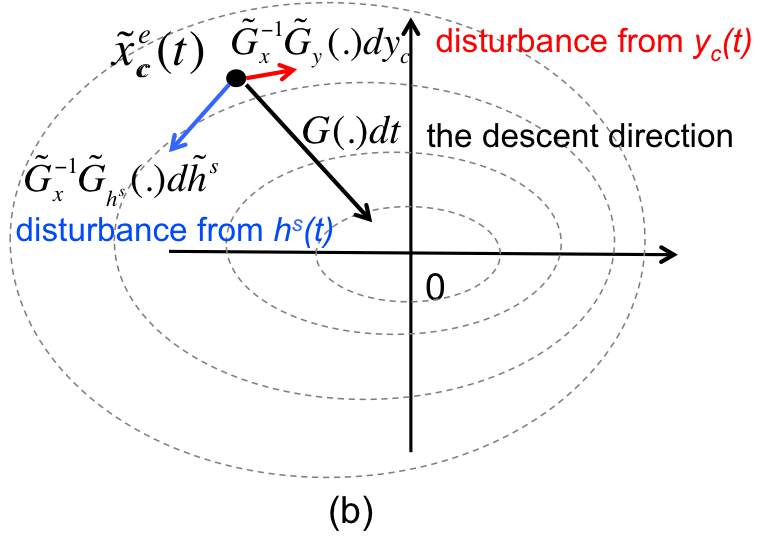}
\par\end{centering}

\begin{centering}
\includegraphics[scale=0.5]{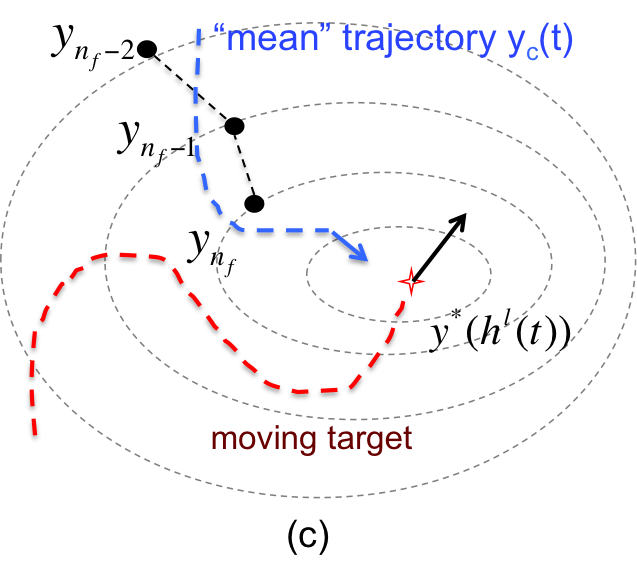} $\qquad\qquad$\includegraphics[scale=0.5]{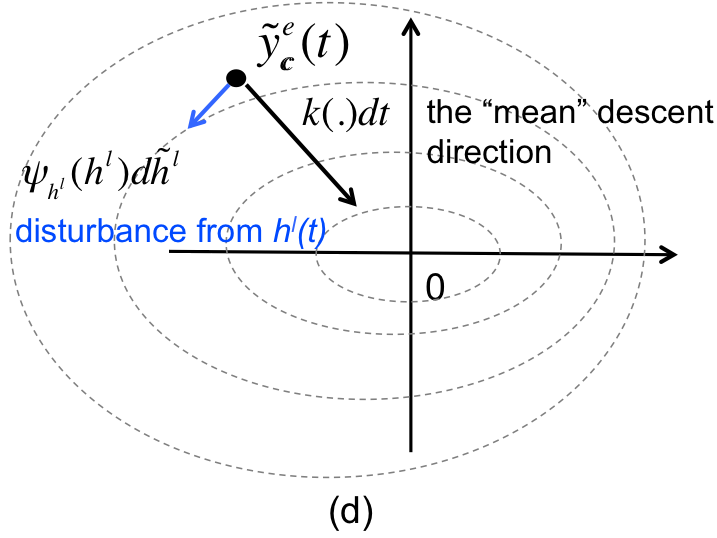}
\par\end{centering}

\caption{\label{fig:disturbance}Illustrations of the convergence of the iterations
$(x_{n_{s}},y_{n_{f}})$ and the virtual error dynamic systems $\widetilde{x}_{c}^{e}(t)$
and $\widetilde{y}_{c}^{e}(t)$. (a) The algorithm iteration $x_{n_{s}}$
tracks the target $\hat{x}(y(t),h(t))$, which is moving due to the
time-varying $y(t)$ and $h(t)$. (b) The corresponding error state
$\widetilde{x}_{c}^{e}(t)$ approaches to $0$ under the mapping $G(\centerdot)$.
However, the exogenous disturbance $\widetilde{G}_{x}^{-1}\widetilde{G}_{h^{s}}(\centerdot)d\widetilde{h}^{s}$
and $\widetilde{G}_{x}^{-1}\widetilde{G}_{y}(\centerdot)dy_{c}$ drag
it away from the origin. (c) The algorithm iteration $y_{n_{f}}$
tracks $y^{*}(h^{l}(t))$ following the ``mean''virtual trajectory
$y_{c}(t)$. (d) The corresponding error state $\widetilde{y}_{c}^{e}(t)$
approaches to $0$ following the virtual direction $k(\centerdot)$.
However, the exogenous disturbance $\psi_{h^{l}}(\widetilde{h}^{l})d\widetilde{h}^{l}$
drags it away from the origin. }
 
\end{figure}

Note that, from Theorem \ref{thm:Connection-discrete-alg-MCTS} and
Corollary \ref{cor:convergence-slow-varying}, when $d\widetilde{h}^{l}=0$,
the error dynamic system (\ref{eq:sde-ye-1}) is asymptotically stable
and the error state $\widetilde{y}_{c}^{e}(t)$ converges to 0. In
addition, when $d\widetilde{h}^{s}=0$, the system (\ref{eq:sde-xe-1})
is stable and the error state $\widetilde{x}_{c}^{e}(t)$ converges
to $0$ as well. However, with the presence of the \emph{exogenous
disturbance} $d\widetilde{h}^{s}$ and $d\widetilde{h}^{l}$, the
error states $\widetilde{x}_{c}^{e}(t)$ and $\widetilde{y}_{c}^{e}(t)$
are continuously disturbed and may fail to converge to the origin.
This is illustrated in Fig. \ref{fig:disturbance}.

Such observation motivates us to offset the exogenous disturbance
to reduce the tracking error under time-varying $h^{s}(t)$ and $h^{l}(t)$.
We start by introducing compensation terms to the MCTS (\ref{eq:ode-xc})-(\ref{eq:ode-yc})
as follows, 
\begin{eqnarray}
dx_{c} & = & G(x_{c},y_{c},\widetilde{h}^{s},\widetilde{h}^{l})dt\underbrace{-\widehat{\widetilde{G}_{x}^{-1}\widetilde{G}_{h^{s}}}(\centerdot)d\widetilde{h}^{s}}_{\mbox{\scriptsize compensation for }\widetilde{h}^{s}(t)}\underbrace{-\widehat{\widetilde{G}_{x}^{-1}\widetilde{G}_{y}(\centerdot)}dy_{c}}_{\mbox{\scriptsize\ compensation for }y(t)}\label{eq:comp-xc}\\
dy_{c} & = & N_{s}^{-1}k(y_{c},\widetilde{h}^{l})dt\underbrace{+\widehat{\psi_{h^{l}}(\widetilde{h}^{l})}d\widetilde{h}^{l}}_{\mbox{\scriptsize\ compensation for }\widetilde{h}^{l}(t)}\label{eq:comp-yc}
\end{eqnarray}
where $-\widehat{\widetilde{G}_{x}^{-1}\widetilde{G}_{h^{s}}}(\centerdot)d\widetilde{h}^{s}$
and $-\widehat{\widetilde{G}_{x}^{-1}\widetilde{G}_{y}(\centerdot)}dy_{c}$
are compensation terms to offset the disturbance in (\ref{eq:sde-xe-1}),
and $\widehat{\psi_{h^{l}}(\widetilde{h}^{l})}d\widetilde{h}^{l}$
is a compensation term to offset the disturbance in (\ref{eq:sde-ye-1}).
Here, for a simple discussion, we drop the reflection terms $dZ_{x}$
and $dZ_{y}$, since the projections that drive the reflections are
always conservative%
\footnote{The reflection is from the constraints that form the convex domain
$\mathcal{X}(y)\times\mathcal{Y}$. As the constraint domain is to
restrict the iteration trajectory, it always helps the convergence.%
}. As a result, the dynamic system in (\ref{eq:sde-xe-1})-(\ref{eq:sde-ye-1})
becomes 
\begin{eqnarray}
d\widetilde{x}_{c}^{e} & = & G(x_{c},y_{c},\widetilde{h}^{s},\widetilde{h}^{l})dt+\left(\widetilde{G}_{x}^{-1}\widetilde{G}_{h^{s}}(\centerdot)-\widehat{\widetilde{G}_{x}^{-1}\widetilde{G}_{h^{s}}(\centerdot)}\right)d\widetilde{h}^{s}\label{eq:sde-comp-x}\\
 &  & \qquad+\left(\widetilde{G}_{x}^{-1}\widetilde{G}_{y}(\centerdot)-\widehat{\widetilde{G}_{x}^{-1}\widetilde{G}_{y}(\centerdot)}\right)dy_{c}\nonumber \\
d\widetilde{y}_{c}^{e} & = & N_{s}^{-1}k(y_{c},\widetilde{h}^{l})dt+\left(\widehat{\psi_{h^{l}}(\widetilde{h}^{l})}-\psi_{h^{l}}(\widetilde{h}^{l})\right)d\widetilde{h}^{l}.\label{eq:sde-comp-y}
\end{eqnarray}

Define the disturbance components as $\varphi_{x}^{h}(\centerdot)\triangleq-\widetilde{G}_{x}^{-1}\widetilde{G}_{h^{s}}(\centerdot)$,
$\varphi_{x}^{y}(\centerdot)\triangleq-\widetilde{G}_{x}^{-1}\widetilde{G}_{y}(\centerdot)$
and $\varphi_{y}^{h}(\centerdot)\triangleq\psi_{h^{l}}(\centerdot)$.
Consider $\hat{\varphi}_{x}^{h}(x;\centerdot)$, $\hat{\varphi}_{x}^{y}(x;\centerdot)$
and $\hat{\varphi}_{y}^{h}(y;\centerdot)$ as the \emph{compensation
estimators} for the disturbance components $ $$\varphi_{x}^{h}(\centerdot)$,
$\varphi_{x}^{y}(\centerdot)$ and $\varphi_{y}^{h}(\centerdot)$.
The corresponding compensated mixed timescale algorithm is given by,
\begin{eqnarray}
x_{n_{s}} & = & \mathcal{P}_{\mathcal{X}}\bigg[x_{n_{s}-1}+\gamma G(x_{n_{s}-1},y_{n_{f}};h^{s}(n_{f}\tau),h^{l}(n_{f}\tau))\label{eq:comp-xn}\\
 &  & \qquad\qquad+\hat{\varphi}_{x}^{h}(x_{n_{s}-1};\centerdot)(\triangle h^{s})_{n_{s}}+\hat{\varphi}_{x}^{y}(x_{n-1};\centerdot)(\triangle y)_{n_{s}}\bigg]\nonumber \\
y_{n_{f}} & = & \mathcal{P}_{\mathcal{Y}}\left[y_{n_{f}-1}+\gamma K(x_{n_{s}},y_{n_{f}-1};h^{s}(n_{f}\tau),h^{l}(n_{f}\tau))+\hat{\varphi}_{y}^{h}(y_{n_{f}-1};\centerdot)(\triangle h^{l})_{n_{f}}\right]\label{eq:comp-yn}
\end{eqnarray}
where $(\triangle h^{s})_{n_{s}}=h^{s}(\lfloor\frac{n_{s}}{N_{s}}\rfloor\tau)-h^{s}(\lfloor\frac{n_{s}-1}{N_{s}}\rfloor\tau)$,
$(\triangle y)_{n_{s}}=y_{\lfloor\frac{n_{s}}{N_{s}}\rfloor}-y_{\lfloor\frac{n_{s}-1}{N_{s}}\rfloor}$,
and $(\triangle h^{l})_{n_{f}}=h^{l}(n_{f}\tau)-h^{l}(n_{f}\tau-\tau)$.
The compensation terms are non-zero on the frame boundary.

\subsection{Derivation of the Compensation Estimators}

\label{sub:derivatio-comp-terms}

Obviously, if we can precisely estimate the disturbance, its impact
to the convergence can be totally suppressed and the tracking errors
$\widetilde{x}_{c}^{e}$ and $\widetilde{y}_{c}^{e}$ go to zero.
Unfortunately, we cannot obtain perfect estimations of the disturbance
terms $-\widetilde{G}_{x}^{-1}\widetilde{G}_{h^{s}}(\centerdot)$,
$-\widetilde{G}_{x}^{-1}\widetilde{G}_{y}(\centerdot)$, and $\psi_{h^{l}}(h^{l})$,
because they require closed form expressions of the target $(\hat{x}(y,h^{s},h^{l}),y^{*}(h^{l}))$.
In this section, we derive approximate compensation terms using Lagrange
duality theory \cite{Bertsekas:1999bs,Boyd:2004kx} and implicit function
theorem in calculus.

\subsubsection{Compensation for the Short-term Iteration $x_{n_{s}}$ }

Consider the optimality condition \cite{Boyd:2004kx} for the inner
problem (\ref{eq:the-problem-inner}), 
\[
\mathcal{G}(\theta_{x},\lambda_{x};\theta_{y},h^{s},h^{l})=\left[\begin{array}{c}
\nabla_{x}F(\theta_{x},\theta_{y};h^{s},h^{l})-\sum_{i}\lambda_{x,i}\nabla w_{i}(\theta_{x},\theta_{y};\centerdot)\\
\{\lambda_{x,i}w_{i}(\theta_{x},\theta_{y};\centerdot)\}_{i=1}^{W}
\end{array}\right]=0
\]
where $\lambda_{x,i}\geq0$ and $w_{i}(\theta_{x},\theta_{y};\centerdot)\leq0$,
for all $1\leq i\leq W$. From the Lagrangian duality theory, $\mathcal{G}(x;\theta_{y},h^{s},h^{l})=0$
has a unique solution $\hat{x}=(\hat{\theta}_{x},\hat{\lambda}_{x})$
for $\hat{\lambda}_{x}\geq0$, and the dynamics of $\hat{x}(t)$ should
satisfy $\mathcal{G}_{x}(\hat{x};\centerdot)\frac{d\hat{x}}{dt}+\mathcal{G}_{h^{s}}(\hat{x};\centerdot)\frac{dh^{s}}{dt}+\mathcal{G}_{y}(\hat{x};\centerdot)\frac{dy_{c}}{dt}=0$.
Suppose the matrix $\mathcal{G}_{x}(x;\centerdot)$ is invertible.
Using the implicit function theorem, the compensation estimators can
be given by 
\begin{equation}
\hat{\varphi}_{x}^{h}(x;\centerdot)=-\mathcal{G}_{x}^{-1}\mathcal{G}_{h^{s}}(x;\centerdot),\;\mbox{and}\;\hat{\varphi}_{x}^{y}(x;\centerdot)=-\mathcal{G}_{x}^{-1}\mathcal{G}_{y}(x;\centerdot).\label{eq:compensation-terms-x}
\end{equation}

\subsubsection{Compensation for the Long-term Iteration $y_{n}$}

From the Lagrange duality theory, the optimality condition for the
outer problem (\ref{eq:the-problem-outer}) is given by 
\begin{equation}
\mathcal{T}(\theta_{y},\lambda_{y};h^{l})=\left[\begin{array}{c}
\nabla_{y}\mathbb{E}F(\theta_{x},\theta_{y};h^{s},h^{l})-\sum_{j}\lambda_{y,j}\nabla q_{j}(\theta_{y};h^{l})\\
\{\lambda_{y,j}q_{j}(\theta_{y};h^{l})\}_{j=1}^{J}
\end{array}\right]=0\label{eq:compensation-equation-y-ideal}
\end{equation}
where $\lambda_{y,j}\geq0$ and $q_{j}(\theta_{y};\centerdot)\leq0$,
for all $1\leq j\leq J$. Similarly, $\mathcal{T}(y;\centerdot)=0$
has a unique solution $y^{*}=(\theta_{y}^{*},\lambda_{y}^{*})$ for
$\lambda_{y}^{*}\geq0$, and the dynamics of $y^{*}(t)$ should satisfy
$\mathcal{T}_{y}(y^{*};\centerdot)\frac{dy^{*}}{dt}+\mathcal{T}_{h^{l}}(y^{*};\centerdot)\frac{dh^{l}}{dt}=0$.
Also, using implicit function theorem, an ideal compensation estimator
can be given by $-\mathcal{T}_{y}^{-1}\mathcal{T}_{h^{l}}(y;\centerdot)$.

Note that closed form expression is usually not available for $\mathcal{T}(y;\centerdot)$,
due to the expectation $\mathbb{E}F(\centerdot)$. Alternatively,
define $\hat{\mathcal{T}}(\theta_{y},\lambda_{y};h^{s},h^{l})\triangleq\left[\begin{array}{c}
\nabla_{y}F(\theta_{x},\theta_{y};h^{s},h^{l})-\sum_{j}\lambda_{y,j}\nabla q_{j}(\theta_{y};\centerdot)\\
\{\lambda_{y,j}q_{j}(\theta_{y};\centerdot)\}_{j=1}^{J}
\end{array}\right]$. Then for given $h^{l}$, $\hat{\mathcal{T}}_{y}(y;\centerdot)$
is an unbiased estimator of $\mathcal{T}_{y}(y;\centerdot)$, since
$\mathbb{E}\hat{\mathcal{T}}(y;\centerdot)=\mathcal{T}(y;\centerdot)$.
Similarly, $\hat{\mathcal{T}}_{h^{l}}(y;\centerdot)$ is a unbiased
estimator of $\mathcal{T}_{h^{l}}(y;\centerdot)$. As a result, the
compensation estimator for the long-term iteration can be given by
\begin{equation}
\hat{\varphi}_{y}^{h}(y;\centerdot)=-\hat{\mathcal{T}}_{y}^{-1}\mathcal{\hat{T}}_{h^{l}}(y;\centerdot).\label{eq:compensation-term-y}
\end{equation}

\subsection{Performance of the Compensation Algorithm}

Although the compensation estimators derived in (\ref{eq:compensation-terms-x})
and (\ref{eq:compensation-term-y}) are from approximation, we can
show that under some technical conditions, the tracking errors $\widetilde{x}_{c}^{e}$
and $\widetilde{y}_{c}^{e}$ from continuous-time trajectories $x_{c}(t)$
and $y_{c}(t)$ can be significantly eliminated. 
\begin{lyxThmQED}
[Tracking performance of the compensation algorithm]\label{thm:conv-comp}
Assume CSI timescale separation for $h^{l}(t)$ and $h^{s}(t)$, i.e.,
$\epsilon\ll a_{H}$. Suppose that $\hat{\varphi}_{x}^{h}(\centerdot)$,
$\hat{\varphi}_{x}^{y}(\centerdot)$ and $\hat{\varphi}_{y}^{h}(y;\centerdot)$
take the forms in (\ref{eq:compensation-terms-x}) and (\ref{eq:compensation-term-y}),
and they are Lipschitz continuous, i.e., there exist positive constants
$\hat{L}_{x}^{h},\hat{L}_{x}^{y},\hat{L}_{y}^{h},\beta_{y}^{h}<\infty$,
such that $\|\hat{\varphi}^{h}(x;\centerdot)-\hat{\varphi}^{h}(\hat{x};\centerdot)\|\leq\hat{L}_{x}^{h}\|x-\hat{x}\|$,
$\|\hat{\varphi}^{y}(x;\centerdot)-\hat{\varphi}^{y}(\hat{x};\centerdot)\|\leq\hat{L}_{x}^{y}\|x-\hat{x}\|$
and $\mathbb{E}\|\hat{\varphi}_{y}^{h}(y;\centerdot)-\varphi_{y}^{h}(\centerdot)\|\leq\hat{L}_{y}^{h}\|y-y^{*}\|+\beta_{y}^{h}$,
for all $x\in\mathcal{X}(y)$, $y\in\mathcal{Y}$. Then, if 
\begin{eqnarray*}
 & \mbox{(i)} & \alpha_{y}\gamma-\epsilon\tau\hat{L}_{y}^{h}>0,\;\mbox{and}\\
 & \mbox{(ii)} & \alpha_{x}-\frac{a_{H}\tau}{\sqrt{2\pi}N_{s}\gamma}\hat{L}_{x}^{h}-\hat{L}_{x}^{y}N_{s}^{-1}l_{y}\frac{\epsilon\tau\beta_{y}^{h}}{\alpha_{y}\gamma-\epsilon\tau\hat{L}_{y}^{h}}-\frac{a_{H}\tau}{N_{s}\gamma}\left(\hat{L}_{x}^{h}\right)^{2}>0,
\end{eqnarray*}
the tracking error $\widetilde{x}_{c}^{e}$ converges to $0$ in probability,
and the the tracking error for $\widetilde{y}_{c}^{e}$ is upper bounded
by $\mathbb{E}\|\widetilde{y}_{c}^{e}\|\leq\frac{\epsilon\tau\beta_{y}^{h}}{\alpha_{y}\gamma-\epsilon\tau\hat{L}_{y}^{h}}$.
\end{lyxThmQED}
\begin{proof}
Please refer to Appendix \ref{app:thm-conv-comp} for the proof.
\end{proof}

In the case when $h^{l}$ is static, the convergence result is given
in the following corollary.
\begin{lyxCorQED}
[Tracking performance under static $h^{l}$]\label{cor:conv-comp-static-hl}
Suppose $\hat{\varphi}_{x}^{h}(\centerdot)$ and $\hat{\varphi}_{x}^{y}(\centerdot)$
are given by (\ref{eq:compensation-terms-x}) and Lipschitz continuous
as specified in Theorem \ref{thm:conv-comp}. Then if $\alpha_{x}-\frac{a_{H}\tau}{\sqrt{2\pi}N_{s}\gamma}\hat{L}_{x}^{h}-\hat{L}_{x}^{y}N_{s}^{-1}l_{y}\frac{\epsilon\tau\beta_{y}^{h}}{\alpha_{y}\gamma-\epsilon\tau\hat{L}_{y}^{h}}-\frac{a_{H}\tau}{N_{s}\gamma}\left(\hat{L}_{x}^{h}\right)^{2}>0$,
the tracking error $\widetilde{x}_{c}^{e}$ converges to $0$ in probability.
\end{lyxCorQED}

Corollary \ref{cor:conv-comp-static-hl} is obtained by setting the
timescale parameter $\epsilon=0$ in Theorem \ref{thm:conv-comp}.
It corresponds to the case 1 scenario studied in Section \ref{sub:case-1-static-h^l}.
However, the performance with compensation is stronger because it
does not require the short-term CSI timescale to be extremely slower
than the algorithm timescale (i.e., $a_{H}\tau\ll\gamma$) in Theorem
\ref{thm:Connection-discrete-alg-MCTS} and Corollary \ref{cor:convergence-slow-varying}
for the convergence.
\begin{remrk}
[Interpretation of the results] Theorem \ref{thm:conv-comp} and
Corollary \ref{cor:conv-comp-static-hl} show the performance advantage
of the compensation algorithm under time-varying CSI. Specifically,
we have the following observations.

\begin{itemize}

\item\emph{ Compensation for the long-term control}: When the bias
$\beta_{y}^{h}$ of the compensation estimator goes to $0$, the tracking
error $\widetilde{y}_{c}^{e}=y_{c}-y^{*}$ of the long-term variable
$y$ converges to $0$ as well. Note that the bias $\beta_{y}^{h}$
comes from using $\nabla_{y}F(\centerdot)$ to estimate $\nabla_{y}\mathbb{E}F(\centerdot)$
in (\ref{eq:compensation-equation-y-ideal}). To reduce the bias,
we can use a Monte-Carlo method to estimate $\widehat{\nabla_{y}\mathbb{E}F(\centerdot)}=\frac{1}{M}\sum_{m=1}^{M}\nabla_{y}F(x_{m},y_{0})$,
by observing many realizations of $\nabla_{y}F(x(t),y(t_{0});\centerdot)$
in the inner timescale.

\item\emph{ Compensation for the short-term control}: Theorem \ref{thm:conv-comp}
implies that when the short timescale CSI $h^{s}(t)$ does not change
too fast (moderate $a_{H}$), the compensation algorithm can keep
track with the stationary point target. This is a much weaker condition
for the conventional convergence result, which requires $a_{H}\tau\ll\gamma$,
i.e., the algorithm must iterate much faster than the CSI dynamics.
Moreover, one can reduce the frame duration $\tau$, increase the
number of slots $N_{s}$ per frame, or increase the step size $\gamma$
to satisfy condition (ii) for enhancing the tracking of the inner
iteration $x_{n_{s}}$, at the cost of larger signaling overhead,
higher computational complexity and larger steady state error $\mathcal{O}(\gamma)$
as discussed in Section \ref{sub:stability-algorithm}.

\end{itemize}~\hfill\IEEEQED
\end{remrk}

Note that, even though there can be zero convergence errors for $x_{c}$
and $y_{c}$ in (\ref{eq:comp-xc}) and (\ref{eq:comp-yc}), the discrete-time
iterations (\ref{eq:comp-xn}) and (\ref{eq:comp-yn}) still have
$\mathcal{O}(\gamma)$ steady state error due to the constant step
size $\gamma$ used for the tracking. Nevertheless, Theorem \ref{thm:conv-comp}
and Corollary \ref{cor:conv-comp-static-hl} indicate that the proposed
compensation algorithm has an eminent convergence capability under
time-varying$ $ $h^{s}(t)$ and $h^{l}(t)$.

\section{An Application Example: Resource Allocations in Wireless Multi-hop
Relay Network}

\label{sec:Example}

The mixed timescale optimization approach has vast applications in
wireless communication networks. In the following, we consider a particular
example of joint flow control and power allocation in wireless relay
network described in Section \ref{sub:example}. From this example,
we demonstrate the compensation algorithm and apply the theoretical
results for the convergence analysis.

\subsection{The Two-Timescale Algorithm}

\label{sub:ex-two-timescale-alg}

We apply the stochastic primal-dual method in (\ref{eq:alg-stochastic-primal-dual})
to derive the iterative algorithm $x_{n_{s}}$ and $y_{n_{f}}$ in
this example. Denote $x=(\mathbf{p},\lambda)$, where $\lambda=(\lambda_{1},\dots,\lambda_{W})$
is the Lagrange multiplier. For the example problem in (\ref{eq:ex-objective}),
we can form the Lagrange function as 
\begin{equation}
L(\mathbf{p},\lambda,\mathbf{r};\mathbf{h})=\sum_{j\in\mathcal{E}}\log(r_{j})-V\sum_{k\in\mathcal{L}}p_{k}-\sum_{j=1}^{J}\lambda_{j}\left[\sum_{k\in\mathcal{S}_{j}}c_{k}(\mathbf{r})-\log\left(1+\sum_{k\in\mathcal{S}_{j}}|h_{k}|^{2}p_{k}\right)\right]\label{eq:ex-Lagrangian}
\end{equation}
where $W$ is the total number of constraints, and $\mathcal{S}_{j}\subset\mathcal{L}^{+}(m)$
for some $m\in\mathcal{R}\cup\{0\}$.

\subsubsection{Iteration for the short-term variable}

The optimality condition (KKT condition \cite{Boyd:1987dz}) for the
inner problem is given by 
\[
\mathcal{G}(\mathbf{p},\lambda;\mathbf{r},\mathbf{h})=\left[\begin{array}{c}
\frac{\partial}{\partial\mathbf{p}}L(\mathbf{p},\lambda,\mathbf{r})\\
\left\{ \lambda_{j}\left[\sum_{k\in\mathcal{S}_{j}}c_{k}(\mathbf{r})-\log\left(1+\sum_{k\in\mathcal{S}_{j}}|h_{k}|^{2}p_{k}\right)\right]\right\} _{j=1}^{W}
\end{array}\right]=\mathbf{0}.
\]
Following the adaptive compensation algorithm in Section \ref{sec:adaptving-algorithm},
the iteration of the short-term variable is given by 
\begin{equation}
\mathbf{p}(n_{s}+1)=\mathcal{P}_{\mathbf{p}}\left[\mathbf{p}(n_{s})+\gamma\frac{\partial}{\partial\mathbf{p}}L\left(\mathbf{p}(n_{s}),\lambda(n_{s});\mathbf{r}(n_{f})\right)+\hat{\Psi}_{p}(\centerdot)\right]\label{eq:ex-alg-power}
\end{equation}
\begin{equation}
\lambda(n_{s}+1)=\mathcal{P}_{\lambda}\left[\lambda(n_{s})-\gamma\frac{\partial}{\partial\lambda}L\left(\mathbf{p}(n_{s}),\lambda(n_{s});\mathbf{r}(n_{f})\right)+\hat{\Psi}_{\lambda}(\centerdot)\right]\label{eq:ex-alg-power-lambda}
\end{equation}
where the projection $\mathcal{P}_{\mathbf{p}}\left(\centerdot\right)$
and $\mathcal{P}_{\lambda}\left(\centerdot\right)$ are to restrict
the elements to be non-negative. The term $(\frac{\partial}{\partial\mathbf{p}}L(\centerdot),\frac{\partial}{\partial\lambda}L(\centerdot))$
corresponds to the iteration mapping $G(\centerdot)$ in (\ref{eq:alg-xn})
(and (\ref{eq:comp-xn})). The compensations $\hat{\Psi}_{p}(\centerdot)$
and $\hat{\Psi}_{\lambda}(\centerdot)$ can be derived as 
\[
\left(\begin{array}{c}
\hat{\Psi}_{p}(\centerdot)\\
\hat{\Psi}_{\lambda}(\centerdot)
\end{array}\right)=-\mathcal{G}_{(\mathbf{p},\lambda)}^{-1}\mathcal{G}_{\mathbf{h}}(\mathbf{p}(n_{s}),\centerdot)\triangle\mathbf{h}(n_{s})-\mathcal{G}_{(\mathbf{p},\lambda)}^{-1}\mathcal{G}_{y}((\mathbf{p}(n_{s}),\centerdot)\triangle\mathbf{r}(n_{s})
\]
where $\triangle\mathbf{h}(n_{s})=\mathbf{h}(\lfloor\frac{n_{s}}{N_{s}}\rfloor\tau)-\mathbf{h}(\lfloor\frac{n_{s}-1}{N_{s}}\rfloor\tau)$
and $\triangle\mathbf{r}(n_{s})=\mathbf{r}(\lfloor\frac{n_{s}}{N_{s}}\rfloor)-\mathbf{r}(\lfloor\frac{n_{s}-1}{N_{s}}\rfloor)$.

\subsubsection{Iteration for the long-term variable}

We first derive an augmented Lagrange function $L_{1}(\centerdot)$
by substituting the equality constraints (\ref{eq:ex-outer-constraint-flow-balance})
into the Lagrangian $L(\centerdot)$ in (\ref{eq:ex-Lagrangian}).
The optimality condition for the outer problem is given by 
\[
\mathcal{T}(\mathbf{r};\mathbf{h}^{l})=\frac{\partial}{\partial\mathbf{r}}\mathbb{E}L_{1}(\mathbf{p}(n_{s}),\lambda(n_{s}),\mathbf{r}(n_{f}))=0.
\]
The update of the long-term variable is given by 
\begin{equation}
\mathbf{r}(n_{f}+1)=\mathcal{P}_{\mathbf{r}}\left[\mathbf{r}(n_{f})+\gamma\frac{\partial}{\partial\mathbf{r}}L_{1}\left(\mathbf{p}(n_{s}),\lambda(n_{s}),\mathbf{r}(n_{f})\right)+\hat{\Psi}_{r}(\centerdot)\right]\label{eq:ex-alg-rate}
\end{equation}
where the projection $\mathcal{P}_{\mathbf{r}}\left(\centerdot\right)$
is to restrict $\mathbf{r}$ to be non-negative. The iteration (\ref{eq:ex-alg-rate})
corresponds to the long-term variable update for $y_{n_{f}}$ in (\ref{eq:alg-yn}),
and the term $\frac{\partial}{\partial\mathbf{r}}L_{1}\left(\mathbf{p}(n_{s});\lambda(n_{s}),\mathbf{r}(n_{f})\right)$
corresponds to the stochastic estimator $K(\centerdot)$ in (\ref{eq:alg-yn}).
The compensation $\hat{\Psi}_{r}(\centerdot)$ can be derived as 
\[
\hat{\Psi}_{r}(\mathbf{r}(n_{f}),\mathbf{h}^{l}(n_{f}\tau))=-\hat{\mathcal{T}}_{r}^{-1}\mathcal{\hat{T}}_{h^{l}}(\mathbf{r}(n_{f});\mathbf{h}^{l}(n_{f}))(\mathbf{h}^{l}(n_{f}\tau)-\mathbf{h}^{l}((n_{f}-1)\tau))
\]
where $\hat{\mathcal{T}}(\mathbf{r};\mathbf{h}^{l})=\frac{\partial}{\partial\mathbf{r}}L_{1}(\mathbf{p},\lambda,\mathbf{r};\mathbf{h}^{l})$
and the path loss variable $\mathbf{h}^{l}(t)$ can be measured by
averaging the CSI $\mathbf{h}(t)$ over a certain time window.

Table \ref{tab:alg-association} summaries the algorithm association
between the example and the mixed timescale model.

\begin{table}
\begin{centering}
\begin{tabular}{|c|c|c|c|}
\hline 
\multicolumn{2}{|c|}{Components in the example} & \multicolumn{2}{c|}{Corresponding components in the model}\tabularnewline
\hline 
\hline 
$(\mathbf{p}(n_{s}),\lambda(n_{s}))$ & Primal-dual inner iteration (\ref{eq:ex-alg-power})-(\ref{eq:ex-alg-power-lambda}) & $x_{n_{s}}$ & Short-term iterative sequence (\ref{eq:alg-xn})\tabularnewline
\hline 
$\mathbf{r}(n_{f})$ & Primal outer iteration (\ref{eq:ex-alg-rate}) & $y_{n_{f}}$ & Long-term iterative sequence (\ref{eq:alg-yn})\tabularnewline
\hline 
$(\hat{\Psi}_{p}(\centerdot),\hat{\Psi}_{\lambda}(\centerdot))$ & Compensation for the inner iteration & $-\hat{\varphi}_{x}^{h}dh^{s}-\hat{\varphi}_{x}^{y}dy$ & Compensation for the inner iteration\tabularnewline
\hline 
$\hat{\Psi}_{r}$ & Compensation for the outer iteration & $\hat{\varphi}_{y}^{h}dh^{l}$ & Compensation for the outer iteration\tabularnewline
\hline 
$\mathbb{R}_{+}^{|\mathcal{L}|}\times\mathbb{R}_{+}^{W}$ & Projection domain for the inner iteration & $\mathcal{X}(y)$ & Projection domain for $x_{n_{s}}$\tabularnewline
\hline 
$\mathbb{R}_{+}^{|\mathcal{L}|}$ & Projection domain for the outer iteration & $\mathcal{Y}$ & Projection domain for $y_{n_{f}}$\tabularnewline
\hline 
\end{tabular}
\par\end{centering}

\caption{\label{tab:alg-association} Algorithm associations between the example
and the mixed timescale system model.}
\end{table}

\subsection{Implementation Considerations}

With the iteration timescale decomposition, we can consider two implementation
scenarios of the two-timescale algorithm: a) \emph{distributive implementation},
and b) \emph{hybrid implementation}. 

\begin{figure}
\begin{centering}
\subfigure[Distributive implementation]{\includegraphics[width=0.7\columnwidth]{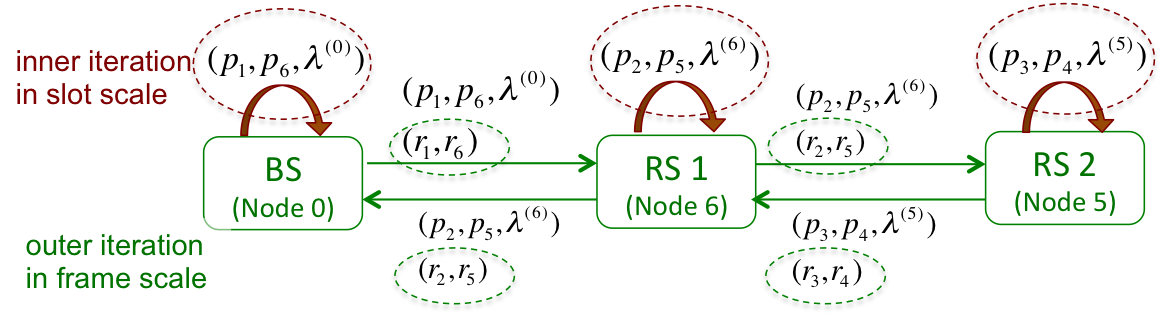}}\\ 
\subfigure[Hybrid implementation]{\includegraphics[width=0.7\columnwidth]{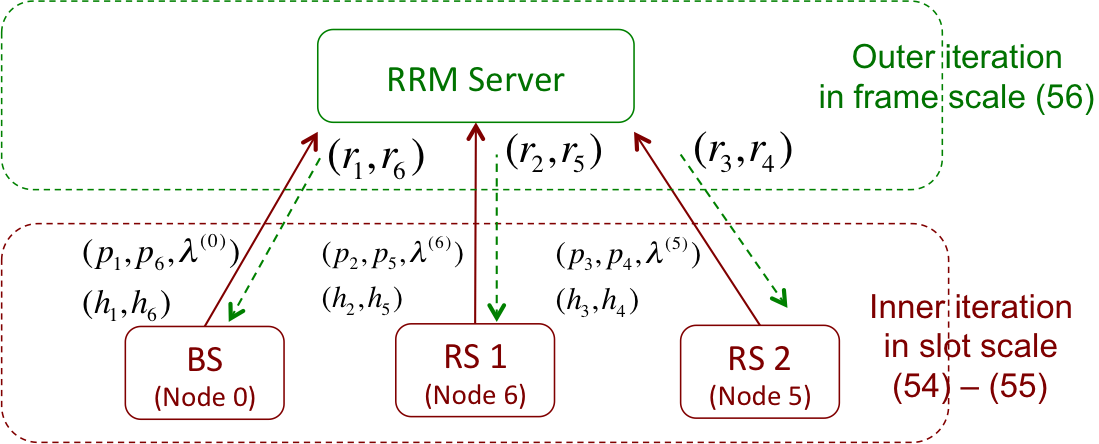}}
\par\end{centering}

\caption{\label{fig:implementation} Example algorithm implementations and
message passing for the network topology in Fig. \ref{fig:topology-relay-network}.}
\end{figure}

Under distributive implementation, at each frame, each BS and RS node
$m\in\mathcal{R}\cup\{0\}$ acquires the local CSI $\{h_{j}\}_{j\in\mathcal{L}^{+}(m)}$
and exchange the local control variables $\{p_{j}\}_{j\in\mathcal{L}^{+}(m)}$,
$\{\lambda_{i}^{(m)}\}$ and $\{r_{j}\}_{j\in\mathcal{L}^{+}(m)}$
with neighbor nodes. It then updates the long-term flow control $\{r_{j}\}_{j\in\mathcal{L}^{+}(m)}$
once according to the outer iteration (\ref{eq:ex-alg-rate}), and
updates the power control variables $\{p_{j}\}_{j\in\mathcal{L}^{+}(m)}$
and $\{\lambda_{i}^{(m)}\}$ in each time slot according to the inner
iterations (\ref{eq:ex-alg-power}) and (\ref{eq:ex-alg-power-lambda}).
As an illustrative example, Fig. \ref{fig:implementation}. a) demonstrates
the message passing under distributive implementation and the network
topology in Fig. \ref{fig:topology-relay-network}.

Under hybrid implementation, there is a RRM server coordinating the
message passing and the outer loop iterations in the network as illustrated
in Fig. \ref{fig:topology-relay-network}.. At the beginning of each
frame, each BS and RS node $m\in\mathcal{R}\cup\{0\}$ obtains long-term
flow control $\{r_{j}\}_{j\in\mathcal{L}^{+}(m)}$ from the RRM server
and acquires the local CSI $\{h_{j}\}_{j\in\mathcal{L}^{+}(m)}$.
It then updates the local power control variables $\{p_{j}\}_{j\in\mathcal{L}^{+}(m)}$
and $\{\lambda_{i}^{(m)}\}$ according to the inner iterations (\ref{eq:ex-alg-power})
and (\ref{eq:ex-alg-power-lambda}) in each time slot within the frame.
At the end of the frame, it passes the local variables $\{p_{j}\}$
and $\{\lambda_{i}^{(m)}\}$ together with the local CSI $\{h_{j}\}$
for $j\in\mathcal{L}^{+}(m)$ to the RRM server. By collecting the
short term variables $\mathbf{p}$ and $\lambda$ as well as the global
CSI $\mathbf{h}$, the RRM server updates the long-term flow control
$\mathbf{r}$ using the outer iteration (\ref{eq:ex-alg-rate}) and
feeds back to the BS and RSs at the beginning of the next frame. Fig.
\ref{fig:implementation}.b) illustrates the message passing under
hybrid implementation and the network topology in Fig. \ref{fig:topology-relay-network}.

Note that as the inner iterations (\ref{eq:ex-alg-power}) and (\ref{eq:ex-alg-power-lambda})
require only local CSI, they can be iterated for a finite number of
steps $N_{s}>1$ at each frame to catch up with the fast timescale
CSI variations. On the other hand, since the outer iteration (\ref{eq:ex-alg-rate})
requires global coordination which involves signaling latency, it
can only be updated once at each frame. However, as the long term
flow control $\mathbf{r}$ adapts to CSI statistics (i.e., the long
term CSI $\mathbf{h}^{l}$), it does not require a fast iteration
and is not sensitive to signaling latency.

Considering the computational complexity, the two-timescale algorithm
with variable partitions reduces the computational cost at the central
controller by distributing the computation of the cross-layer network
utility optimization to different nodes locally. Table \ref{tab:Complexity-CPU-time}
gives a comparison on the computational complexity in terms of CPU
time over one frame under the example in Section \ref{sub:example}.
The inner iteration is assumed to update for $N_{s}=30$ steps in
each frame under two-timescale algorithms. The one-timescale centralized
algorithm consumes more CPU time, which may not be a good choice,
since the transmission power control is delay-sensitive.

\begin{table}
\begin{centering}
\begin{tabular}{|c|c|c|}
\hline 
 & RRM Server (ms) & Each BS (RS) (ms)\tabularnewline
\hline 
\hline 
Two-timescale distributive algorithm  & - & 0.601\tabularnewline
\hline 
Two-timescale hybrid algorithm  & 0.0511 & 0.533\tabularnewline
\hline 
One-timescale centralized algorithm & 1.26 & -\tabularnewline
\hline 
\end{tabular}
\par\end{centering}

\caption{\label{tab:Complexity-CPU-time}Computational complexity in terms
of CPU time of the one-timescale centralized algorithm and two-timescale
algorithms over one frame. The simulation was done on a MATLAB platform
running on a desktop computer with a 2.8 GHz single core CPU. }
\end{table}

\subsection{The Convergence Analysis}

In this subsection, we derive the upper bound of the tracking error
of the two-timescale iterations (\ref{eq:ex-alg-power})-(\ref{eq:ex-alg-rate})
using the theoretical results developed in Section \ref{sec:convergence-VSDS}.
We derive the convergence rate parameters $\alpha_{x}$ and $\alpha_{y}$
as follows.
\begin{lyxThmQED}
[Local convergence speed]\label{thm:ex-convergence-speed} Denote
\[
M_{L}(\centerdot)=\left(\begin{array}{cc}
\frac{\partial^{2}}{\partial\mathbf{p}\partial\mathbf{p}}L(\centerdot) & \frac{\partial^{2}}{\partial\mathbf{p}\partial\lambda}L(\centerdot)\\
-\frac{\partial^{2}}{\partial\lambda\partial\mathbf{p}}L(\centerdot) & \mathbf{0}_{J\times J}
\end{array}\right).
\]
Then $M_{L}(\centerdot)$ is negative definite for all $(\mathbf{p}^{*},\mathbf{r}^{*})$
and $\lambda^{*}(\mathbf{p}^{*})$, under all $\mathbf{h}$. In addition,
given any optimal points $\mathbf{w}=(\mathbf{p}^{*},\mathbf{r}^{*})$,
we have the local convergence rate $\alpha_{x}(\mathbf{w})\geq-\lambda_{\max}(\frac{1}{2}(M_{L}(\mathbf{w})+M_{L}^{T}(\mathbf{w})))$,
$\alpha_{y}(\mathbf{w})=-\lambda_{\max}\left(\frac{\partial^{2}}{\partial\mathbf{r}\partial\mathbf{r}}L(\mathbf{w})\right)$,
where $\lambda_{\max}(A)$ denotes the maximum eigenvalue of matrix
$A$.
\end{lyxThmQED}
\begin{proof}
Please refer to Appendix \ref{app:pf-thm-ex-conv-speed} for the proof.
\end{proof}

Using Theorem \ref{thm:ex-convergence-speed}, a lower bound of the
global convergence rate can be obtained by%
\footnote{In fact, the domain $\mathcal{X}(y)\times\mathcal{Y}$ may not be
compact, and $\alpha_{x}$ (or $\alpha_{y}$) may then be degenerated.
However, in practice, the control variables $x$ and $y$ (corresponding
to power and flow data rate here) cannot go unbounded. Therefore,
one can identify a confident domain $\overline{\mathcal{X}}(y)\times\overline{\mathcal{Y}}\subseteq\mathcal{X}(y)\times\mathcal{Y}$
and $\overline{\mathcal{H}}\subseteq\mathcal{H}$, which are compact,
to estimate the lower bound of the convergence rate $\alpha_{x}$
(or $\alpha_{y}$).%
} $\alpha_{x}=\inf\{\alpha_{x}(\mathbf{\bm{\omega}}):\bm{\omega}\in\mathcal{X}(y)\times\mathcal{Y},\forall y\in\mathcal{Y},\forall\mathbf{h}\}$
and $\alpha_{y}=\inf\{\alpha_{y}(\bm{\omega}):\bm{\omega}\in\mathcal{X}(y)\times\mathcal{Y},\forall y\in\mathcal{Y},\forall\mathbf{h}\}$.

Given the results in Theorem \ref{thm:ex-convergence-speed}, the
the condition for algorithm stability and tracking error bound then
directly follows from the results in Theorem \ref{thm:sufficient-cond-stability-alg}
and \ref{thm:upper-bound-error}, respectively.

Note that we can always enhance the convergence and increase $\alpha_{x}$
and $\alpha_{y}$ by introducing a carefully chosen positive definite
scaling matrix $\Gamma$ in the iterations. However, the computation
of the scaling matrix may increase the complexity for the inner iteration
and require additional signaling overhead for the outer iteration.

\section{Numerical Results}

\label{sec:numerical-results}

In this section, we simulate the tracking performance of the mixed
timescale algorithm for the example cross-layer stochastic optimization
problem studied in Section \ref{sub:example} and Section \ref{sec:Example}.
We demonstrate the performance advantage of the mixed timescale algorithm
over one-timescale algorithms under the CSI model discussed in Section
\ref{sub:CSI-model}. In addition, we show that the proposed two-timescale
compensation algorithm in Section \ref{sec:adaptving-algorithm} significantly
reduces the tracking error under time-varying CSI.

We consider the wireless heterogeneous relay network described in
Section \ref{sub:example}. Specifically, the network has $1$ macro
BS, $2$ RSs and $4$ mobile users who want to transmit data flows
to the macro BS. The BSs are static and the mobiles are moving around
with a speed at most $v_{\max}=100$ km/h. The mobility is according
to the Levy walk mobility model in Section \ref{sub:network-mobility-model}
with parameter $D_{\min}=75$ m, $\iota=1.8$ and $c_{0}=D_{\min}^{\iota}$.
There are $6$ wireless links as illustrated in Fig. \ref{fig:topology-relay-network},
and it is assumed that the network topology does not change during
the simulation. Correspondingly, the long-term CSI timescale parameter
is $\epsilon=6\times10^{-4}$ sec$^{-1}$. The control objective is
to determine the flow rate congestion control $\mathbf{r}$ and power
allocation $\mathbf{p}$ according to the proportional fair utility
in (\ref{eq:ex-objective}). The frame duration is $\tau=1$ ms, and
the inner iterations (\ref{eq:ex-alg-power}) and (\ref{eq:ex-alg-power-lambda})
are updated for $N_{s}=30$ steps in each frame. 

We consider the following baseline schemes:
\begin{itemize}
\item \textbf{Baseline 1 - One-timescale centralized algorithm based on
real-time global CSI} \cite{Chiang2005,Chen:2006nx}: The central
controller (RRM server in Fig. \ref{fig:topology-relay-network})
solves the deterministic version (dropping the expectation) of the
problem (\ref{eq:ex-objective}) at each time slot. The controller
collects real-time global CSI (GCSI) $\mathbf{h}(t)$ at each time
slot and computes the optimal flow rate congestion control $\mathbf{r}(\mathbf{h}(t))$
and power allocation $\mathbf{p}(\mathbf{h}(t))$ that adapt to each
realization of $\mathbf{h}(t)$. 
\item \textbf{Baseline 2 - One-timescale centralized algorithm based on
statistical CSI} \cite{Papandriopoulos:2008kl}: For every $T_{s}=100$
ms, the central controller solves a relaxed version of the stochastic
optimization problem (\ref{eq:ex-objective}), where the link capacity
constraint (\ref{eq:ex-constraint-capacity}) is replaced by the probability
outage constraint $\mbox{Pr}\left[\eqref{eq:ex-constraint-capacity}\mbox{ is not satisfied}\right]\leq\Theta_{\text{out}}$,
and the flow rate congestion control $\mathbf{r}$ and power allocation
$\mathbf{p}$ adapt to the statistics of the GCSI $\mathbf{h}(t)$.
\item \textbf{Baseline 3 - Two-timescale stochastic gradient without compensations}:
The algorithm iterations are based on stochastic gradient in (\ref{eq:alg-stochastic-projected-x})
and (\ref{eq:alg-stochastic-projected-y}) in solving Problem 4. 
\end{itemize}

Note that the baseline 1 suffers from huge computational complexity,
as it searches for the optimal solution at each time slot, which is
not scalable to large networks. Moreover, baseline 1 is very sensitive
to signaling latency for the message passing throughout the network%
\footnote{In the current practical communication networks, such as LTE, the
backhaul latency is typically around $10$-$20$ ms \cite{TR36814}.%
}. On the other hand, baseline 2 is not sensitive to the signaling
latency but it is too conservative as it does not exploit the local
real-time CSI knowledge at the BS and the RSs. Hence, baseline 1 and
baseline 2 are for performance benchmark only.

\subsection{Performance of the Mixed Timescale Algorithms}

Due to the exogenous stochastic variation of $h^{s}(t)$ and $h^{l}(t)$,
the instantaneous link capacity constraint in (\ref{eq:ex-constraint-capacity})
and (\ref{eq:ex-constraint-capacity-2}) may not be satisfied for
every iteration outputs. To quantify the associated performance penalty,
we define the constraint outage probability as follows 
\[
\mathbb{P}_{\text{out}}=\frac{1}{N_{T}}\sum_{n=1}^{N_{T}}\sum_{j\in\mathcal{L}}1\left\{ r_{j}\notin\mathcal{C}_{m}^{\text{cap}}(\mathbf{p}(n),\mathbf{h}(n)),\forall m:j\in\mathcal{L}^{+}(m)\right\} 
\]
where $N_{T}$ is the total number of transmission frames, $1\{\centerdot\}$
is the indicator function, and $\mathcal{C}_{m}^{\text{cap}}(\mathbf{p},\mathbf{h})$
is the multi-access channel (MAC) capacity region at receiver node
$m$, and is specified by (\ref{eq:MUD-MAC}). 

\begin{figure}
\begin{centering}
\includegraphics[width=0.8\columnwidth]{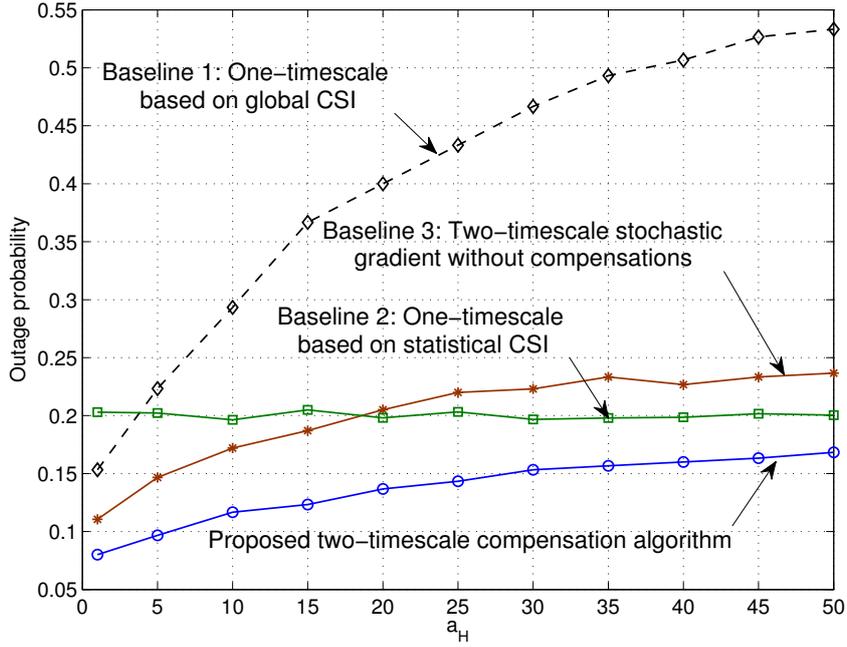}
\par\end{centering}

\caption{\label{fig:sim-outage} The constraint outage probability under different
CSI fading parameters $a_{H}$ and $\epsilon=6\times10^{-4}$. The
signaling latency is $\tau=5$ ms.}
\end{figure}

Note that, $a_{H}=50$ corresponds to around $10$ ms channel coherence
time \cite{Tse2005:fundamental:Wireless} and $a_{H}=1$ yields over
200 ms channel coherence time. Fig. \ref{fig:sim-outage} shows the
constraint outage probability under different CSI fading parameters
$a_{H}$ and $\epsilon=6\times10^{-4}$. The constraint outage probability
increases when the channel is changing faster, but the proposed two-timescale
compensation algorithm has the least constraint outage probability
compared with other baselines under $5$ ms signaling latency and
various channel fading rates.

\begin{figure}
\begin{centering}
\subfigure[Assuming no signaling latency]{\includegraphics[width=0.49\columnwidth]{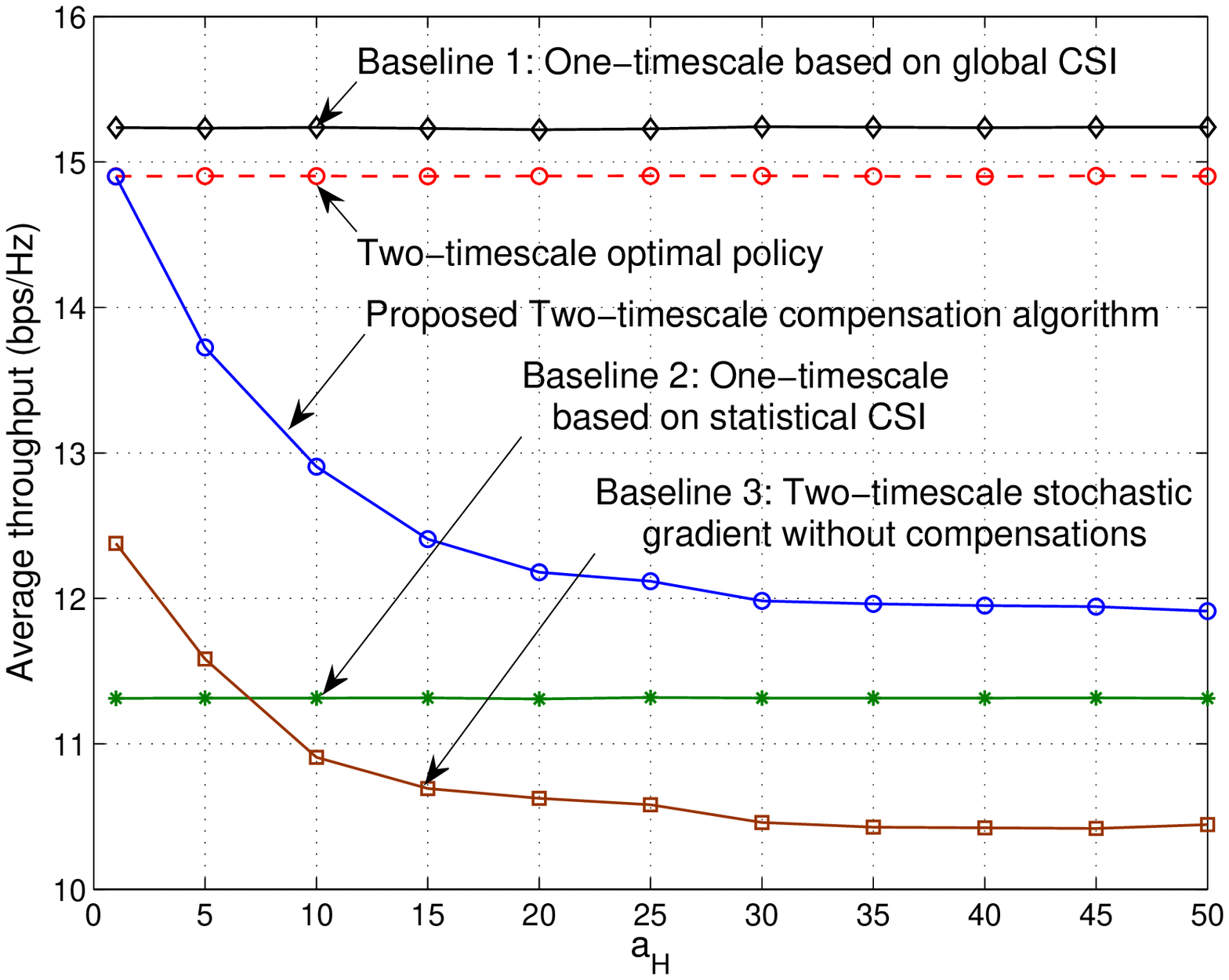}}\subfigure[$\tau=5$ ms signaling latency]{\includegraphics[width=0.49\columnwidth]{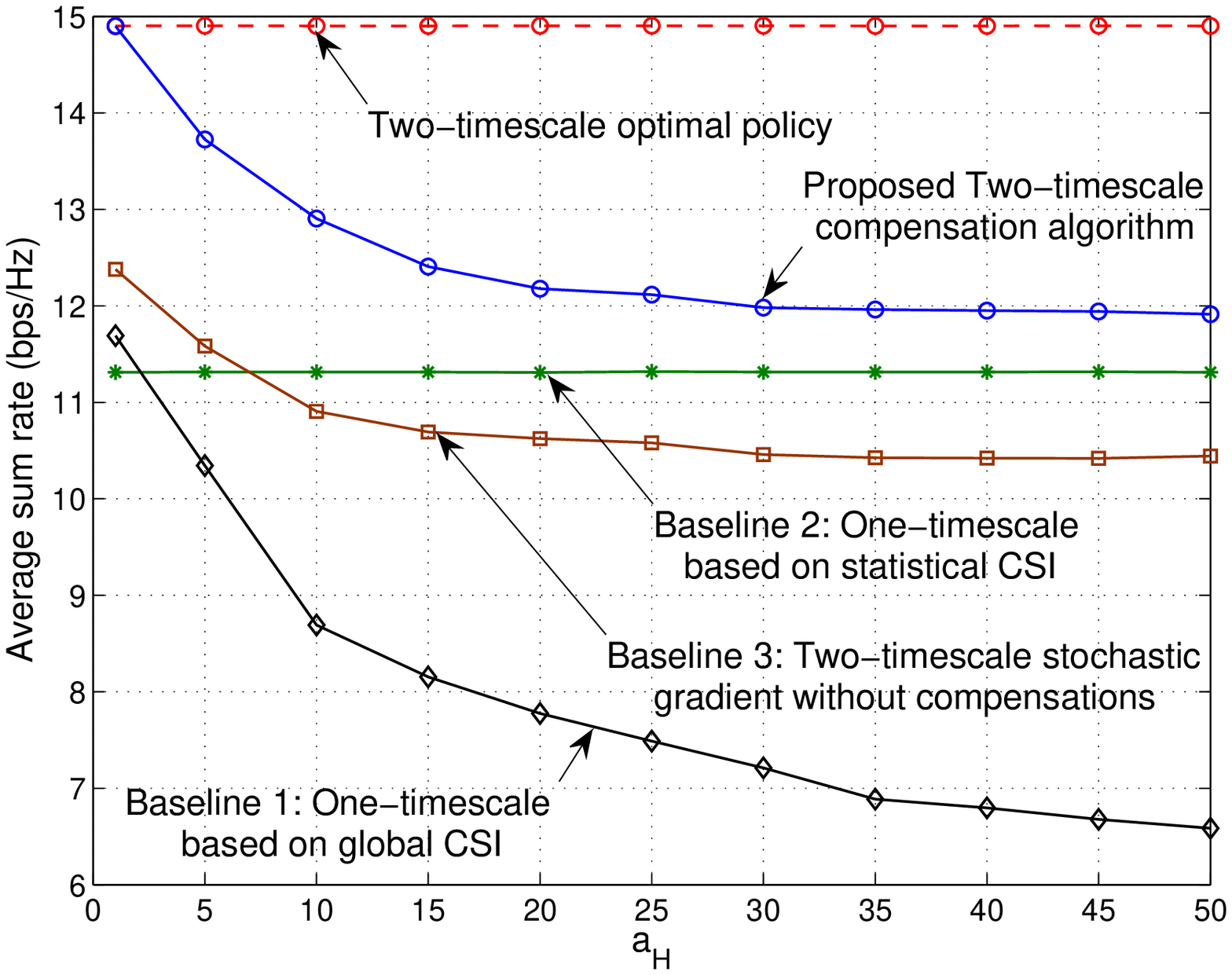}}
\par\end{centering}

\caption{\label{fig:sim-sum-rate} Throughput performance of the different
schemes under $\epsilon=6\times10^{-4}$ and average SNR $11$ dB.
Note that, $a_{H}=50$ corresponds to around $10$ ms channel coherence
time \cite{Tse2005:fundamental:Wireless} and $a_{H}=1$ yields over
200 ms channel coherence time. }
\end{figure}

\begin{figure}
\begin{centering}
\includegraphics[width=0.8\columnwidth]{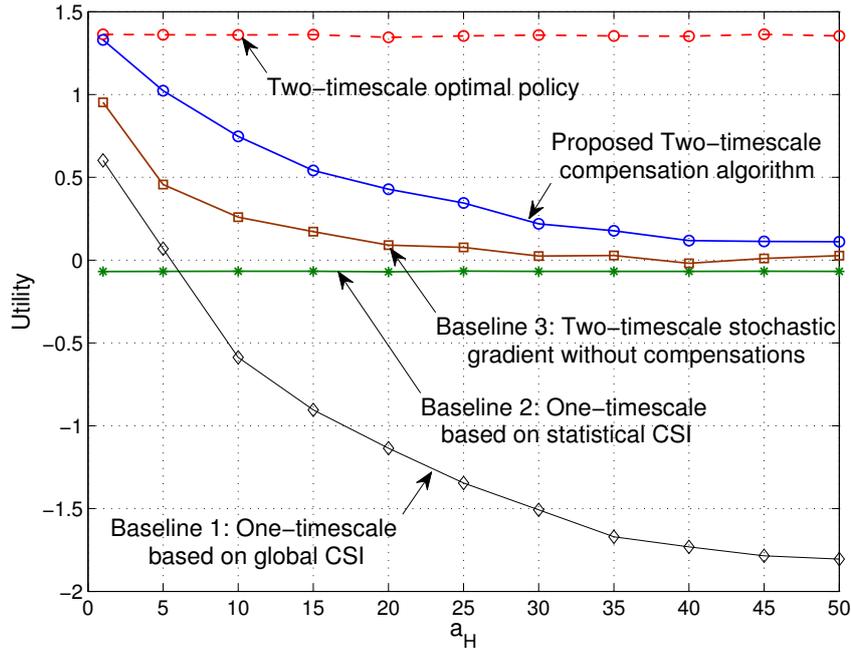}
\par\end{centering}

\caption{\label{fig:sim-utility} Proportional fair utility of the different
schemes under 11 dB average SNR, $5$ ms signaling latency and the
long timescale CSI parameter $\epsilon=6\times10^{-4}$. The proposed
algorithm performs much better than all the baseline schemes.}
\end{figure}

Fig. \ref{fig:sim-sum-rate}.a) gives the throughput performance assuming
no signaling latency. Baseline 1 yields the best performance, but
it is highly sensitive to signaling latency, as shown in Fig. \ref{fig:sim-sum-rate}.b),
where $5$ ms signaling latency is considered. In Fig. \ref{fig:sim-sum-rate}.b),
as the CSI varies faster, the throughput performance of all the schemes
decrease, except for baseline 2. However, baseline 2 does not exploit
the short-term transmission opportunity and achieves only moderate
performance. As a comparison, the proposed two timescale compensation
algorithm has the best performance and is robust to signaling latency.
Fig. \ref{fig:sim-utility} demonstrates the corresponding proportional
fair utility for the different schemes under signaling latency of
$5$ ms. The proposed algorithm performs much better than all the
other schemes. 

\subsection{Tracking Performance of the Adaptive Compensation Algorithm}

We evaluate the tracking performance of the two-timescale compensation
algorithm over the baseline stochastic gradient algorithm.

\begin{figure}
\begin{centering}
\includegraphics[width=0.8\columnwidth]{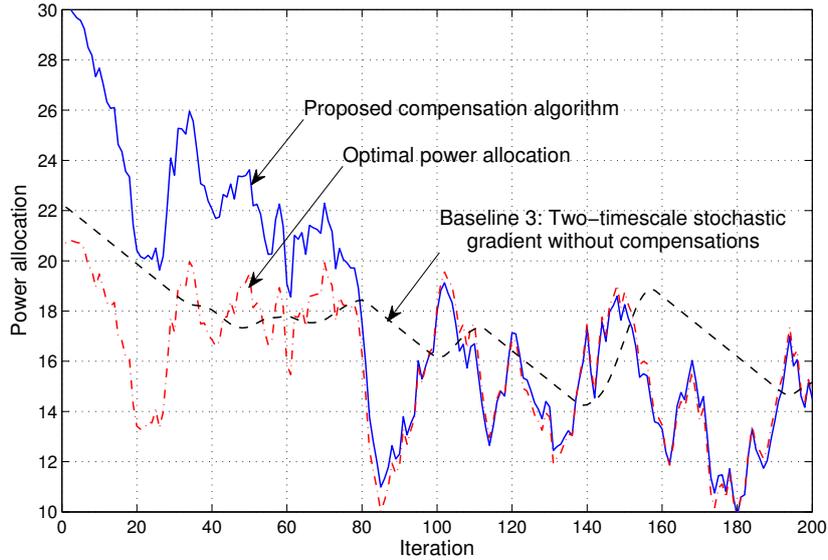}
\par\end{centering}

\caption{\label{fig:sim-tracking} A snapshot of algorithm trajectories of
the proposed compensation algorithm and the stochastic gradient algorithm
without compensations under short timescale CSI fading parameter $a_{H}=10$
and long timescale parameter $\epsilon=6\times10^{-4}$. The trajectories
represent the online power allocation policy $p_{5}$. The proposed
compensation algorithm quickly converges to the optimal trajectory
of the inner iteration, while the baseline algorithm fails to track
the optimal target and yields much larger tracking errors.}
\end{figure}

Fig. \ref{fig:sim-tracking} shows a snapshot of the algorithm trajectories
of the proposed two-timescale compensation algorithm and the baseline
stochastic gradient tracking algorithm without compensations, under
short timescale CSI fading rate $a_{H}=10$ and long timescale parameter
$\epsilon=6\times10^{-4}$. The trajectories represent the online
power allocation policy $p_{5}$. The proposed compensation algorithm
quickly converges to the optimal trajectory of the inner iteration,
while the baseline algorithm fails to track the optimal target and
yields much larger tracking errors.

\section{Conclusions}

\label{sec:conclusions}

In this paper, we have analyzed the convergence behavior of a mixed
timescale cross-layer stochastic optimization driven by multi-timescale
CSI. The CSI dynamic is modeled by an auto-regressive process in the
short timescale (small-scale fading), and a mobility-driven dynamic
process in the long timescale (large-scale fading). We partitioned
the control variables into short-term control variables and long-term
control variables, and studied the convergence of the corresponding
mixed timescale stochastic iterative algorithm. We derived a VSDS
and showed that studying the algorithm convergence is equivalent to
studying the stochastic stability of the VSDS. Using Lyapunov stochastic
stability analysis, we derived a sufficient condition for the algorithm
to be stable. In addition, we derived a tracking error upper bound
in terms of the parameters of the mixed timescale CSI process. Based
on these results, we proposed an adaptive compensation algorithm for
enhancing the tracking performance. The analysis framework and the
proposed algorithms were applied to an application example in a wireless
heterogeneous network. Numerical results matched with the theoretical
insights and demonstrated significant performance gain of the proposed
compensation algorithms over the baselines.

\appendices

\section{Derivations of the Reflection Terms $z_{x}$ and $z_{y}$}

\label{app:derivation-reflection-term}

Taking a small step $\triangle t$, the ODE dynamics (\ref{eq:ode-xc})-(\ref{eq:ode-yc})
can be written as 
\begin{eqnarray*}
x_{c}(t+\triangle t) & = & x_{c}(t)+G(x_{c}(t),y_{c}(t),h^{s}(t),h^{l}(t))\triangle t+z_{x}(t)\triangle t\\
 & = & \mathcal{P}_{\mathcal{X}(y)}\left[x_{c}(t)+G(x_{c}(t),y_{c}(t),h^{s}(t),h^{l}(t))\triangle t\right]\\
y_{c}(t+\triangle t) & = & y_{c}(t)+k(y_{c}(t),h^{l}(t))\triangle t+z_{y}(t)\triangle t\\
 & = & \mathcal{P}_{\mathcal{Y}}\left[y_{c}(t)+k(y_{c}(t),h^{l}(t))\triangle t\right].
\end{eqnarray*}
Consider that the convex domains $\mathcal{X}(y)$ and $\mathcal{Y}$
can be specified by a set of constraints $\omega_{i}(x,y;h)\leq0$,
$i=1,\dots,W$, and $q_{i}(y;h^{l})\leq0$, $i=1,\dots,J$, respectively.
Then the Euclidean projections are equivalent to find the points $x_{0}(\triangle t)$
and $y_{0}(\triangle t)$, which solve the following minimization
problems 
\begin{eqnarray}
\min_{x} & \frac{1}{2}\|x-\left(x_{c}(t)+G(x_{c}(t),y_{c}(t),\centerdot)\triangle t\right)\|_{2}^{2}\label{eq:reflection-minimization-x}\\
\mbox{subject to} & \omega_{i}(x,y;\centerdot)\leq0,\qquad\forall i=1,\dots,W.\nonumber 
\end{eqnarray}
and 
\begin{eqnarray}
\min_{y} & \frac{1}{2}\|y-\left(y_{c}(t)+k(y_{c}(t),\centerdot)\triangle t\right)\|_{2}^{2}\label{eq:reflection-minimization-y}\\
\mbox{subject to} & q_{i}(y;\centerdot)\leq0,\qquad\forall i=1,\dots,J.\nonumber 
\end{eqnarray}

The corresponding Lagrange functions are given by 
\[
L^{(x)}(x,y,\lambda_{x})=\frac{1}{2}\|x-\left[x_{c}(t)+G(\centerdot)\triangle t\right]\|_{2}^{2}+\sum_{i=1}^{W}\lambda_{x,i}=\omega_{i}(x,y;\centerdot)
\]
and 
\[
L^{(y)}(y,\lambda_{y})=\frac{1}{2}\|y-\left[y_{c}(t)+k(\centerdot)\triangle t\right]\|_{2}^{2}+\sum_{i=1}^{J}\lambda_{y,i}q_{i}(y;\centerdot).
\]
The KKT condition \cite{Bertsekas:1999bs,Boyd:2004kx} for the problem
(\ref{eq:reflection-minimization-x}) on the $x$ variable is given
by 
\begin{eqnarray}
x_{0}(\triangle t)-\left[x_{c}(t)+G(x_{c}(t),y_{c}(t),\centerdot)\triangle t\right]+\sum_{i=1}^{W}\lambda_{x,i}^{*}\nabla_{x}\omega_{i}(x_{0}(\triangle t),y_{c}(t);\centerdot) & = & 0\label{eq:reflection-term-x-KKT-1}\\
\lambda_{x,i}^{*}\omega_{i}(x_{0}(\triangle t),y_{c}(t);\centerdot) & = & 0,\;\forall i.\label{eq:reflection-term-x-KKT-2}
\end{eqnarray}
Solving (\ref{eq:reflection-term-x-KKT-1})-(\ref{eq:reflection-term-x-KKT-2}),
we obtain $x_{0}(\triangle t)$. Similarly, by writing the KKT condition
for (\ref{eq:reflection-minimization-y}), we can obtain $y_{0}(\triangle t)$.
Then the reflection terms are given by 
\[
z_{x}(t)=\lim_{\triangle t\to0}\frac{x_{0}(\triangle t)-\left[x_{c}(t)+G(x_{c}(t),y_{c}(t),\centerdot)\triangle t\right]}{\triangle t}
\]
and 
\[
z_{y}(t)=\lim_{\triangle t\to0}\frac{y_{0}(\triangle t)-\left[y_{c}(t)+k(y_{c}(t),\centerdot)\triangle t\right]}{\triangle t}.
\]

\section{Sketch Proof of Theorem \ref{thm:Connection-discrete-alg-MCTS}}

\label{app:thm-connection-alg-MCTS}

We ignore the transient states for $x_{n_{s}}$ and $y_{n_{f}}$,
and just focus on their stationary states. 

For the inner iteration $x_{n_{s}}$ in (\ref{eq:alg-xn}), consider
a large enough $n_{s}$. Since $\mu_{n_{f}}$ is decreasing, we have
$\mu_{n_{f}}\ll\gamma$. From the timescale condition $a_{H}\tau\ll\gamma$
and the step size condition $\mu_{n_{f}}\ll\gamma$, the iteration
(\ref{eq:alg-xn}) finds the partial optimum $\hat{x}(y,h(n_{f}\tau))$
for each $h(t)=(h_{1}(t),\dots,h_{N}(t))$, where $h_{j}(t)=h_{j}^{l}h_{j}^{s}(t)$.
This can be shown under Assumption \ref{asm:problem-property-P0}
and \ref{asm:Property-iteration} given a sufficiently small step
size $\gamma$ \cite{Bertsekas:1999bs,Boyd:2004kx}. Note that the
partial stationary point $\hat{x}(y,h(n_{f}\tau))$ of (\ref{eq:alg-xn})
is also the partial equilibrium point $x_{c}(t)$ of (\ref{eq:ode-xc}).
Therefore, we have established that $x_{n_{s}(t)}\to x_{c}(t)$, where
$n_{s}(t)=\lfloor tN_{s}/\tau\rfloor$. 

For the outer iteration $y_{n_{f}}$ in (\ref{eq:alg-yn}), with sufficient
large $n_{f}$, we have 
\begin{eqnarray}
\mathbb{E}\left[K(x_{n_{s}-1},y_{n_{f}-1};h^{s}(n_{f}\tau),h^{l}\big|h^{l}\right] & = & \mathbb{E}\left[\Gamma_{y}\nabla_{y}F(\hat{x}(y_{n_{f}-1},\centerdot),y_{n_{f}-1};h^{s}(n_{f}\tau),h^{l})\big|h^{l}\right]\nonumber \\
 & = & \mathbb{E}\left[\Gamma_{y}\nabla_{y}\mathcal{P}_{1}(y,h^{s},h^{l})\big|h^{l}\right]\nonumber \\
 & = & \Gamma_{y}\int\nabla_{y}\mathcal{P}_{1}(y,\omega,h^{l})dF_{h^{s}}(\omega)\nonumber \\
 & = & \Gamma_{y}\nabla_{y}\int\mathcal{P}_{1}(y,\omega,h^{l})dF_{h^{s}}(\omega)\nonumber \\
 & = & \Gamma_{y}\nabla_{y}\mathbb{E}\left[\mathcal{P}_{1}(y,h^{s},h^{l})\big|h^{l}\right]\nonumber \\
 & \triangleq & \overline{K}(y,h^{l})\label{eq:K-and-overline_K}
\end{eqnarray}
where $F_{h^{s}}(\omega)$ is the cumulative distribution function
of the conditional probability $\mbox{Pr}(h^{s}=\omega\big|h^{l})=\mbox{Pr}(h^{s}=\omega)$,
and the interchange of the integration and the differentiation is
because the integration is bounded (i.e., the expectation of $F(\centerdot)$
is bounded). We take conditional expectation here because $y$ adapts
to each realization of $h^{l}$. Therefore, we have the gradient estimator
$K(\centerdot)=\overline{K}(y,h^{l})+\xi(\centerdot)$, where $\xi(\centerdot)$
is some ``noise'' and $\mathbb{E}\xi(\centerdot)=0$. Using the
stochastic approximation \cite{benveniste1990adaptive,Kushner2003vn},
$y_{n}$ converges to $y^{*}$ almost surely under the assumed step
size rule for $\mu_{n}$. 

From Assumption \ref{asm:Property-iteration}, the matrix $\nabla_{y}k(y,h^{l})=\nabla_{y}\mathbb{E}[K(\centerdot)]=\mathbb{E}\left[\nabla_{y}K(\centerdot)\right]$
is a negative definite matrix. Therefore, the system $\dot{y}_{c}=\nabla_{y}k(y_{c},h^{l})$
is asymptotically stable \cite{Khalil1996} at a unique stationary
point $y_{c}^{*}(h^{l})$, where $\dot{y}_{c}^{*}(h^{l})=\nabla_{y}k(y_{c},h^{l})=0$.
According to the definition of $k(y,h^{l})$, we have $\overline{K}(y,h^{l})\equiv k(y,h^{l})$.
This implies that the stationary point $y_{n}$ of (\ref{eq:alg-yn})
is just the equilibrium $y_{c}^{*}$ of (\ref{eq:ode-yc}), i.e.,
$y^{*}=y_{c}^{*}$. Then we have established the asymptotic result
for $\lim\sup_{t\to\infty}\mbox{Pr}\left\{ \|y_{n_{f}(t)}-y_{c}(t)\|>\eta\right\} =0$,
where $n_{f}(t)=\lfloor t/\tau\rfloor$.

\section{Proof of Lemma \ref{lem:Dynamics-moving-equilibrium}}

\label{app:pf-lem-dynamic-moving}

The above results are obtained from the implicit function theorem.
From the optimality condition $\widetilde{G}(\hat{x}_{c},y_{c},h^{s},h^{l})=0$
and the implicit function theorem, we have 
\begin{eqnarray*}
d\hat{x}_{c}(y_{c},\centerdot) & = & \widetilde{G}_{x}^{-1}(\hat{x}_{c}(y_{c},\centerdot),y_{c},h^{s},h^{l})\bigg[\widetilde{G}_{h^{s}}(\hat{x}_{c}(y_{c},\centerdot),y_{c},h^{s},h^{l})dh^{s}\\
 &  & \qquad\qquad+\widetilde{G}_{h^{l}}(\hat{x}_{c}(y_{c},\centerdot),y_{c},h^{s},h^{l})dh^{l}+\widetilde{G}_{y}(\hat{x}_{c}(y_{c},\centerdot),y,h^{s},h^{l})dy\bigg].
\end{eqnarray*}
But since $dh^{l}\ll dh^{s}$ due to the small variation of $h^{l}(t)$
(controlled by $\epsilon\ll a_{H}$), the term $G_{h^{l}}(\centerdot)dh^{l}$
is comparatively small. Ignoring this term, equation (\ref{eq:sde-moving-optimum-x})
yields.

\section{Proof of Theorem \ref{thm:Weak-convergence} }

\label{app:proof-weak-conv}

Theorem \ref{thm:Weak-convergence} can be obtained from the weak
convergence results in \cite{benveniste1990adaptive,Kushner2003vn,Buche:2002bs}.
In the following, we sketch briefly how we can apply those results. 

Recall that the algorithm is implemented on the timescale $t_{n_{f}}=n_{f}\tau$,
where $n_{f}$ is the frame index and $\tau$ is the frame duration.
To establish the VSDS, we define the virtual algorithm timescales
as follows.
\begin{lyxDefQED}
[Virtual algorithm timescale] The \emph{virtual algorithm timescale
on frame} is a mapping from the frame index $n_{f}$ to a real number
$s_{n_{f}}=n_{f}N_{s}\gamma$. The \emph{virtual algorithm timescale
on slot} is a mapping from the slot index $n_{s}$ to a real number
$\widetilde{s}_{n_{s}}=n_{s}\gamma$.
\end{lyxDefQED}

Under the virtual algorithm timescale, the iteration indices are related
as $n_{s}(s)=\lfloor s/\gamma\rfloor$ and $n_{f}(s)=\lfloor s/(N_{s}\gamma)\rfloor$.
The virtual algorithm times is just a scaled implementation time,
and their relationship is given by $t_{n_{f}}=N_{s}n_{f}\gamma\frac{\tau}{N_{s}\gamma}=s_{n_{f}}\frac{\tau}{N_{s}\gamma}$.

Accordingly, denote the virtual CSI state as $\widetilde{h}^{s}(s)$.
Since $dt_{n_{f}}=\frac{\tau}{N_{s}\gamma}ds_{n_{f}}$, the timescale
of the virtual CSI dynamics $\widetilde{h}^{s}(s)$ can be aligned
with the virtual algorithm timescale $s_{n_{f}}$ by adding a gain
parameter $\frac{\tau}{N_{s}\gamma}$ to (\ref{eq:sde-hs}) as, 
\begin{equation}
d\widetilde{h}^{s}=-\frac{1}{2}\frac{a_{H}\tau}{N_{s}\gamma}\widetilde{h}^{s}ds+\sqrt{\frac{a_{H}\tau}{N_{s}\gamma}}dW_{s}.\label{eq:sde-hs-VSDS}
\end{equation}
$\widetilde{h}^{s}(s)$ has the same shape as $h^{s}(t)$, but with
a different timescale.

The same trick applies to the long-term virtual CSI dynamics $\widetilde{h}^{l}(s)$,
as $d\widetilde{h}_{j}^{l}=-\frac{\tau}{N_{s}\gamma}c_{0}\iota D_{j}(s)^{-\iota-1}v_{j}(s)ds$,
$\forall j$. In a vector form, we have 
\begin{equation}
d\widetilde{h}^{l}=-\frac{\tau}{N_{s}\gamma}H_{L}(t)ds\label{eq:sde-hl-VSDS}
\end{equation}
where $H_{L}(t)$ is an $N\times N$ diagonal matrix, with the $j$-th
diagonal element being $c_{0}\iota D_{j}(t)^{-\iota-1}v_{j}(t)$.
The virtual CSI timescale separation parameter becomes $\widetilde{\epsilon}=\frac{\tau}{N_{s}\gamma}\epsilon$.

We use a localization method \cite{Buche:2002bs} and consider the
algorithm trajectories (\ref{eq:alg-xn})-(\ref{eq:alg-yn}) and (\ref{eq:ode-xc})-(\ref{eq:ode-yc})
start from $(x_{0},y_{0})$ at time $s=s_{0}=0$, which lies in the
neighborhood of $(x^{*}(\widetilde{h}^{s},\widetilde{h}^{l}),y^{*}(\widetilde{h}^{l}))$.
We have

\begin{eqnarray}
 &  & x_{n_{s}}-x_{c}(\widetilde{s}_{n_{s}})\nonumber \\
 & = & x_{n_{s}-1}-x_{c}(\widetilde{s}_{n_{s}-1})\nonumber \\
 &  & \quad+\gamma\bigg[G(x_{n_{s}-1},y_{n_{f}},\widetilde{h}^{s}(s_{n_{f}}),\widetilde{h}^{l}(s_{n_{f}}))+z_{x,n_{s}}-G(x_{c}(\widetilde{s}_{n_{s}-1}),y_{c}(s_{n_{f}}),\widetilde{h}^{s}(s_{n_{f}}),\widetilde{h}^{l}(s_{n_{f}}))-z_{x}(\widetilde{s}_{n_{s}})\bigg]\label{eq:app-weak-conv-xn-1}\\
 &  & \quad-\left[x_{c}(\widetilde{s}_{n_{s}})-x_{c}(\widetilde{s}_{n_{s}-1})-\gamma G(x_{c}(\widetilde{s}_{n_{s}-1}),y_{c}(s_{n_{f}}),\widetilde{h}^{s}(s_{n_{f}}),\widetilde{h}^{l}(s_{n_{f}}))-z_{x}(\widetilde{s}_{n_{s}})\right]\label{eq:app-weak-conv-xn-2}
\end{eqnarray}
where using Taylor expansion, 
\begin{eqnarray*}
\mbox{term }\eqref{eq:app-weak-conv-xn-1} & = & \gamma G_{x}\left(x_{c}(\widetilde{s}_{n_{s}-1}),y_{c}(s_{n_{f}}),\widetilde{h}^{s}(s_{n_{f}}),\widetilde{h}^{l}(s_{n_{f}})\right)\left(x_{n_{s}-1}-x_{c}(\widetilde{s}_{n_{s}-1})\right)\\
 &  & \qquad\qquad-\gamma G_{y}\left(x_{c}(\widetilde{s}_{n_{s}-1}),y_{c}(s_{n_{f}}),\widetilde{h}^{s}(s_{n_{f}}),\widetilde{h}^{l}(s_{n_{f}})\right)\left(y_{n_{f}}-y_{c}(s_{n_{f}})\right)+o(\gamma).
\end{eqnarray*}
Here, it is reasonable to consider $z_{x,n_{s}}-z_{x}(\widetilde{s}_{n_{s}})=0$,
since if the partial equilibrium $\hat{x}_{c}\in\mathring{\mathcal{X}}$,
we eventually have $z_{x,n_{s}}=z_{x}(\widetilde{s}_{n_{s}})=0$.
If $\hat{x}_{c}\in\partial\mathcal{X}$, both trajectories eventually
search along the boundary, and $z_{x,n_{s}}=z_{x}(\widetilde{s}_{n_{s}})$
for a large enough $n_{s}$. 

The term (\ref{eq:app-weak-conv-xn-2}) is just a first order Taylor
expansion of the continuous trajectory (\ref{eq:ode-xc}) at the point
$x_{c}(\widetilde{s}_{n_{s}-1})$. By taking $\widetilde{x}_{c}^{\gamma}(\widetilde{s}_{n_{s}})=\frac{1}{\sqrt{\gamma}}\left(x_{n_{s}}-x_{c}(\widetilde{s}_{n_{s}})\right)$
and $\widetilde{y}_{c}^{\gamma}(s_{n_{f}})=\frac{1}{\sqrt{\gamma}}\left(y_{n_{f}}-y_{c}(s_{n_{f}})\right)$,
we have 
\begin{eqnarray*}
\widetilde{x}_{c}^{\gamma}(\widetilde{s}_{n}) & = & \widetilde{x}_{c}^{\gamma}(0)+\gamma\sum_{j=0}^{n-1}\bigg[G_{x}\left(x_{c}(\widetilde{s}_{j}),y_{c}(s_{\underline{j}}),\widetilde{h}^{s}(s_{\underline{j}}),\widetilde{h}^{l}(s_{\underline{j}})\right)\widetilde{x}_{c}^{\gamma}(\widetilde{s}_{j})\\
 &  & \qquad\qquad\qquad+G_{y}\left(x_{c}(\widetilde{s}_{j}),y_{c}(s_{\underline{j}}),\widetilde{h}^{s}(s_{\underline{j}}),\widetilde{h}^{l}(s_{\underline{j}})\right)\widetilde{y}_{c}^{\gamma}(s_{\underline{j}})\bigg]+o(\gamma)\\
 & = & \widetilde{x}_{c}^{\gamma}(0)+\int_{0}^{\widetilde{s}_{n}}\left(G_{x}(\centerdot)\widetilde{x}_{c}^{\gamma}+G_{y}(\centerdot)\widetilde{y}_{c}^{\gamma}\right)ds+o(\gamma)
\end{eqnarray*}
where $\underline{j}\triangleq\lfloor j/N_{s}\rfloor$ frame index
of the outer iteration when the inner iteration is at the $j$-th
slot. Equivalently, 
\begin{equation}
d\widetilde{x}_{c}^{\gamma}=G_{x}(\centerdot)\widetilde{x}_{c}^{\gamma}ds+G_{y}(\centerdot)\widetilde{y}_{c}^{\gamma}ds+o(\gamma).\label{eq:app-weak-conv-dxc-gamma}
\end{equation}

Similarly, we derive the dynamic $\widetilde{y}_{c}^{\gamma}(s)$
as follows.

\begin{eqnarray}
 &  & y_{n_{f}}-y_{c}(s_{n_{f}})\nonumber \\
 & = & y_{n_{f}-1}-y_{c}(s_{n_{f}-1})\nonumber \\
 &  & \quad+N_{s}\gamma\left[N_{s}^{-1}K\left(x_{\overline{n_{f}}},y_{n_{f}-1};\centerdot\right)+z_{y,n_{f}}-N_{s}^{-1}K\left(\hat{x}_{c}(s_{n_{f}-1}),y_{c}(s_{n_{f}-1});\centerdot\right)-\hat{z}_{y}(s_{n_{f}})\right]\label{eq:app-weak-conv-yn-1}\\
 &  & \quad+\gamma\left[K\left(\hat{x}_{c}(s_{n_{f}-1}),y_{c}(s_{n_{f}-1});\centerdot\right)+\hat{z}_{y}(s_{n_{f}})-k(y_{c}(s_{n_{f}-1}),\widetilde{h}^{l}(s_{n_{f}}))-z_{y}(s_{n_{f}})\right]\label{eq:app-weak-conv-yn-2}\\
 &  & \quad-\left[y_{c}(s_{n_{f}})-y_{c}(s_{n_{f}-1})-\gamma k(y_{c}(s_{n_{f}-1}),\widetilde{h}^{l}(s_{n_{f}}))-\gamma z_{y}(s_{n_{f}})\right]\label{eq:app-weak-conv-yn-3}
\end{eqnarray}
where $\overline{n_{f}}\triangleq N_{s}n_{f}$ is the slot index of
the inner iteration when the outer iteration is at the $n_{f}$-th
frame . The\\$\mbox{term }\eqref{eq:app-weak-conv-yn-1}$ 
\begin{eqnarray*}
 & = & (N_{s}\gamma)N_{s}^{-1}\bigg[K\left(x_{\overline{n_{f}}},y_{n_{f}-1};\centerdot\right)-K\left(\hat{x}_{c}(s_{n_{f}-1}),y_{n_{f}-1};\centerdot\right)\\
 &  & \qquad\qquad+K\left(\hat{x}_{c}(s_{n_{f}-1}),y_{n_{f}-1};\centerdot\right)-K\left(\hat{x}_{c}(s_{n_{f}-1}),y_{c}(s_{n_{f}-1});\centerdot\right)\bigg]\\
 & = & (N_{s}\gamma)N_{s}^{-1}K_{x}\left(\hat{x}_{c}(s_{n_{f}-1}),y_{c}(s_{n_{f}-1});\centerdot\right)\left(x_{\overline{n_{f}}-1}-x_{c}(s_{n_{f}-1})+x_{c}(s_{n_{f}-1})-\hat{x}_{c}(s_{n_{f}-1})\right)\\
 &  & \qquad\qquad+(N_{s}\gamma)N_{s}^{-1}K_{y}\left(\hat{x}(s_{n_{f}-1}),y_{c}(s_{n_{f}-1});\centerdot\right)(y_{n_{f}-1}-y_{c}(s_{n_{f}-1}))+o(\gamma)\\
 & = & (N_{s}\gamma)N_{s}^{-1}K_{x}\left(\hat{x}_{c}(s_{n_{f}-1}),y_{c}(s_{n_{f}-1});\centerdot\right)\left(\widetilde{x}_{c}^{\gamma}+\widetilde{x}_{c}^{e,\gamma}\right)\\
 &  & \qquad\qquad+(N_{s}\gamma)N_{s}^{-1}K_{y}\left(\hat{x}_{c}(s_{n_{f}-1}),y_{c}(s_{n_{f}-1});\centerdot\right)\widetilde{y}_{c}^{\gamma}+o(\gamma)
\end{eqnarray*}
by the first order Taylor expansion of function $K(\centerdot)$ at
$(\hat{x}_{c}(s_{n_{f}-1}),y_{c}(s_{n_{f}-1}))$. Note that, since
$\triangle s=s_{n_{f}}-s_{n_{f}-1}=N_{s}\gamma$, 
\begin{equation}
\frac{1}{\sqrt{\gamma}}\mbox{term }\eqref{eq:app-weak-conv-yn-1}=N_{s}^{-1}K_{x}(\hat{x}_{c},y_{c},\centerdot)(\widetilde{x}_{c}^{\gamma}+\widetilde{x}_{c}^{e,\gamma})\triangle s+N_{s}^{-1}K_{y}(\hat{x}_{c},y_{c},\centerdot)\widetilde{y}_{c}^{\gamma}\triangle s+o(\gamma).\label{eq:app-weak-conv-yn4}
\end{equation}

Consider $y^{*}(\widetilde{h}^{l})\in\mathring{\mathcal{Y}}$ is in
the interior of the domain. (The case when $y^{*}(\widetilde{h}^{l})$
is on the boundary will be discussed later). Then it is reasonable
to consider $y_{n_{f}}\in\mathring{\mathcal{Y}}$, with probability
1. We consider the following process for the term (\ref{eq:app-weak-conv-yn-2}),
\[
S^{\gamma}(s):=\gamma\sum_{j=1}^{n_{f}(s)}\left[K(\hat{x}_{c}(s_{j}),y_{c}(s_{j});\centerdot)-k(y_{c}(s_{j});\widetilde{h}^{l})\right].
\]
Choosing a sufficiently small $\delta>0$, we have 
\[
S^{\gamma}(s+\delta)-S^{\gamma}(s)\approx\gamma\sum_{j=n_{f}(s)+1}^{n_{f}(s+\delta)}\left[K(\hat{x}_{c}(s_{j}),y;\centerdot)-k(y;\widetilde{h}^{l})\right].
\]
where $y=y_{c}(s)$. The above is an approximation since we use $y_{c}(s_{j})\approx y_{c}(s)$
for $n_{f}(s)<j\leq n_{f}(s+\delta)$. However, the approximation
is asymptotically accurate for sufficiently small $\delta$ and $\gamma$.
The central limit theorem for the state dependent process suggests
that $\hat{\Sigma}_{s,\delta}^{-1/2}\left(S^{\gamma}(s+\delta)-S^{\gamma}(s)\right)$
weakly converge to a normal random variable, where $\hat{\Sigma}_{s,\delta}$
is the covariance matrix of $S^{\gamma}(s+\delta)-S^{\gamma}(s)$
\cite{benveniste1990adaptive}. This implies that 
\begin{equation}
\frac{1}{\sqrt{\gamma}}\left(S^{\gamma}(s+\delta)-S^{\gamma}(s)\right)\to\int_{s}^{s+\delta}\widetilde{\Sigma}_{s,\delta}^{\frac{1}{2}}dW_{u}\label{eq:app-weak-conv-yn-5}
\end{equation}
where $W_{u}$ is a standard Winner process and 
\begin{eqnarray*}
\widetilde{\Sigma}(y_{c}(s_{0});\widetilde{h}^{l}) & = & \gamma\sum_{j=-\infty}^{+\infty}\mbox{cov}\left[K\left(\hat{x}_{c}(s_{j}),y_{c}(s_{j});\widetilde{h}(s_{j})\right),K\left(\hat{x}_{c}(s_{0}),y_{c}(s_{0});\widetilde{h}(s_{0})\right)\big|\widetilde{h}^{l}\right]
\end{eqnarray*}
is the covariance matrix of the estimator $K(\centerdot)$ under the
virtual long-term CSI state $\widetilde{h}^{l}$. In addition, we
have 
\begin{eqnarray*}
\hat{\Sigma}(y_{c};\widetilde{h}^{l}) & \triangleq & \sum_{j=-\infty}^{+\infty}\mbox{cov}\left[K\left(\hat{x}_{c}(t_{j}),y_{c}(t_{j});h(t_{j})\right),K\left(\hat{x}_{c}(t_{0}),y_{c}(t_{0});h(t_{0})\right)\big|\widetilde{h}^{l}\right]\\
 & = & \frac{N_{s}}{\tau}\widetilde{\Sigma}(y_{c};\widetilde{h}^{l})
\end{eqnarray*}
to be the covariance matrix of the estimator $K(\centerdot)$ under
the CSI state $h$, since the virtual algorithm timescale $s_{n}$
is $\frac{\tau}{N_{s}\gamma}$ times denser than the implementation
timescale $t_{n}$. Therefore, from (\ref{eq:app-weak-conv-yn4})
and (\ref{eq:app-weak-conv-yn-5}), we obtain 
\[
d\widetilde{y}_{c}^{\gamma}=N_{s}^{-1}K_{x}(\hat{x}_{c},y_{c},\centerdot)(\widetilde{x}_{c}^{\gamma}+\widetilde{x}_{c}^{e,\gamma})ds+N_{s}^{-1}K_{y}(\hat{x}_{c},y_{c},\centerdot)\widetilde{y}_{c}^{\gamma}ds+\frac{\tau}{N_{s}}\hat{\Sigma}^{\frac{1}{2}}(y_{c};\widetilde{h}^{l})dW_{s}+o(\gamma).
\]

Note that, in a finite horizon case for $s\in[0,T_{s}]$, by letting
$\gamma\to0$, one can drop the $o(\gamma)$ term and obtain the convergence
results $(\widetilde{x}_{c}^{\gamma},\widetilde{y}_{c}^{\gamma})\to(\widetilde{x}_{c},\widetilde{y}_{c})$,
where $(\widetilde{x}_{c},\widetilde{y}_{c})$ is the solution to
(\ref{eq:sde-t_xc})-(\ref{eq:sde-t_yc}). In addition, using the
sophisticated techniques in \cite{benveniste1990adaptive,Kushner2003vn},
one can further prove the convergence in the infinite horizon case
for $s\in[0,\infty)$ and obtain the following,
\begin{eqnarray}
d\widetilde{x}_{c} & = & G_{x}(x_{c},y_{c},\widetilde{h}^{s},\widetilde{h}^{l})\widetilde{x}_{c}ds+G_{y}(x_{c},y_{c},\widetilde{h}^{s},\widetilde{h}^{l})\widetilde{y}_{c}ds\label{eq:sde-t_xc}\\
d\widetilde{y}_{c} & = & N_{s}^{-1}K_{x}(\hat{x}_{c},y_{c},\centerdot)(\widetilde{x}_{c}+\widetilde{x}_{c}^{e})ds+N_{s}^{-1}K_{y}(\hat{x}_{c},y_{c},\centerdot)\widetilde{y}_{c}ds\label{eq:sde-t_yc}\\
 &  & \qquad+\sqrt{\tau N_{s}^{-1}}\Sigma^{\frac{1}{2}}(y_{c},\widetilde{h}^{l})dW_{s}+dZ_{y}.\nonumber 
\end{eqnarray}
where $\Sigma^{\frac{1}{2}}(y_{c},\widetilde{h}^{l})=\hat{\Sigma}^{\frac{1}{2}}(y_{c},\widetilde{h}^{l})$
in the case for $y^{*}(\widetilde{h}^{l})\in\mathring{\mathcal{Y}}$.

Moreover, changing the MCTS (\ref{eq:ode-xc})-(\ref{eq:ode-yc})
into to the virtual algorithm time $s$, we get 
\begin{eqnarray}
dx_{c} & = & G(x_{c},y_{c,}\widetilde{h}^{s},\widetilde{h}^{l})ds+dZ_{x}\label{eq:ode-t_xc}\\
dy_{c} & = & N_{s}^{-1}k(y_{c},\widetilde{h}^{l})ds+dZ_{y}\label{eq:ode-t_yc}
\end{eqnarray}
where $dZ_{x}$ and $dZ_{y}$ are reflection terms. Notice that $d\widetilde{x}_{c}^{e}=dx_{c}-d\hat{x}_{c}$
and $d\widetilde{y}_{c}^{e}=dy_{c}-dy^{*}$. We obtain the following
error dynamic system 
\begin{eqnarray}
d\widetilde{x}_{c}^{e} & = & G(x_{c},y_{c},\widetilde{h}^{s},\widetilde{h}^{l})ds+dZ_{x}+\widetilde{G}_{x}^{-1}\widetilde{G}_{h^{s}}(\centerdot)d\widetilde{h}^{s}+\widetilde{G}_{x}^{-1}\widetilde{G}_{y}(\centerdot)dy_{c}\label{eq:sde-t_xe}\\
d\widetilde{y}_{c}^{e} & = & N_{s}^{-1}k(y_{c},\widetilde{h}^{l})ds+dZ_{y}-\psi_{h^{l}}(\widetilde{h}^{l})d\widetilde{h}^{l}\label{eq:sde-t_ye}
\end{eqnarray}

Consider the SDEs (\ref{eq:sde-t_xc})-(\ref{eq:sde-t_yc}) and (\ref{eq:sde-t_xe})-(\ref{eq:sde-t_ye}),
and the virtual CSI dynamics (\ref{eq:sde-hs-VSDS})-(\ref{eq:sde-hl-VSDS}).
Rearranging the terms and changing the virtual time notation $s$
to $t$, we obtain the VSDS in (\ref{eq:VSDS}). This proves the claimed
results.
\begin{remrk}
[The case $y^{*}(\widetilde{h}^{l})$ on the boundary] When $y^{*}(\widetilde{h}^{l})$
is on the boundary, we can follow the argument in \cite{Buche:2002bs}
to find out the behavior of $\widetilde{y}_{c}(s)$. Note that the
corresponding effect is only on the diffusion term $\Sigma^{\frac{1}{2}}(y_{c},\widetilde{h}^{l})dW_{s}$,
where $\Sigma(y_{c},\widetilde{h}^{l})=\Sigma_{0}(y_{c})\hat{\Sigma}(y_{c},\widetilde{h}^{l})$,
and $\Sigma_{0}(y)=\mbox{diag}\left(\{\sigma_{i}^{0}(y)\}_{i=1}^{N_{y}}\right)$,
which is defined in the following. Consider the $i$-th component
of $y^{*}(\widetilde{h}^{l})$ is on the boundary. There are two cases.
Case i), the $i$-th component of the drift $k(y^{*};h^{l})$ is non-zero,
which means there must be a reflection force $z_{y}^{(i)}\neq0$ on
$y_{n_{f}}^{(i)}$ and $y_{c}^{(i)}$ to keep them from reaching out
of the boundary. Then obviously, upon reaching $y^{*(i)}$, $y_{n_{f}}^{(i)}$
is not likely to be disturbed by the noise from the estimator $K(\centerdot)$
{[}unless the noise is larger than the drift $k^{(i)}(y^{*};h^{l})${]},
and the $i$-th component of the diffusion term $\Sigma^{\frac{1}{2}}(y_{c},\widetilde{h}^{l})dW_{s}$
should be zero. Hence $\sigma_{i}^{0}=0$. Case ii), the $i$-th component
of the drift $k(y^{*};h^{l})$ is zero, which means the reflection
force $z_{y}^{(i)}$ depends on the disturbance noise. According to
\cite{Buche:2002bs}, we can simply consider $\sigma_{i}^{0}=1$,
just as the case when $y^{*}(\widetilde{h}^{l})$ is in the interior
of $\mathcal{Y}$. ~\hfill\IEEEQED
\end{remrk}

\section{Proof of Theorem \ref{thm:sufficient-cond-stability-alg}}

\label{app:pf-thm-sufficient-cond}

\subsection{The Lyapunov Drift of the VSDS}

It is equivalent to show the VSDS $u(t)$ is stochastically stable.
We first give the following lemma.
\begin{lyxLemQED}
[Ito's lemma]\label{lem:lyapunov-drift} Consider a stochastic process
$u(t)$ given by the following SDE, $du=f(u)dt+g(u)dW_{t}$ and a
function $V(u)\in\mathbb{R}_{+}$. The Lyapunov drift operator on
$V(\centerdot)$ can be written as $\widetilde{\mathcal{L}}V=\frac{\partial V}{\partial u}f(u)+\mbox{tr}\left[g(u)^{T}\frac{\partial^{2}V}{\partial uu^{T}}g(u)\right].$
\end{lyxLemQED}

We first consider that the optimal solution $y^{*}$ is in the interior
of $\mathcal{Y}$, which means the reflection term $dZ_{y}\equiv0$
and the matrix $\Sigma^{0}$ (defined in Appendix \ref{app:proof-weak-conv})
is an identity matrix. Then using Lemma \ref{def:lyapunov-drift},
the Lyapunov drift of the stochastic process $u(t)$ can be written
as 
\begin{eqnarray}
\widetilde{\mathcal{L}}V(u) & = & \left[\begin{array}{c}
\widetilde{x}_{c}\\
\widetilde{y}_{c}
\end{array}\right]^{T}\left[\begin{array}{cc}
G_{x}(x_{c},y_{c},\centerdot) & G_{y}(x_{c},y_{c},\centerdot)\\
N_{s}^{-1}K_{x}(\hat{x}_{c},y_{c},\centerdot) & N_{s}^{-1}K_{y}(\hat{x}_{c},y_{c},\centerdot)
\end{array}\right]\left[\begin{array}{c}
\widetilde{x}_{c}\\
\widetilde{y}_{c}
\end{array}\right]\label{eq:LV-z(t)}\\
 &  & \qquad+\widetilde{y}_{c}^{T}N_{s}^{-1}K_{x}(\hat{x}_{c},y_{c},\centerdot)\widetilde{x}_{c}^{e}+\left(\widetilde{x}_{c}^{e}\right)^{T}G(x_{c},\centerdot)+\left(\widetilde{x}_{c}^{e}\right)^{T}G_{x}^{-1}G_{h^{s}}(-\frac{1}{2})\frac{a_{H}\tau}{N_{s}\gamma}\widetilde{h}^{s}\nonumber \\
 &  & \qquad+\left(\widetilde{x}_{c}^{e}\right)^{T}G_{x}^{-1}G_{y}N_{s}^{-1}k(y_{c},\widetilde{h}^{l})+\left(\widetilde{y}_{c}^{e}\right)^{T}N_{s}^{-1}k(y_{c},\widetilde{h}^{l})\nonumber \\
 &  & \qquad+\frac{\tau}{N_{s}\gamma}\left(\widetilde{y}_{c}^{e}\right)^{T}\psi_{h^{l}}(\widetilde{h}^{l})H_{L}-\frac{1}{2}\frac{a_{H}\tau}{N_{s}\gamma}\left(\widetilde{h}^{s}\right)^{T}\widetilde{h}^{s}+\frac{1}{2}\mbox{tr}\left(\tau N_{s}^{-1}\Sigma(y_{c})\right)\nonumber \\
 &  & \qquad+\frac{1}{2}\mbox{tr}\left[\frac{a_{H}\tau}{N_{s}\gamma}\left(G_{x}^{-1}G_{h^{s}}\right)^{T}\left(G_{x}^{-1}G_{h^{s}}\right)+\frac{a_{H}\tau}{N_{s}\gamma}\mathbf{I}\right]\nonumber 
\end{eqnarray}
where $\hat{x}_{c}$ denotes the partial optimum $\hat{x}_{c}(y_{c},h^{s},h^{l};t)$.

\subsection{The Upper Bound of the Drift}

We bound each term from the above equation in the following. 

First of all, from Assumption \ref{asm:Property-iteration}, we can
show that 
\begin{eqnarray}
Q_{c} & \triangleq & [\begin{array}{cc}
\widetilde{x}_{c}^{T} & \widetilde{y}_{c}^{T}\end{array}]\left[\begin{array}{cc}
G_{x}(x_{c},y_{c},h) & G_{y}(x_{c},y_{c}h)\\
N_{s}^{-1}K_{x}(x_{c},y_{c},h) & N_{s}^{-1}K_{y}(x_{c},y_{c},h)
\end{array}\right]\left[\begin{array}{c}
\widetilde{x}_{c}\\
\widetilde{y}_{c}
\end{array}\right]\nonumber \\
 & \leq & -N_{s}^{-1}\alpha(\|\widetilde{x}_{c}\|^{2}+\|\widetilde{y}_{c}\|^{2}).\label{eq:pf-Qc-1}
\end{eqnarray}

Secondly, the mapping $G(x_{c},\centerdot)$ on the $x$ part has
the property 
\begin{eqnarray}
(\widetilde{x}_{c}^{e})^{T}G(x_{c,}\centerdot) & = & (\widetilde{x}_{c}^{e})^{T}\int_{0}^{1}G_{x}(x_{c}^{*}+\xi\widetilde{x}_{c}^{e},\centerdot)\widetilde{x}_{c}^{e}d\xi\nonumber \\
 & \leq & -\alpha_{x}\|\widetilde{x}_{c}^{e}\|^{2}.\label{eq:app-lem-conv-static-h-ay-xe}
\end{eqnarray}

Thirdly, the mean mapping $k(y_{c},\widetilde{h}^{l})$ on the $y$
part satisfies 
\begin{eqnarray}
(\widetilde{y}_{c}^{e})^{T}k(y_{c},\widetilde{h}^{l}) & = & (\widetilde{y}_{c}^{e})^{T}\lim_{n\to\infty}\mathbb{E}\left[K(y_{c},\widetilde{h}^{l})\right]\nonumber \\
 & = & \lim_{n\to\infty}\mathbb{E}\left[(\widetilde{y}_{c}^{e})^{T}\int_{0}^{1}K_{y}(y_{c}^{*}+\xi\widetilde{y}_{c}^{e},\widetilde{h}^{l})\widetilde{y}_{c}^{e}d\xi\right]\nonumber \\
 & \leq & \lim_{n\to\infty}\mathbb{E}\left[\int_{0}^{1}-\alpha_{y}\|\widetilde{y}_{c}^{e}\|^{2}d\xi\right]\nonumber \\
 & = & -\alpha_{y}\|\widetilde{y}_{c}^{e}\|^{2}.\label{eq:app-lem-conv-static-h-ay-ye}
\end{eqnarray}

In addition, since $K(y^{*},\widetilde{h}^{l})=\mathbf{0}$ due to
the property of the stationary point $y^{*}\in\mathring{\mathcal{Y}}$,
we have
\begin{eqnarray}
\|k(y_{c},\widetilde{h}^{l})\| & = & \lim_{n\to\infty}\mathbb{E}\|K(y_{c},\widetilde{h}^{l})\|\nonumber \\
 & = & \lim_{n\to\infty}\mathbb{E}\|K(y_{c},\widetilde{h}^{l})-K(y_{c}^{*},\widetilde{h}^{l})\|\nonumber \\
 & \leq & \lim_{n\to\infty}\mathbb{E}l_{y}\|y_{c}-y_{c}^{*}\|\nonumber \\
 & = & l_{y}\|\widetilde{y}_{c}^{e}\|\label{eq:app-k-bound}
\end{eqnarray}
where the inequality is from the Lipschitz property in Assumption
\ref{asm:Property-iteration}. 

Finally, using the above result, we can find an upper bound for the
Lyapunov drift in (\ref{eq:LV-z(t)}) as 
\begin{eqnarray}
\mathcal{\widetilde{\mathcal{L}}}V & \leq & -N_{s}^{-1}\alpha\left(\|\widetilde{x}_{c}\|^{2}+\|\widetilde{y}_{c}\|^{2}\right)+\frac{l_{x}}{N_{s}}\|\widetilde{y}_{c}\|\|\widetilde{x}_{c}^{e}\|-\alpha_{x}\|\widetilde{x}_{c}^{e}\|^{2}+\frac{1}{2}v_{H}\|\widetilde{x}_{c}^{e}\|\|\widetilde{h}^{s}\|\frac{a_{H}\tau}{N_{s}\gamma}\label{eq:LV-upper-bound-1}\\
 &  & \qquad+\frac{v_{y}l_{y}}{N_{s}}\|\widetilde{x}_{c}^{e}\|\|\widetilde{y}_{c}^{e}\|-\frac{\alpha_{y}}{N_{s}}\|\widetilde{y}_{c}^{e}\|^{2}+\frac{\epsilon\tau}{N_{s}\gamma}\varpi\|\widetilde{y}_{c}^{e}\|-\frac{1}{2}\frac{a_{H}\tau}{N_{s}\gamma}\|\widetilde{h}^{s}\|^{2}+C_{0}(y^{*}(\widetilde{h}^{l}))\nonumber \\
 & \triangleq & \phi(\chi)\label{eq:LV-upper-bound-2}
\end{eqnarray}
where $C_{0}(y^{*}(\widetilde{h}^{l}))=\frac{\tau}{2N_{s}}\mbox{tr}\left(\Sigma(y^{*})\right)+\frac{1}{2}\frac{a_{H}\tau}{N_{s}\gamma}v_{H}^{2}N+\frac{1}{2}\frac{a_{H}\tau}{N_{s}\gamma}N$
is from the trace terms in (\ref{eq:LV-z(t)}). The parameter $N$
is the dimension of the parameters $h^{s}$ and $h^{l}$. $\phi:\chi\mapsto\mathbb{R}$
is the \emph{upper bound drift function} and $\chi=(\|\widetilde{x}_{c}\|,\|\widetilde{y}_{c}\|,\|\widetilde{x}_{c}^{e}\|,\|\widetilde{y}_{c}^{e}\|,\|\widetilde{h}^{s}\|)$
is the vector measuring the deviations of each virtual state from
the origin. 

Note that $\phi(\chi)$ is a quadratic function and we can write it
as 
\[
\phi(\chi)=-\chi^{T}A\chi+b^{T}\chi+C_{0}
\]
where 
\begin{equation}
A=\left[\begin{array}{ccccc}
\frac{\alpha}{N_{s}} &  &  & \dots & 0\\
 & \frac{\alpha}{N_{s}} & -\frac{l_{x}}{2N_{s}} &  & \vdots\\
 & -\frac{l_{x}}{2N_{s}} & \alpha_{x} & -\frac{v_{y}l_{y}}{2N_{s}} & -\frac{v_{H}a_{H}\tau}{4N_{s}\gamma}\\
\vdots &  & -\frac{v_{y}l_{y}}{2N_{s}} & \frac{a_{y}}{N_{s}} & 0\\
0 & \dots & -\frac{v_{H}a_{H}\tau}{4N_{s}\gamma} & 0 & \frac{1}{2}\frac{a_{H}\tau}{N_{s}\gamma}
\end{array}\right]\label{eq:Lyapunov-upper-bound-drfit-matrix-A}
\end{equation}
$b=[0,0,0,\frac{\epsilon\varpi\tau}{N_{s}\gamma},0]^{T}$ and $C_{0}=\frac{1}{2N_{s}}\left[\tau\mbox{tr}\left(\Sigma(y^{*})\right)+\frac{a_{H}\tau}{\gamma}N(1+v_{H}^{2})\right]$.

According to Lemma \ref{lem:stoch-stability-region-z}, a sufficient
condition for the VSDS $u(t)$ to be mean square stable is that the
function $\widetilde{\mathcal{L}}V(u)\leq\phi(\chi)$ can be further
upper bounded by $\phi(\chi)\leq-f(\chi)+C_{f}$, where $f(\chi)=c\chi^{T}\chi$,
for some constant $c>0$ and $C_{f}<\infty$. This is equivalent to
verifying if the function $\phi(\chi)+f(\chi)=-\chi^{T}(A-c\mathbf{I})\chi+b^{T}\chi+C_{0}$
is bounded above. Therefore, we only need to check the positive definite
property of the coefficient matrix $A$. To do this, we can calculate
each of the leading principle minors of $A$, and make them positive.
These calculations lead to the sufficient condition (\ref{eq:rho-sufficient-condition})
in Theorem \ref{thm:sufficient-cond-stability-alg}.

\begin{remrk}
[The case $y^{*}(\widetilde{h}^{l})$ on the boundary] When the
optimal solution $y^{*}$ is on the boundary of the constraint domain,
we may have non-zero $dZ_{y}$ and a non-identity matrix $\Sigma_{0}(y_{c})$.
We may have the following two modifications in the above flow. (i)
The term $\widetilde{y}_{c}^{T}\left[K_{x}(\centerdot)\left(\widetilde{x}_{c}+\widetilde{x}_{c}^{e}\right)+K_{y}(\centerdot)\widetilde{y}_{c}\right]$
in (\ref{eq:LV-z(t)}) and (\ref{eq:pf-Qc-1}) now becomes $\widetilde{y}_{c}^{T}\left[K_{x}(\centerdot)\left(\widetilde{x}_{c}+\widetilde{x}_{c}^{e}\right)+K_{y}(\centerdot)\widetilde{y}_{c}+dZ_{y}\right]$.
Since the term about the $x$ variable $K_{x}(\centerdot)\left(\widetilde{x}_{c}+\widetilde{x}_{c}^{e}\right)$
does not contribute to the reflection $dZ_{y}$, we only need to evaluate
$\widetilde{y}_{c}^{T}\left[K_{y}(\centerdot)\widetilde{y}_{c}+dZ_{y}\right]$.
Note that, when there is a non-zero reflection on the $i$-th component
of $y_{c}$, we must have $\widetilde{y}_{c}^{(i)}=0$ for most of
the time. Thus we still have $\widetilde{y}_{c}^{T}\left[K_{y}(\centerdot)\widetilde{y}_{c}+dZ_{y}\right]\widetilde{y}_{c}\leq-\alpha_{y}\|\widetilde{y}_{c}\|^{2}$,
as we did in (\ref{eq:pf-Qc-1}). (ii) Consider the covariance matrix
in (\ref{eq:sde-t_yc}). Its trace must be smaller than the case when
$y^{*}(\widetilde{h}^{l})$ is in the interior, since there are some
zero diagonal elements in the matrix $\Sigma_{0}(y^{*})$ {[}see Appendix
\ref{app:proof-weak-conv}{]}. Thus the constant $C_{0}$ defined
above is still an upper bound. Combining the cases (i) and (ii), the
optimal solution $y^{*}$ being on the boundary does not change the
upper bound of the drift as in (\ref{eq:LV-upper-bound-1}).~\hfill\IEEEQED
\end{remrk}

\section{Proof of Theorem \ref{thm:upper-bound-error}}

\label{app:thm-upper-bound}

Choose a function $f(\chi)=c\chi^{T}A\chi$ for some constant $0<c<1$.
We have $f(\chi)\geq c\lambda_{\min}(A)\|\chi\|^{2}$, where $\lambda_{\min}(A)$
denotes the smallest eigenvalue of matrix $A$. 

Under the sufficient condition in Theorem \ref{thm:sufficient-cond-stability-alg}
and from (\ref{eq:LV-upper-bound-2}), we have the stochastic Lyapunov
drift be upper bounded by 
\[
\widetilde{\mathcal{L}}V(u)\leq\phi(\chi(u))=-\chi^{T}A\chi+b^{T}\chi+C_{0}\leq-f(\chi)+C_{1}
\]
where $C_{1}=C_{0}+\frac{C_{b}}{1-c}$, and $C_{b}=\frac{1}{4}b^{T}A^{-1}b=\left(\frac{\epsilon\varpi\tau}{N_{s}\gamma}\right)^{2}\frac{4N_{s}^{-1}\alpha\alpha_{x}-N_{s}^{-2}l_{x}^{2}-N_{s}^{-2}l_{y}^{2}v_{y}^{2}}{N_{s}^{-1}\alpha(8\alpha_{x}-a_{H}v_{H}^{2})-2N_{s}^{-2}l_{x}^{2}-2N_{s}^{-2}l_{y}^{2}v_{y}^{2}}<\infty$. 

Then, using Lemma \ref{lem:stoch-stability-region-z}, we have 
\begin{eqnarray*}
\lim\sup_{t\to\infty}\frac{1}{t}\int_{0}^{t}\mathbb{E}\|u(\tau)\|^{2}d\tau & \leq & \frac{1}{c\lambda_{\min}(A)}\times\lim\sup_{t\to\infty}\frac{1}{t}\int_{0}^{t}\left(C_{0}+\frac{C_{b}}{1-c}\right)d\tau\\
 & = & \frac{1}{c\lambda_{\min}(A)}\left(\overline{C}_{0}+\frac{C_{b}}{1-c}\right)
\end{eqnarray*}
where $\overline{C}_{0}=\frac{1}{2N_{s}}\left[\tau\overline{\Sigma}+\frac{a_{H}\tau}{\gamma}N(1+v_{H}^{2})\right]$,
and $\overline{\Sigma}=\lim\sup_{t\to\infty}\frac{1}{t}\int_{0}^{t}\mbox{tr}\left(\Sigma(y^{*}(h^{l}(\tau)))\right)d\tau$
is the time-averaged covariance matrix of the estimator $K(\centerdot)$.
Choosing $c$ to minimize the above upper bound, we obtain $c^{*}=\frac{\overline{C}_{0}+C_{b}}{2\overline{C}_{0}}$.
With the observation that $C_{b}(\epsilon^{2})$ can typically be
very small, due to the small timescale separation parameter $\epsilon\ll1$.
Then we choose $c=\frac{1}{2}$ for a reasonable tight upper bound
of the tracking error.

We now derive the term $\lambda_{\min}(A)$, by using the eigenvalue
lower bound $\lambda_{\min}(A)\geq\frac{\left|\det\left(A\right)\right|}{2^{n/2-1}\|A\|_{F}}$
from \cite{Piazza:2002oq}, where $\|\centerdot\|_{F}$ denotes the
Frobenius norm and $n$ is the dimension of $A$. We apply a trick
to obtain a good eigenvalue bound by letting $A_{0}=N_{s}A$. Then
$\lambda_{\min}(A)=N_{s}^{-1}\lambda_{\min}(A_{0})\geq N_{s}^{-1}\frac{\left|\det\left(A_{0}\right)\right|}{2^{3/2}\|A_{0}\|_{F}}=N_{s}^{-1}\frac{\rho}{\eta}$,
where $\rho=\left|\det\left(A_{0}\right)\right|=\frac{\alpha a_{H}\tau}{16\gamma}\left(8N_{s}\alpha\alpha_{x}\alpha_{y}-\alpha\alpha_{y}\frac{a_{H}\tau}{\gamma}v_{H}^{2}-2\alpha_{y}l_{x}^{2}-2\alpha l_{y}^{2}v_{y}^{2}\right)=\mathcal{O}(N_{s}\alpha^{2}\alpha_{x}\alpha_{y})$
and $\eta=2^{3/2}\|A_{0}\|_{F}=\mathcal{O}\left(\sqrt{N_{s}^{2}\alpha_{x}^{2}+\alpha^{2}}\right)$
. Therefore, we have 
\begin{eqnarray*}
\lim\sup_{t\to\infty}\frac{1}{t}\int_{0}^{t}\mathbb{E}\|u(\tau)\|^{2}d\tau & \leq & \frac{1}{\frac{1}{2}N_{s}^{-1}\frac{\rho}{\eta}}\left(\frac{1}{2N_{s}}\left[\tau\overline{\Sigma}+\frac{a_{H}\tau}{\gamma}N(1+v_{H}^{2})\right]+C_{b}(\epsilon^{2})\right)\\
 & = & \frac{\eta}{\rho}\left(\tau\overline{\Sigma}+C\right)
\end{eqnarray*}
where $C=\frac{a_{H}\tau}{\gamma}N(1+v_{H}^{2})+4C_{b}(\epsilon)=\frac{a_{H}\tau}{\gamma}N(1+v_{H}^{2})+\mathcal{O}(\epsilon^{2}\varpi^{2}\tau^{2}\gamma^{-2})$.

Since $\|\widetilde{x}_{c}\|^{2}+\|\widetilde{x}_{c}^{e}\|^{2}+\|\widetilde{y}_{c}\|^{2}+\|\widetilde{y}_{c}^{e}\|^{2}\leq\|u\|^{2}$,
we have $e_{x}+e_{y}\leq\frac{\eta}{\rho}\left(\frac{1}{2}\overline{\Sigma}+C\right)$.

\section{Proof of Theorem \ref{thm:conv-comp}}

\label{app:thm-conv-comp}

Using the compensated MCTS (\ref{eq:comp-xc})-(\ref{eq:comp-yc})
and the CSI dynamics (\ref{eq:sde-hs})-(\ref{eq:sde-hl}), the dynamic
system (\ref{eq:sde-comp-x})-(\ref{eq:sde-comp-y}) can be written
as 
\begin{eqnarray}
d\widetilde{x}_{c}^{e} & = & G(x_{c,}y_{c};\centerdot)dt+\left(\varphi_{x}^{h}(\centerdot)-\hat{\varphi}_{x}^{h}(\centerdot)\right)(-\frac{a_{H}\tau}{2N_{s}\gamma}\widetilde{h}^{s})dt+\left(\varphi_{x}^{y}(\centerdot)-\hat{\varphi}_{x}^{y}(\centerdot)\right)N_{s}^{-1}k(y_{c},\widetilde{h}^{l})dt\nonumber \\
 &  & \qquad+\left(\varphi_{x}^{y}(\centerdot)-\hat{\varphi}_{x}^{y}(\centerdot)\right)\hat{\varphi}_{y}^{h}(y_{c},\widetilde{h}^{s},\widetilde{h}^{l})d\widetilde{h}^{l}+\left(\varphi_{x}^{h}(\centerdot)-\hat{\varphi}_{x}^{h}(\centerdot)\right)\sqrt{\frac{a_{H}\tau}{N_{s}\gamma}}dW_{t}\label{eq:app-thm-conv-comp-sde-x}\\
d\widetilde{y}_{c}^{e} & = & N_{s}^{-1}k(y_{c},\widetilde{h}^{l})dt+\left(\varphi_{y}^{h}(\widetilde{h}^{l})-\hat{\varphi}_{y}^{h}(y_{c},\widetilde{h}^{s},\widetilde{h}^{l})\right)d\widetilde{h}^{l}\label{eq:app-thm-conv-comp-sde-y}
\end{eqnarray}

Consider two Lyapunov functions $V_{1}(\widetilde{x}_{c}^{e})=\frac{1}{2}\left(\widetilde{x}_{c}^{e}\right)^{T}\widetilde{x}_{c}^{e}$
and $V_{2}(\widetilde{y}_{c}^{e})=\frac{1}{2}(\widetilde{y}_{c}^{e})^{T}\widetilde{y}_{c}^{e}$
defined along the trajectory of the virtual system (\ref{eq:app-thm-conv-comp-sde-x})-(\ref{eq:app-thm-conv-comp-sde-y}). 

Define $\xi_{y}^{h}(y_{c},\widetilde{h}^{s},\widetilde{h}^{l})\triangleq\|\varphi_{y}^{h}(\widetilde{h}^{l})-\hat{\varphi}_{y}^{h}(y_{c},\widetilde{h}^{s},\widetilde{h}^{l})\|-\mathbb{E}\|\varphi_{y}^{h}(\widetilde{h}^{l})-\hat{\varphi}_{y}^{h}(y_{c},\widetilde{h}^{s},\widetilde{h}^{l})\|$.
Note that $\xi_{y}^{h}(\centerdot)$ acts like a ``noise'' term
that depends on $\widetilde{h}^{s}$ and satisfies $\mathbb{E}\xi_{y}^{h}(\centerdot)=0$.
Using Lemma \ref{lem:lyapunov-drift}, the Lyapunov drift of $V_{2}(\centerdot)$
is given by 
\begin{eqnarray}
\widetilde{\mathcal{L}}V_{2}(\centerdot) & = & (\widetilde{y}_{c}^{e})^{T}N_{s}^{-1}k(y_{c},\widetilde{h}^{l})+(\widetilde{y}_{c}^{e})^{T}\left(\varphi_{y}^{h}(\widetilde{h}^{l})-\hat{\varphi}_{y}^{h}(y_{c},\widetilde{h}^{s},\widetilde{h}^{l})\right)\frac{d\widetilde{h}^{l}}{dt}\nonumber \\
 & \leq & -\frac{\alpha_{y}}{N_{s}}\|\widetilde{y}_{c}^{e}\|^{2}+\|\widetilde{y}_{c}^{e}\|\left(\hat{L}_{y}^{h}\|\widetilde{y}_{c}^{e}\|+\beta_{y}^{h}+\xi_{y}^{h}(\centerdot)\right)\frac{\tau}{N_{s}\gamma}\epsilon(D_{\min},v_{\max})\nonumber \\
 & = & -\left(\widetilde{\alpha}_{y}-\widetilde{\epsilon}\hat{L}_{y}^{h}\right)\|\widetilde{y}_{c}^{e}\|^{2}+\widetilde{\epsilon}\beta_{y}^{h}\|\widetilde{y}_{c}^{e}\|+\widetilde{\epsilon}\xi_{y}^{h}(\centerdot)\|\widetilde{y}_{c}^{e}\|\nonumber \\
 & \leq & -c\|\widetilde{y}_{c}^{e}\|+\widetilde{\epsilon}\xi_{y}^{h}(\centerdot)\|\widetilde{y}_{c}^{e}\|+\frac{(\widetilde{\epsilon}\beta_{y}^{h}+c)^{2}}{4(\widetilde{\alpha}_{y}-\widetilde{\epsilon}\hat{L}_{y}^{h})}
\end{eqnarray}
where $\epsilon(D_{\min},v_{\max})=2D_{\min}^{-\iota-1}v_{\max}$,
$\widetilde{\alpha}_{y}$ and $\widetilde{\epsilon}=\frac{\tau}{N_{s}\gamma}\epsilon$
according to the mobility model and CSI model in Section \ref{sub:two-timescale-control},
$c>0$ is an arbitrary constant, and the first inequality is from
(\ref{eq:app-lem-conv-static-h-ay-ye}).

Notice that $\mathbb{E}\left[\widetilde{\epsilon}\xi_{y}^{h}(\centerdot)\|\widetilde{y}_{c}^{e}\|+\frac{(\widetilde{\epsilon}\beta_{y}^{h}+c)^{2}}{4(\widetilde{\alpha}_{y}-\widetilde{\epsilon}\hat{L}_{y}^{h})}\right]=\frac{(\widetilde{\epsilon}\beta_{y}^{h}+c)^{2}}{4(\widetilde{\alpha}_{y}-\widetilde{\epsilon}\hat{L}_{y}^{h})}$.
Using Lemma \ref{lem:stoch-stability-region-z}, we obtain the upper
bound of the tracking error $\|\widetilde{y}_{c}^{e}\|$ as 
\begin{equation}
\mathbb{E}\|\widetilde{y}_{c}^{e}\|\leq\frac{(\widetilde{\epsilon}\beta_{y}^{h}+c)^{2}}{4c(\widetilde{\alpha}_{y}-\widetilde{\epsilon}\hat{L}_{y}^{h})}.\label{eq:app-thm-conv-comp-ye-0}
\end{equation}
A tight upper bound can be given by minimizing (\ref{eq:app-thm-conv-comp-ye-0})
over $c>0$. Then we obtain 
\begin{equation}
\mathbb{E}\|\widetilde{y}_{c}^{e}\|\leq\frac{\widetilde{\epsilon}\beta_{y}^{h}}{\widetilde{\alpha}_{y}-\widetilde{\epsilon}\hat{L}_{y}^{h}}=\frac{\epsilon\tau\beta_{y}^{h}}{\alpha_{y}\gamma-\epsilon\tau\hat{L}_{y}^{h}}.\label{eq:app-thm-conv-comp-ye-1}
\end{equation}

To study $V_{1}(\centerdot)$, define $\xi_{h}=\|\widetilde{h}^{s}\|-\mathbb{E}\|\widetilde{h}^{s}\|$,
where $\mathbb{E}\|\widetilde{h}^{s}\|=\sqrt{\frac{2}{\pi}}$ according
to the CSI model in (\ref{eq:sde-hs}). Define $\xi_{y}=\|\widetilde{y}_{c}^{e}\|-\mathbb{E}\|\widetilde{y}_{c}^{e}\|$.
Note that $\xi_{h}$ and $\xi_{y}$ only depend on $\widetilde{h}^{s}$
and $\widetilde{y}_{c}^{e}$, respectively, and $\mathbb{E}\xi_{h}=\mathbb{E}\|\widetilde{y}_{c}^{e}\|=0$.

We then derive the Lyapunov drift of $V_{1}(\centerdot)$ as follows,
\begin{eqnarray}
\widetilde{\mathcal{L}}V_{1}(\centerdot) & \approx & (\widetilde{x}_{c}^{e})^{T}G(\centerdot)+(\widetilde{x}_{c}^{e})^{T}\left(\varphi_{x}^{h}(\centerdot)-\hat{\varphi}_{x}^{h}(\centerdot)\right)\left(-\frac{a_{H}\tau}{2N_{s}\gamma}\widetilde{h}^{s}\right)+(\widetilde{x}_{c}^{e})^{T}\left(\varphi_{x}^{y}(\centerdot)-\hat{\varphi}_{x}^{y}(\centerdot)\right)N_{s}^{-1}k(y_{c},\widetilde{h}^{l})\nonumber \\
 &  & \qquad+\left(\varphi_{x}^{y}(\centerdot)-\hat{\varphi}_{x}^{y}(\centerdot)\right)\hat{\varphi}_{y}^{h}(y_{c},\centerdot)d\widetilde{h}_{l}+\mbox{tr}\left[\left(\varphi_{x}^{h}(\centerdot)-\hat{\varphi}_{x}^{h}(\centerdot)\right)\frac{a_{H}\tau}{N_{s}\gamma}\left(\varphi_{x}^{h}(\centerdot)-\hat{\varphi}_{x}^{h}(\centerdot)\right)^{T}\right]\nonumber \\
 & \leq & -\alpha_{x}\|\widetilde{x}_{c}^{e}\|^{2}+\hat{L}_{x}^{h}\|\widetilde{x}_{c}^{e}\|^{2}\frac{a_{H}\tau}{2N_{s}\gamma}(\sqrt{\frac{2}{\pi}}+\xi_{h})+\hat{L}_{x}^{y}\|\widetilde{x}_{c}^{e}\|^{2}N_{s}^{-1}l_{y}(\mathbb{E}\|\widetilde{y}_{c}^{e}\|+\xi_{y})\nonumber \\
 &  & \qquad+\frac{a_{H}\tau}{N_{s}\gamma}\left(\hat{L}_{x}^{h}\right)^{2}\|\widetilde{x}_{c}^{e}\|^{2}\label{eq:app-thm-conv-comp-LV1-1}\\
 & \leq & -\left(\alpha_{x}-\frac{a_{H}\tau}{\sqrt{2\pi}N_{s}\gamma}\hat{L}_{x}^{h}-\hat{L}_{x}^{y}N_{s}^{-1}l_{y}\frac{\epsilon\tau\beta_{y}^{h}}{\alpha_{y}\gamma-\epsilon\tau\hat{L}_{y}^{h}}-\frac{a_{H}\tau}{N_{s}\gamma}\left(\hat{L}_{x}^{h}\right)^{2}\right)\|\widetilde{x}_{c}^{e}\|^{2}\nonumber \\
 &  & \qquad+\hat{L}_{x}^{h}\|\widetilde{x}_{c}^{e}\|^{2}\frac{a_{H}\tau}{2N_{s}\gamma}\xi_{h}+\hat{L}_{x}^{y}\|\widetilde{x}_{c}^{e}\|^{2}l_{y}\xi_{y}\label{eq:app-thm-conv-comp-LV1-2}
\end{eqnarray}
(c.f. equations (\ref{eq:app-lem-conv-static-h-ay-xe}) and (\ref{eq:app-k-bound})
for the first inequality), where in (\ref{eq:app-thm-conv-comp-LV1-1}),
the $d\widetilde{h}^{l}$ term is dropped, since $d\widetilde{h}^{l}$
is much smaller than the $d\widetilde{h}^{s}$.

Note that the last two terms in (\ref{eq:app-thm-conv-comp-LV1-2})
have mean $0$ due to the ``noise'' terms $\xi_{h}$ and $\xi_{y}$.
Therefore, according to Lemma \ref{lem:stoch-stability-region-z},
if $\alpha_{x}-\frac{a_{H}\tau}{\sqrt{2\pi}N_{s}\gamma}\hat{L}_{x}^{h}-\hat{L}_{x}^{y}N_{s}^{-1}l_{y}\frac{\epsilon\tau\beta_{y}^{h}}{\alpha_{y}\gamma-\epsilon\tau\hat{L}_{y}^{h}}-\frac{a_{H}\tau}{N_{s}\gamma}\left(\hat{L}_{x}^{h}\right)^{2}>0$,
then $\widetilde{x}_{c}^{e}$ converges to $0$ in probability.

\section{Proof of Theorem \ref{thm:ex-convergence-speed}}

\label{app:pf-thm-ex-conv-speed}

We have the iteration mappings $G(\centerdot)=\left(\begin{array}{cc}
\frac{\partial}{\partial\mathbf{p}}L(\centerdot), & -\frac{\partial}{\partial\lambda}L(\centerdot)\end{array}\right)$ and $K(\centerdot)=\frac{\partial}{\partial\mathbf{r}}L(\centerdot)$.
In addition, we have $G_{x}(\centerdot)=M_{L}(\centerdot)=\left(\begin{array}{cc}
\frac{\partial^{2}}{\partial\mathbf{p}\partial\mathbf{p}}L(\centerdot) & \frac{\partial^{2}}{\partial\mathbf{p}\partial\lambda}L(\centerdot)\\
-\frac{\partial^{2}}{\partial\lambda\partial\mathbf{p}}L(\centerdot) & \mathbf{0}_{J\times J}
\end{array}\right)$ and $K_{y}(\centerdot)=\frac{\partial^{2}}{\partial\mathbf{r}\partial\mathbf{r}}L(\centerdot)$.
From the convex assumption, $K_{y}(\centerdot)$ is negative definite.
It follows that $\alpha_{y}(\centerdot)=-\lambda_{\max}\left(\frac{\partial^{2}}{\partial\mathbf{r}\partial\mathbf{r}}L(\centerdot)\right)$.

Note that, the Hessian of the Lagrange function $\frac{\partial^{2}}{\partial\mathbf{p}\partial\mathbf{p}}L(\centerdot)$
is negative definite. Then according to \cite[Proposition 4.4.2]{Bertsekas:1999bs},
the matrix $M_{L}(\centerdot)$ is also negative definite.

To derive the convergence for the $x$ part, we consider a Lyapunov
function $V(x_{e})=\frac{1}{2}x_{e}^{T}x_{e}$, where $x_{e}=x-x^{*}$.
From a modification of Ito's formula in Lemma \ref{lem:lyapunov-drift},
it follows that $\mathcal{\widetilde{L}}V=\frac{1}{2}(x_{e}^{T}dx_{e}+dx_{e}^{T}x_{e})=\frac{1}{2}\left(x_{e}^{T}G(\centerdot)+G(\centerdot)^{T}x_{e}\right)\leq\frac{1}{2}\left(x_{e}^{T}G_{x}x_{e}+x_{e}^{T}G_{x}x_{e}\right)=\frac{1}{2}x_{e}^{T}\left(G_{x}+G_{x}^{T}\right)x_{e}\leq-\lambda_{\max}(\frac{1}{2}(G_{x}+G_{x}^{T}))$.

\bibliographystyle{IEEEtran}
\bibliography{My_Reference}

\end{document}